\documentclass[prd,aps,showpacs,tightenlines, 10pt, reprint,nofootinbib,superscriptaddress,floatfix,twocolumn]{revtex4-2}

\usepackage[utf8]{inputenc} 
\usepackage{acronym}
\usepackage{amsmath}
\usepackage{amsthm}  
\usepackage{amssymb}
\usepackage{mathtools}
\usepackage{phaistos}
\usepackage{graphicx}
\usepackage[section]{placeins}
\usepackage{orcidlink}
\usepackage{physics} 
\usepackage{IEEEtrantools}  
\usepackage{xspace}
\usepackage{cancel}
\usepackage[normalem]{ulem} 
\usepackage{array}
\newcolumntype{C}[1]{>{\centering\let\newline\\\arraybackslash\hspace{0pt}}m{#1}}
\usepackage{cellspace}
\usepackage{comment}
\usepackage{epsfig}
\usepackage{multirow} 
\usepackage{color}
\usepackage[export]{adjustbox} 
\usepackage{floatrow}
\usepackage{float}
\usepackage[font=small,labelfont=bf,justification=Justified]{caption}
\usepackage{subcaption}
\usepackage{makecell}
\usepackage[titles]{tocloft}
\usepackage{titlesec}
\usepackage{verbatim}
\usepackage{csquotes}

\newfloatcommand{capbtabbox}{table}[][\FBwidth]
\usepackage{tabularx}
\usepackage{units}
\usepackage[T1]{fontenc}
\usepackage[normalem]{ulem} 
\usepackage{slashed}
\usepackage{bbold}
\usepackage[absolute,overlay]{textpos}

\hypersetup{
    colorlinks,
    linkcolor={red!60!black},
    citecolor={blue!50!black},
    urlcolor={blue!80!black}
}
\allowdisplaybreaks

\advance\cftsecnumwidth 0.5em\relax
\advance\cftsubsecindent 0.5em\relax
\advance\cftsubsecnumwidth 0.5em\relax

\newcommand{\rmi}[1]{{\mbox{\scriptsize #1}}}
\newcommand{\rmii}[1]{{\mbox{\tiny\rm{#1}}}}
\newcommand{\MPl}{M_\mathrm{Pl}}
\newcommand{\Neff}{N_\mathrm{eff}}
\newcommand{\MeV}{\mathrm{MeV}}
\newcommand{\GeV}{\mathrm{GeV}}

\newcommand{\rsc}{r_\rmi{sc}}
\newcommand{\rsci}{r_{\rmi{sc},i}}
\newcommand{\fphi}{f_\phi}
\newcommand{\mphi}{m_\phi}

\newcommand{\GammaphitoSM}{\Gamma_{\phi\to\gamma\gamma}}

\newcommand{\Tosc}{T_\rmi{osc}}
\newcommand{\Hosc}{H_\rmi{osc}}
\newcommand{\Hoscaa}{H_\rmi{osc,aa}}
\newcommand{\rscmin}{r_\rmi{sc}^\rmi{min}}

\newcommand{\Toscaa}{T_\rmi{osc,aa}}
\newcommand{\Omegaphiosc}{\Omega_{\phi,\mathrm{osc}}}
\newcommand{\Omegaphioscaa}{\Omega^{\mathrm{aa}}_{\phi,\mathrm{osc}}}
\newcommand{\rhophiosc}{\rho_{\phi,\mathrm{osc}}}
\newcommand{\rhophistar}{\rho_{\phi,\star}}
\newcommand{\rhophiafter}{\rho_{\phi}^\rmii{after}}
\newcommand{\rhoradafter}{\rho_\rmii{rad}^\rmii{after}}
\newcommand{\rhoradmd}{\rho_\rmii{rad}^\rmii{md}}
\newcommand{\rhophimd}{\rho_\phi^\rmii{md}}

\newcommand{\rhorad}{\rho_\rmii{rad}}
\newcommand{\mD}{m_\rmii{D}}
\newcommand{\astar}{a_\star}
\newcommand{\aosc}{a_\rmi{osc}}
\newcommand{\adecay}{a_\rmi{rh}}
\newcommand{\amd}{a_\rmi{md}}

\newcommand{\taustar}{\tau_\star}
\newcommand{\tauosc}{\tau_\rmi{osc}}
\newcommand{\epssup}{\epsilon_\rmi{sup}}
\newcommand{\Omegaphistar}{\Omega_{\phi,\star}}
\newcommand{\Omegaphitoday}{\Omega_{\phi,0}}
\newcommand{\OmegaGWtildestar}{\tilde{\Omega}_{\rmii{GW},\star}}

\newcommand{\ktildestar}{\tilde{k}_\star}
\newcommand{\ftildeGWstar}{\tilde{f}_{\rmii{GW},\star}}

\newcommand{\OmegaGWTildeToday}{\tilde{\Omega}_{\rmii{GW},0}}
\newcommand{\alphamin}{\alpha_\rmi{min}}
\newcommand{\Trh}{T_\rmi{rh}}

\newcommand{\gepsrh}{g_{\epsilon}^\rmi{rh}}
\newcommand{\gsrh}{g_{s}^\rmi{rh}}
\newcommand{\gepsstar}{g_{\epsilon}^\star}

\newcommand{\Hrh}{H_\rmi{rh}}

\newcommand{\kclose}{k_\rmi{close}}

\newcommand{\TBBN}{T_\rmii{BBN}}
\newcommand{\HBBN}{H_\rmii{BBN}}

\titleformat{\section}
  {\centering\normalfont\fontsize{10}{15}\bfseries}{\thesection}{1em}{}
\titleformat{\subsection}
  {\centering\normalfont\fontsize{10}{15}\bfseries}{\thesubsection}{1em}{}

\begin{document}

\begin{textblock*}{\textwidth}[1,0](200mm,10mm)
\noindent
\flushright
\text{MITP-25-028}
\end{textblock*}

\vspace*{0.1cm}
\title{Supercooled Audible Axions}
\author{Christopher Gerlach}
\email{cgerlach@uni-mainz.de} 
\affiliation{PRISMA$^+$ Cluster of Excellence \& Mainz Institute for Theoretical Physics, Johannes Gutenberg-Universität Mainz, 55099 Mainz, Germany}

\author{Daniel Schmitt}
\email{dschmitt@itp.uni-frankfurt.de} 
\affiliation{Institute for Theoretical Physics, Goethe University, 60438 Frankfurt am Main, Germany}

\author{Pedro Schwaller}
\email{pedro.schwaller@uni-mainz.de} 
\affiliation{PRISMA$^+$ Cluster of Excellence \& Mainz Institute for Theoretical Physics, Johannes Gutenberg-Universität Mainz, 55099 Mainz, Germany}

\date{\today}
\begin{abstract}
In the audible axion mechanism, axion-like particles source primordial gravitational waves via their coupling to a dark Abelian gauge field.
The original setup, however, relies on a large axion decay constant and coupling to produce sizable signals.
In this article, we show that
delaying the onset of axion oscillations opens up the testable parameter space and reduces the required coupling to $\alpha \gtrsim 1$.
Furthermore, we investigate the emission of gravitational waves via the axion coupling to the Standard Model photon in the presence of Schwinger pair production, generating a strong signal in the $\mu$Hz or ultra-high frequency range.
Cosmological constraints and gravitational wave projections are provided for both scenarios. 
\end{abstract}
\maketitle

\section{Introduction}

The axion, originally introduced to solve the strong CP problem via the Peccei-Quinn mechanism \cite{Peccei:1977hh,Peccei:1977ur,Weinberg:1977ma,Wilczek:1977pj}, has become a popular dark matter (DM) candidate \cite{Preskill:1982cy,Abbott:1982af,Dine:1982ah}. While the QCD axion parameter space is subject to strong constraints, 
its generalized brother---the axion-like particle (ALP)---remains less restricted, as its decay constant and mass are independent model parameters.
Additional motivation for axions or ALPs arises from string theory \cite{Witten:1984dg,Svrcek:2006yi,Arvanitaki:2009fg,Marsh:2015xka} and models of inflation \cite{Freese:1990rb,Pajer:2013fsa,Adshead:2015pva,Daido:2017wwb,Takahashi:2021tff,Domcke:2019qmm,Kitajima:2021bjq}. While there is a wide range of ongoing axion or ALP searches \cite{OHare:2024nmr}, gravitational waves (GWs) may provide new insights.

The first detection of gravitational waves from astrophysical origin \cite{LIGOScientific:2016aoc} initiated unbroken interest in possible GW sources. Due to their almost undisturbed propagation, gravitational waves offer a valuable glimpse into the early Universe. 
While it remains an unsettled question whether the background detected by pulsar timing array (PTA) collaborations \cite{NANOGrav:2020bcs,NANOGrav:2021flc,Goncharov:2021oub,EPTA:2021crs,Antoniadis:2022pcn,
NANOGrav:2023gor,EPTA:2023fyk,Reardon:2023gzh,Xu:2023wog,
Miles:2024seg} 
is of astrophysical or cosmological origin, many more observatories in a much wider range of frequencies have been proposed to search for signatures from the early Universe. It appears to be a matter of time that some signal (or its absence) is able to constrain early Universe cosmology.

ALPs can generate a stochastic GW background via tachyonic instabilities which source a resonant production of 
photons. 
The considered models 
can be differentiated into inflation and inflationary preheating scenarios \cite{Anber:2009ua,Anber:2012du,Barnaby:2012xt,Domcke:2016bkh,
Greene:1997fu,Kofman:1994rk,Shtanov:1994ce,Kofman:1997yn,Dufaux:2007pt,Maleknejad:2016qjz,Figueroa:2016wxr,Adshead:2018doq,Cuissa:2018oiw,Cyncynates:2023zwj,Cyncynates:2024yxm}, and scenarios where the resonance appears after inflation
\cite{Machado:2018nqk,Chatrchyan:2020pzh,Salehian:2020dsf,  Namba:2020kij,Kite:2020uix,Kitajima:2020rpm,Ratzinger:2020oct,Banerjee:2021oeu,Madge:2021abk,Eroncel:2022vjg,Madge:2023dxc,Su:2025nkl,Hook:2016mqo,Agrawal:2017eqm}. The latter mechanism has been dubbed \textit{audible axion} and has further been investigated in the context of kinetic misalignment~\cite{Co:2019jts,Chang:2019tvx,Co:2021rhi,Madge:2021abk} and relaxion trapping~\cite{Banerjee:2021oeu}.\footnote{Other studies focus on a similar mechanism in the context of alternative theories of gravity \cite{Xu:2024kwy} or more generic oscillating scalars \cite{Ramberg:2022irf,Cui:2023fbg}.}

In the original mechanism \cite{Machado:2018nqk,Machado:2019xuc,Ratzinger:2020oct}, sizable GWs can only be sourced for very large decay constants and a large coupling to a dark gauge field. In this article, we consider delayed evolution of the ALP, inducing a finite period of supercooling in the primordial Universe.
This opens up parameter space for the audible axion due to the decreased Hubble parameter at the onset of oscillations. 
The reduced oscillation temperature furthermore allows for efficient production of Standard Model~(SM) photons and corresponding GW signals, eliminating the necessity for a dark gauge field.

This paper is structured as follows. In the subsequent secs.~\ref{sec:Standard_mechanism} and~\ref{sec:effect_supercooling}, we first review the original setup and introduce the effect of supercooling on the audible axion mechanism. 
In sec.~\ref{sec:dark_photon}, we revisit the production of dark gauge modes and compute the associated GW spectra. In sec.~\ref{sec:SM_photon}, we extend our analysis to the SM photon.

\subsection{The standard audible axion}\label{sec:Standard_mechanism}
The basic setup consists of an axion-like particle (ALP) $\phi$ interacting with a $U(1)$ gauge field $X$ in an expanding spacetime~\cite{Machado:2018nqk,Madge:2021abk}:
\begin{align}
    {\cal S}  = \int &d^4x \sqrt{-g}\left[ \frac{1}{2}\partial_\mu \phi \partial^\mu \phi - V(\phi) \right. \notag\\ 
     &\left. - \frac{1}{4}X_{\mu\nu}X^{\mu\nu} - \frac{\alpha}{4 f_\phi} \phi X_{\mu\nu}\tilde{X}^{\mu\nu} \right],
\end{align}
with ALP decay constant $f_\phi$, dimensionless coupling $\alpha$, potential $V(\phi)$ and 
the field strength and dual field strength tensors $X_{\mu\nu}$ and $\tilde{X}_{\mu\nu}$, respectively. 

Using conformal time $dt = a(\tau) d\tau$, the line element can be written as 
$ds^2 = a(\tau)^2(d\tau^2 - \delta_{ij} dx^i dx^j)$, where $a(\tau)$ is the scale factor. The Hubble rate in conformal coordinates reads $H = a'/a^2$, where the prime denotes derivatives with respect to conformal time $\tau$. Accordingly, we express all momenta in comoving coordinates in the following.

The original audible axion scenario~\cite{Machado:2018nqk,Machado:2019xuc} considers a potential 
\begin{align}
    V(\phi) & = \mphi^2 f_\phi^2 \left( 1 - \cos \left( \frac{\phi}{f_\phi} \right) \right) \,, 
\end{align}
with minimum at $\phi=0$ where $\phi$ has the mass $\mphi$. The field is initially displaced from the origin at $\phi_i = \theta f_\phi$, with an equation of motion 
\begin{align}\label{eq:axion_eom}
    \phi'' + 2 a H \phi' + a^2 \frac{\partial V}{\partial \phi} & = \frac{\alpha}{f_\phi} a^2 \Vec{E}\cdot \Vec{B}\,,
\end{align}
where $\Vec{E}$ and $\Vec{B}$ are the electric and magnetic fields associated with $X_{\mu\nu}$. Hubble friction keeps the field at $\phi_i$ until $\Hoscaa \sim \mphi$, when it starts to oscillate around the origin. During radiation domination~(RD), this happens at a temperature 
\begin{align}\label{eq:T_osc,aa}
    \Toscaa & \equiv \left(\frac{90}{\pi^2 g^\mathrm{osc,aa}_{\epsilon}}\right)^\frac{1}{4} \sqrt{\mphi \MPl}\,,
\end{align}
where $\MPl = 2.4\times 10^{18}~{\rm GeV}$ is the reduced Planck scale and $g^\mathrm{osc,aa}_{\epsilon}$ denotes the number of relativistic degrees of freedom in the thermal bath.
The initial relative energy density of the ALP then reads 
\begin{equation} \label{eq:Omega_phi_osc_aa}
    \Omegaphioscaa = \frac{V(\phi_i)}{3 \Hoscaa^2 \MPl^2} \approx \frac{1}{6} \left(\frac{\theta \fphi}{\MPl}\right)^2 \, ,
\end{equation}
where we have expanded for small misalignment angles, $\rho_{\phi,\rmi{osc}}^{\rmi{aa}} \approx \theta^2 \mphi^2 \fphi^2 /2$.

Once the ALP starts to oscillate, it modifies the dispersion relation of the gauge field. The latter is usually~\cite{Machado:2018nqk,Madge:2021abk} decomposed into modes of definite spatial momentum $\Vec{k}$, whose occupation numbers are given by mode functions $v_\pm(k,\tau)$ for helicities $\pm$, which only depend on $k = |\Vec{k}|$ due to spatial homogeneity. They satisfy an equation of motion~\cite{Machado:2018nqk}
\begin{align}\label{eq:dark_photon_eom}
    v_\pm ''(k,\tau) + \omega^2_\pm (k,\tau) v_\pm(k,\tau) & = 0\,,
\end{align}
with a frequency that is affected by the evolution of $\phi$:
\begin{align}\label{eq:dark_photon_dispersion}
    \omega_\pm^2(k,\tau) = k^2 \mp k \frac{\alpha}{f_\phi}\phi'(\tau)\,.
\end{align}
As explained in detail in~\cite{Machado:2018nqk}, the frequencies can now become imaginary for a range of momenta when $\phi' \neq 0$, causing exponential growth of the occupation numbers of the corresponding modes. The result is a rapid transfer of energy from the ALP to the dark photon (see also \cite{Agrawal:2017eqm}), and the amplification of small quantum fluctuations into macroscopic spatial anisotropies, which source gravitational waves. 

As discussed here, the scenario is rather minimal, depending mainly on the parameters $\mphi$ and $\fphi$, which control the frequency and amplitude of the GW signal, and the coupling $\alpha$, which is required to be rather large, $\alpha \gtrsim 20$, for GW production to be efficient. The GW amplitude is ultimately proportional to the square of the fraction of the total energy density that is carried by the ALPs at $\Tosc$, which leads to a $(\fphi/\MPl)^4$ scaling of the signal. Thus only values of $\fphi$ close to the Planck scale yield observable GWs. Furthermore it is worth noting that the SM photon would not experience a tachyonic instability in the above setup, since its dispersion relation is affected by the presence of a plasma of charged particles, prohibiting efficient tachyonic growth.\footnote{The SM photon always exhibits a tachyonic band, however, the growth rate is suppressed by its Debye mass $\sim eT$. We return to this issue in sec.~\ref{sec:SM_photon}.}

In this work, we explore the possibility of delaying the onset of ALP evolution to temperatures below $\Tosc$, in order to expand the parameter space of the audible axion scenario. This will allow us to
\begin{itemize}
    \item obtain observable GWs for smaller decay constants $\fphi$,
    \item reduce the required magnitude of $\alpha$ to $\alpha \gtrsim 1$, 
    \item consider the scenario where the SM photon plays the role of the gauge field $X_\mu$. 
\end{itemize}

\subsection{Effect of supercooling}\label{sec:effect_supercooling}
To parametrize the delay of oscillations in a model independent way, we introduce the temperature ratio
\begin{equation}\label{eq:r_sc}
    \rsc \equiv \frac{\Tosc}{\Toscaa}\, ,
\end{equation}
where the subscript \enquote{aa} denotes quantities in the original setup discussed above, and $\Tosc$ is the temperature where the ALP actually starts to evolve. While we will mostly remain agnostic regarding the specific mechanism that realizes this delay, we demonstrate in appendix~\ref{app:trapped_misalignment} that huge temperature ratios can be easily realized in models of trapped misalignment~\cite{Higaki:2016yqk,Kawasaki:2017xwt,Nakagawa:2020zjr,DiLuzio:2021pxd,DiLuzio:2021gos,Kitajima:2023pby,DiLuzio:2024fyt}. 

Eq.~\eqref{eq:r_sc} quantifies the amount of supercooling undergone by the axion.
We directly note the enhancement of the relative axion energy density at the onset of oscillation,
\begin{equation}\label{eq:axion_energy_density_supercooling}
    \Omegaphiosc = \rsc^{-4}\,\Omegaphioscaa \,\frac{g^\mathrm{osc,aa}_{\epsilon}}{g^\mathrm{osc}_{\epsilon}} \, ,
\end{equation}
where $\Omegaphioscaa$ is given by eq.~\eqref{eq:Omega_phi_osc_aa} and $\rhophiosc = \rhophiosc^\rmi{aa} = V(\phi_i)$.
By solving $\rhophiosc = \rhorad(T_i)$, we find the temperature ratio where the energy density of the ALP starts to exceed the one in the thermal bath,
\begin{equation} \label{eq:r_sc_thermal_inflation}
    \rsci = \left(\frac{g^\mathrm{osc,aa}_{\epsilon}}{g^\mathrm{osc}_{\epsilon}}\right)^\frac{1}{4}\left(\frac{\theta \fphi}{\sqrt{6}\MPl}\right)^\frac{1}{2} \approx \left(\frac{\fphi}{\MPl}\right)^\frac{1}{2} \, .
\end{equation}
For lower $\Tosc \leq T_i$, the axion dominates the energy density of the Universe, $\Omegaphiosc \to 1$, inducing a period of thermal inflation which ends with the onset of axion oscillations.
Hence, the Hubble parameter at the start of oscillations reads 
\begin{equation}\label{eq:Hubble_osc}
    \Hosc = \max\left\{\mphi\rsc^2 \left(\frac{g^\mathrm{osc}_{\epsilon}}{g^\mathrm{osc,aa}_{\epsilon}}\right)^\frac{1}{2}, \frac{\theta \mphi \fphi}{\sqrt{6} \MPl} \right\}\, ,
\end{equation}
where the second expression applies if the axion is the dominant energy component in the Universe.

In the following, we first apply this setup to ALPs coupled to a dark photon, where delayed oscillations open up the detectable parameter space towards smaller decay constants $\fphi$ and couplings $\alpha$.
We then extend our analysis to ALP-SM photon systems, where a 
large amount of supercooling suppresses the Debye mass of the photon to allow for efficient GW production.

\section{Dark photon case}\label{sec:dark_photon}
Let us start by studying ALPs coupled to a dark photon, which we assume to be secluded from the SM.\footnote{In the absence of couplings to the SM the dark sector does not thermalize, i.e., we assume the initial dark photon distribution to be given by the Bunch-Davies vacuum. 
All temperature scales are hence given with respect to the SM thermal bath.
If the dark sector had a finite temperature, the discussion from sec.~\ref{sec:SM_photon} would apply.}
Furthermore, we assume the dark photon to remain massless.
At the temperature $\Tosc$, the ALP starts to oscillate. At $T_\star$, most of its energy is transferred to the dark photon via the tachyonic instability, producing GWs in the process. The dark photons then behave as dark radiation, contributing to the effective relativistic degrees of freedom $\Neff$. 
The ALP scales like matter, therefore contributes to the DM abundance if its decay rate to dark photons is sufficiently small. 
If the decay rate is large, ALPs decay into dark photons before big bang nucleosynthesis~(BBN), sourcing an additional contribution to $\Neff$.

The dark photon dispersion relation~\eqref{eq:dark_photon_dispersion} dictates the range of unstable modes
\begin{equation}
    0 < k < \frac{\alpha}{\fphi} |\phi'| \, ,
\end{equation}
which is unaltered compared to the original setup without supercooling. 
The fastest growing dark photon mode carries momentum 
\begin{equation}\label{eq:dark_photon_peak_k}
    \tilde{k}_\rmi{osc} = \frac{\alpha}{2\fphi} |\phi'| = \frac{\alpha}{2} \theta \mphi \aosc = \tilde{\omega}_\rmi{osc} \, ,
\end{equation}
at the onset of ALP oscillations~\cite{Machado:2018nqk}.
The dark photon energy density grows with the square of the mode functions, hence $\rho_X \propto \exp(2|\omega|\tau)$. 
Tachyonic amplification becomes inefficient once all unstable modes have growth rates less than half of the oscillation frequency $|\omega| < a\mphi/2$. This yields
\begin{equation}\label{eq:tachyonic_band_closure}
    \frac{a_\rmi{close}}{\aosc} = (\alpha \theta)^\frac{2}{3} \, .
\end{equation}
For details, see appendix~\ref{app:minimal_alpha}.

In the following we first compute an upper bound on the amount of supercooling, before deriving cosmological constraints to ensure a consistent cosmic evolution.
Finally, we compute the GW signal in the presence of supercooling and identify the parameter space observable by future GW experiments.

\subsection{Amount of supercooling}
\label{sec:amount_of_supercooling_dark_photon}
In the case of an axion-dark photon system, eq.~\eqref{eq:r_sc_thermal_inflation} gives an immediate lower bound on $\Tosc$. 
If $\rsc < \rsci$, the Universe enters a phase of axion-driven thermal inflation.
In the absence of decay channels to the SM, the total energy density then would be dominated by either ALPs or dark photons from the time of production until today.

A stronger limit can be derived by considering the experimental bounds on the effective relativistic degrees of freedom $\Neff$. 
Dark photons free-stream from the time of production, scaling as radiation. 
Their contribution to $\Neff$ at the time of recombination reads 
\begin{equation}
    \begin{split}
        \Delta \Neff &= \frac{8}{7} \left(\frac{11}{4}\right)^\frac{4}{3} \frac{\rho_{X}}{\rho_\gamma}\Biggr|_{T=T_\mathrm{rec}}\\ &= \frac{8}{7} \left(\frac{11}{4}\right)^\frac{4}{3} \frac{g^{\star}_{\epsilon}}{g_\gamma} \left(\frac{g_{s}^0}{g_s^{\star}}\right)^\frac{4}{3} \Omega_{\phi,\star} < 0.3\, ,
    \end{split}
\end{equation}
where the limit results from the Planck 2018 dataset~\cite{Planck:2018vyg} and $\rho_\gamma$ denotes the SM photon energy density, with $g_\gamma = 2$.
In the second line, we have replaced the ratio of energy densities at the time of recombination by the ratio at the time of GW production, denoted by the subscript $\star$.
In addition, we have set $\rho_{X,\star} = \rho_{\phi,\star}$, as a substantial fraction of the axion energy density is converted into dark photons. 
This gives a conservative upper bound on the amount of supercooling.
Since the axion follows a matter-like scaling from the onset of oscillations, its abundance is enhanced relative to the radiation-dominated background.
Then, the axion energy density at the time of dark photon production reads
\begin{equation}\label{eq:Omegaphistar}
    \Omegaphistar = \Omegaphiosc \frac{\astar}{\aosc} \, ,
\end{equation}
where $\Omegaphiosc$ is given by eq.~\eqref{eq:axion_energy_density_supercooling}. 
The scale factor ratio parametrizes the time required for the dark photon modes to grow from the vacuum until $\rho_X \sim \rho_\phi$.
This is computed by considering the characteristic growth rate~\eqref{eq:dark_photon_peak_k}, relative to the comoving Hubble parameter $\aosc\Hosc$.
Relegating the explicit derivation to appendix~\ref{app:minimal_alpha}, we simply state the final result

\begin{figure*}
    \centering
    \includegraphics[width=\linewidth]{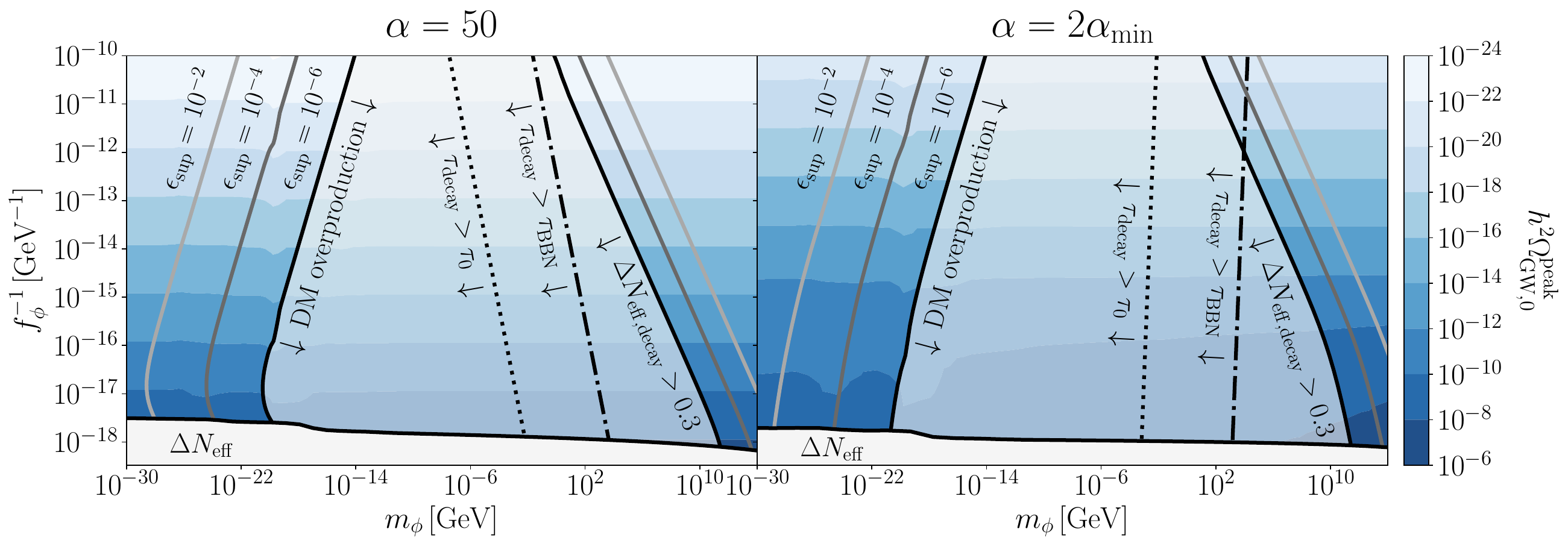}
    \caption{Overview of the available parameter space for supercooled ALP-dark photon systems, imposing $\alpha=50$ (left) and $\alpha=2\alphamin$ (right), respectively. In both panels, we set $\theta=1$. The dotted lines indicate cosmologically stable ALPs, while the dash-dotted lines signal axion decays before BBN. In the white-shaded region, dark photons violate bounds on the effective number of degrees of freedom at BBN, even in the case without supercooling. The solid lines show the parameter spaces consistent with axion dark matter~(eq.~\eqref{eq:eps_sup_DM}) and $\Neff$~(eq.~\eqref{eq:eps_sup_Neff}) constraints for a suppression of the axion abundance $\epssup \in \{10^{-2},10^{-4},10^{-6}\}$. The color contours display the GW peak amplitude~\eqref{eq:GW_peak_estimates_today}, which is significantly enhanced when decreasing $\alpha$.}
    \label{fig:darkphoton_cosmoconstraints}
\end{figure*}
\begin{equation}\label{eq:growth_time_estimate}
    \frac{\astar}{\aosc} = 1 + \frac{\pi}{\alpha \theta} \rsc^2  \ln\left(\frac{128 \pi^2}{\alpha^4 \theta^2}\frac{\fphi^2}{\mphi^2}\right) \, .
\end{equation}
In appendix~\ref{app:numerics} we show an excellent agreement between this estimate and our numerical simulations.
From this expression, we directly see that a small $\rsc$,~i.e., strong supercooling, decreases the growth time due to the suppression of the Hubble parameter.
This in turn will allow us to impose a significantly smaller $\alpha$ compared to the scenario without supercooling.
To this end, we compare eq.~\eqref{eq:growth_time_estimate} with the time when the tachyonic band closes, eq.~\eqref{eq:tachyonic_band_closure}.

Evaluating all quantities at the time of GW production, the lower limit on $\rsc$, corresponding to the maximum length of the supercooling period for ALP-dark photon systems, then reads
\begin{equation}\label{eq:r_sc_max}
    \rscmin \simeq
    1.25 \left(\frac{g_{s}^0}{g_{s}^\star}\right)^\frac{1}{3}\left(\frac{g^{\star}_{\epsilon} g^\mathrm{osc,aa}_{\epsilon}}{g_\gamma g^\mathrm{osc}_{\epsilon}} \right)^\frac{1}{4} \left(\frac{\astar}{\aosc}\right)^\frac{1}{4} \left(\frac{\theta \fphi}{\MPl}\right)^\frac{1}{2}  \,.
\end{equation}
This ensures that the bound on $\Delta \Neff$ is not violated by the dark photons produced from the tachyonic instability.
Since we are interested in the upper limit on the GW amplitude, we set $\rsc = \rscmin$ in the following.

\subsection{Cosmological constraints}\label{sec:dark_photon_cosmo_constraints}
Having imposed the amount of supercooling, let us now derive constraints on the model parameters to ensure a consistent cosmological evolution.
To this end, we divide the parameter space into two regimes: light, cosmologically stable axions that may constitute DM and heavy axions, decaying into dark photons before BBN and thus contributing to the effective relativistic degrees of freedom.
For an axion to decay before BBN, its decay rate must exceed the Hubble parameter,
\begin{equation} \label{eq:GammaPhi_larger_HBBN}
    \Gamma_{\phi\to XX} = \frac{\alpha^2 \mphi^3}{64\pi \fphi^2} > \HBBN \, .
\end{equation}

The axion energy density after dark photon production reads
\begin{equation}\label{eq:Omega_phi_after}
    \Omega_{\phi}^\rmii{after} = \epssup \Omegaphistar\,,
\end{equation}
where $\Omegaphistar$ is given by eq.~\eqref{eq:Omegaphistar} and $\epssup$ parametrizes the energy transfer to dark photons through the tachyonic resonance.
As we restrict ourselves to linearized simulations (cf.~appendix~\ref{app:numerics}), i.e., backreaction effects are not fully included~\cite{Ratzinger:2020oct}, we cannot reliably extract $\epssup$. 
Therefore we treat $\epssup$ as a free parameter, leaving a lattice analysis of this quantity for the future.

To obtain a limit on the light axion parameter space, we demand that the axion energy density, redshifted to today, does not exceed the observed DM abundance. 
In terms of the model parameters, this requires a suppression of the axion abundance of at least
\begin{equation}\label{eq:eps_sup_DM}
    \epssup^\rmii{DM} \lesssim 2.89 \,\times\, 10^{-9} \left(\frac{g_\epsilon^\star}{g_\epsilon^\rmi{osc}} \frac{\astar}{\aosc}\right)^\frac{3}{4} \left(\frac{\mathrm{eV}}{\mphi}\right)^\frac{1}{2} \left(\frac{10^{10}\,\mathrm{GeV}}{\theta \fphi}\right)^\frac{1}{2}\, .
\end{equation}
We refer to appendix~\ref{app:cosmo_constraints} for the derivation.

Regarding heavy axions, we demand that the contribution to the effective degrees of freedom through perturbative decays into dark photons is $\Delta \Neff^\rmi{decay} < 0.3$. 
This yields a limit on the required suppression of the abundance,
\begin{equation}\label{eq:eps_sup_Neff}
    \begin{split}
        \epssup^{\rmi{decay}} \lesssim 
        6.&64 \times 10^{-8} (g_\epsilon^\star)^{-\frac{1}{4}} (g_s^\star)^\frac{1}{3} \\ \times &\left(\frac{\astar}{\aosc}\right)^\frac{3}{4} \frac{\alpha}{\theta^\frac{1}{2}} \,\frac{\mphi}{\GeV}\left(\frac{10^{10}\,\GeV}{\fphi}\right)^\frac{3}{2} \, ,
    \end{split}
\end{equation}
where the derivation is again relegated to appendix~\ref{app:cosmo_constraints}. 
The size of the available parameter space is determined by the competition between the suppression of the relative axion abundance through the tachyonic instability, and the subsequent enhancement until decays into dark photons become efficient.

In fig.~\ref{fig:darkphoton_cosmoconstraints} we display the cosmological bounds. 
Here, different panels correspond to different choices of $\alpha$~(cf.~appendix~\ref{app:minimal_alpha}), while the color contours indicate the GW peak amplitude~(see below).
The dotted (dash-dotted) line denotes the regime where axions do not decay before today~(BBN), while the cosmologically viable parameter spaces are shown by the straight lines for $\epssup \in \{10^{-2},10^{-4},10^{-6}\}$.
Lattice studies of the original mechanism~\cite{Ratzinger:2020oct} imply that the typical axion suppression is $\epssup \sim \mathcal{O}(10^{-2})$ for $\alpha \sim 50-100$. 
In the supercooled case, a much smaller axion-dark photon coupling $\alpha \sim \mathcal{O}(1)$ is sufficient to retain the efficiency of the tachyonic resonance. 
We expect the reduced coupling to affect the backscattering of the gauge modes, possibly leading to $\epssup \ll 10^{-2}$.
Furthermore, additional model building can alleviate constraints from DM overproduction.
A time-varying axion mass, for instance, renders the backscattered modes relativistic after production.
As a consequence, the matter-like scaling sets in at a later time, reducing the final axion abundance.
To exemplify the impact of such effects on the parameter space, we vary $\epssup$ over a large range, relegating a lattice study to future work.

\subsection{Gravitational wave signal}\label{sec:GW_signal_darkphoton}
The exponential amplification of dark photon modes induces anisotropies in the energy-stress tensor, i.e., GWs; see ref.~\cite{Machado:2018nqk} for the full computation of the GW spectrum.
Since a full numerical study of the entire parameter space is unfeasible, we employ the parametrization~\cite{Machado:2019xuc,Madge:2021abk}
\footnote{In~\cite{Madge:2023dxc} a different parametrization, based on lattice simulations~\cite{Ratzinger:2020oct}, is proposed. They qualitatively agree, though the lattice simulations suggest a polynomial (rather than exponential) fall-off in the UV. However it is not clear whether those results can be extrapolated to small $\alpha$, where backreaction effects might be suppressed.}
\begin{equation}\label{eq:GW_fit_template}
    \Omega_{\rmii{GW},0}(f) = \mathcal{A}_s \OmegaGWTildeToday \frac{\left(\tilde{f}/f_s\right)^p}{1 + \left(\tilde{f}/f_s\right)^p \exp\left[\gamma\left(\tilde{f}/f_s - 1\right)\right]} \, ,
\end{equation}
where $\tilde{f} = f/\tilde{f}_0$.
Here, $\tilde{f}_0$ ($\OmegaGWTildeToday$) denotes the peak frequency (amplitude) today.
To fix the free parameters $\mathcal{A}_s$, $f_s$, $\gamma$, and $p$, we conduct benchmark simulations using $\mathcal{O}(10^4)$ dark photon modes and fit the GW peak region~(see appendix~\ref{app:numerics}). 
To this end, let us first derive analytical expressions for $\OmegaGWTildeToday$ and $\tilde{f}_0$.

\begin{figure*}
    \centering
    \includegraphics[width=\linewidth]{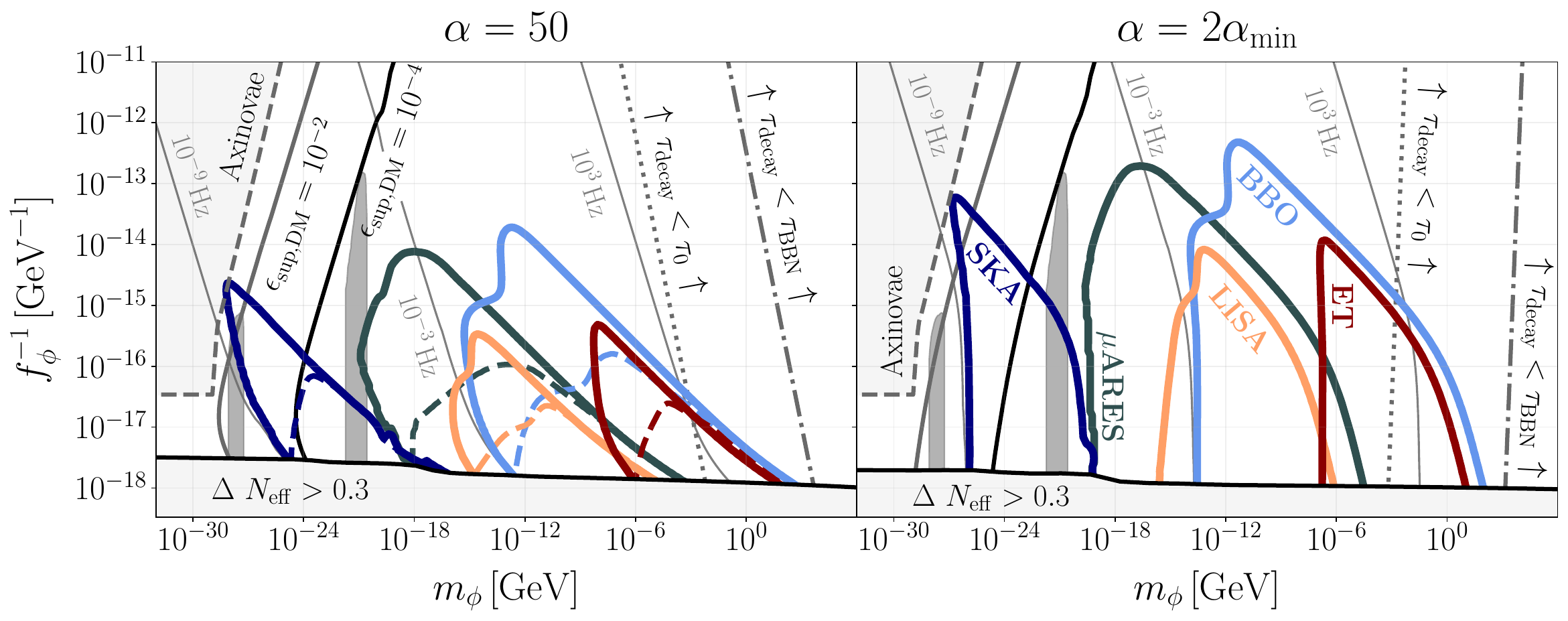}
    \caption{Observational prospects of future GW observatories for ALPs coupled to dark photons, installing $\theta=1$~(both), $\alpha=50$~(left), and $\alpha = 2\alphamin$~(right). The colored solid lines correspond to delayed ALP oscillations, while the dashed lines indicate the original setup without supercooling. Again, the dotted (dash-dotted) lines show the parameter space where the ALP is stable until today~(BBN). Constraints from black hole~(BH) superradiance~\cite{AxionLimits,Baryakhtar:2020gao,Stott:2020gjj,Hoof:2024quk,Unal:2020jiy,Witte:2024drg,Cardoso:2018tly} are displayed by the gray-shaded regions. Delayed oscillations significantly enhance the peak amplitude, extending the observable parameter range by up to two orders of magnitude in $\fphi$. In addition, supercooled oscillations allow for a smaller ALP-dark photon coupling $\alpha$, further improving detection prospects. Note, however, that most of the parameter space requires a large suppression of the ALP abundance through the tachyonic instability to avoid overclosure.}
    \label{fig:darkphoton_detectableparamspace}
\end{figure*}

The peak amplitude $\OmegaGWtildestar$ and frequency $\ftildeGWstar$ at the time of production can be estimated as~\cite{Buchmuller:2013lra,Giblin:2014gra,Machado:2018nqk,Machado:2019xuc,Madge:2021abk}
\begin{equation}\label{eq:GW_peak_estimate_production}
    \begin{gathered}
        \OmegaGWtildestar = c_\rmi{eff}^2 \, \Omegaphistar^2 \left(\frac{H_\star a_\star}{2\ktildestar}\right)^2\,,\\
        \ftildeGWstar = 2 \frac{\ktildestar}{a_\star} = \alpha\theta \mphi \left(\frac{\aosc}{\astar}\right)^\frac{3}{2} \, ,
    \end{gathered}
\end{equation}
where $\aosc/\astar$ is our growth time estimate~\eqref{eq:growth_time_estimate} and $c_\rmi{eff}$ is an efficiency parameter we absorb into $\mathcal{A}_s$. 
Employing eqs.~\eqref{eq:Hubble_osc},~\eqref{eq:dark_photon_peak_k}, and~\eqref{eq:Omegaphistar}, we can express the peak amplitude in terms of the model parameters
\begin{equation}\label{eq:peak_estimate_model_params}
    \OmegaGWtildestar \simeq 
    0.99\frac{(g_s^\star)^\frac{4}{3}}{g_\epsilon^\star} \left(\frac{\fphi}{\MPl}\right)^2 \frac{1}{\alpha^2} \, .
\end{equation}
Here, we have included the best-fit value $\mathcal{A}_s = 268.57$ (cf.~table~\ref{tab:GW_fit_parameters}) in the numerical prefactor.
We observe two differences compared to the original setup~\cite{Machado:2018nqk}.
First, the scaling of the peak amplitude changes from $\fphi^4 \to \fphi^2$ in the case of supercooled oscillations.
This results from the increased relative axion energy density $\Omegaphistar \propto \rsc^{-4}$. 
Concurrently, the tachyonic modes are shifted deeper into the horizon as the Hubble parameter decreases $\propto \rsc^2$, leaving an overall enhancement of the peak amplitude $\propto \rsc^{-4} \sim (\MPl/\fphi)^2$.

Second, the amplitude becomes independent of the misalignment angle $\theta$. 
This is expected since we impose the maximum amount of supercooling $\rscmin$ to not violate $\Delta \Neff$ constraints.
Then, the axion energy density at the time of oscillation does not depend on the initial displacement.
A smaller $\theta$ merely leads to a longer period of supercooling.
Note, however, that $\alpha$ in eq.~\eqref{eq:peak_estimate_model_params} depends implicitly on $\theta$ through the requirement that tachyonic growth occurs before the instability band closes~(see appendices~\ref{app:minimal_alpha} and \ref{app:smaller_theta}). 
Therefore, for a small $\theta \ll 1$, $\alpha$ has to be increased accordingly, effectively decreasing the peak amplitude. 

Finally, the GW signal has to be redshifted to today. 
As a non-standard cosmological evolution is prohibited in the dark photon case, we have~\cite{Kamionkowski:1993fg,Caprini:2018mtu}
\begin{equation}\label{eq:GW_peak_estimates_today}
    \begin{gathered}
        h^2 \OmegaGWTildeToday = 1.67 \times 10^{-5} \left(\frac{100}{g_\epsilon^\star}\right)^\frac{1}{3} \OmegaGWtildestar \, , \\
        \tilde{f}_0 = 1.65 \times 10^{-7} \,\mathrm{Hz} \,\frac{\ftildeGWstar}{H_\star} \frac{T_\star}{\mathrm{GeV}} \left(\frac{g_\epsilon^\star}{100}\right)^\frac{1}{6}\, ,
    \end{gathered}
\end{equation} 
where we have set $g_\epsilon^\star = g_s^\star$ for simplicity.
This gives us a simple expression for the GW peak today,
\begin{equation} \label{eq:peak_freq_today_model_params}
    \begin{gathered}
    h^2 \OmegaGWTildeToday = 7.69\times 10^{-5}\left(\frac{\fphi}{\MPl}\right)^2 \frac{1}{\alpha^2} \,, \\
    \tilde{f}_0 = 
    28.53\,\mathrm{Hz}
    \left(\frac{\aosc}{\astar}\right)^\frac{3}{4} \alpha \theta^\frac{1}{2}  
    \left(\frac{\mphi}{\mathrm{eV}}\right)^\frac{1}{2} \left(\frac{10^{10}\,\mathrm{GeV}}{\fphi}\right)^\frac{1}{2} \, ,
    \end{gathered}
\end{equation}
where we have set $g_\epsilon^\star = g^\mathrm{osc}_{\epsilon,\rmi{aa}} = g_s^\star$. Note that the above expression for the peak frequency includes our fit factor $f_s = 0.46$~(cf.~table~\ref{tab:GW_fit_parameters}).

To study the observational prospects, we employ the parametrization of the GW signal to compute signal-to-noise ratios~(SNR) at the future Square Kilometre Array~(SKA)~\cite{Janssen:2014dka}, $\mu$ARES~\cite{Sesana:2019vho}, Laser Interferometer Space Antenna~(LISA)~\cite{2017arXiv170200786A,Robson:2018ifk,LISACosmologyWorkingGroup:2022jok}, Big Bang Observer~(BBO)~\cite{Crowder:2005nr}, and Einstein Telescope~(ET)~\cite{Sathyaprakash:2012jk}; see ref.~\cite{Breitbach:2018ddu} for the computational details.\footnote{Note that the sensitivity curves of the respective observatories should be seen as optimistic forecasts, as they assume that a cosmological GW background is resolvable in the presence of astrophysical sources.}
We translate these results to sensitivity regions in the $\mphi - \fphi^{-1}$ plane. This is shown in fig.~\ref{fig:darkphoton_detectableparamspace} for $\alpha = 50$~(left) and $\alpha = 2\alphamin$~(right), respectively.
The dashed lines in the left panel correspond to the original setup, i.e., no supercooling.
We observe that delayed axion oscillations significantly enhance the observational prospects across the entire frequency range.
For a fixed $\alpha$, a finite period of supercooling opens up the sensitivity region by $\sim 2$ orders of magnitude towards the small-$\fphi$ region.
At the same time the frequency increases, which is why the colored curves tilt towards smaller masses.

In the right panel, the ALP-dark photon coupling is as low as $\alphamin = \mathcal{O}(1)$. 
This further enhances the GW amplitude, since the spectral peak of the fluctuations is shifted towards the horizon scale.
In this case, LISA will be sensitive to decay constants of $\mathcal{O}(10^{14})\,\GeV$, whereas more futuristic observatories such as BBO may probe the axion parameter space down to $\fphi \sim 10^{12}\,\GeV$. In the low-mass region we have included the conservative constraint from recurrent axinovae \cite{Fox:2023xgx}. Note that these constraints rely on the axion making up all of DM. Whether in our setup sufficiently large overdensities can be produced to collapse into star-like objects later remains for future investigation.

While delayed axion oscillations substantially enhance the GW spectrum, we only find a small parameter space that both is observable and reproduces the correct relic DM abundance.
Prior to dark photon production, the axion carries a considerable fraction of the total energy density. Therefore, a large suppression through the tachyonic resonance is required to not overclose the Universe. 
This is indicated by the gray and black lines in fig.~\ref{fig:darkphoton_detectableparamspace}.
We leave a more careful study of the relic axion abundance for the future.
Let us however mention that extensions of the minimal scenario, such as a time-varying axion mass~\cite{McAllister:2008hb,Silverstein:2008sg,Hebecker:2014eua,McAllister:2014mpa,Blumenhagen:2014gta,Marchesano:2014mla}, can further suppress the relic abundance.

\section{Standard Model photon case}\label{sec:SM_photon}
We now replace the dark gauge field with the SM photon which implies additional complications. 
Due to interactions with the thermal bath, the photon dispersion relation is modified, suppressing the growth rate of tachyonic modes. Even if this suppression is overcome, the generation of large electromagnetic fields inevitably leads to Schwinger pair production of light SM fermions, reducing the efficiency of the tachyonic resonance.
In the following, we demonstrate how supercooled ALP oscillations lift these restrictions, and identify the parameter space where ALP-SM photon systems produce sizable GWs.

\subsection{Treating the thermal mass}\label{sec:thermal_mass}

\begin{figure}
    \centering
    \includegraphics[width=\linewidth]{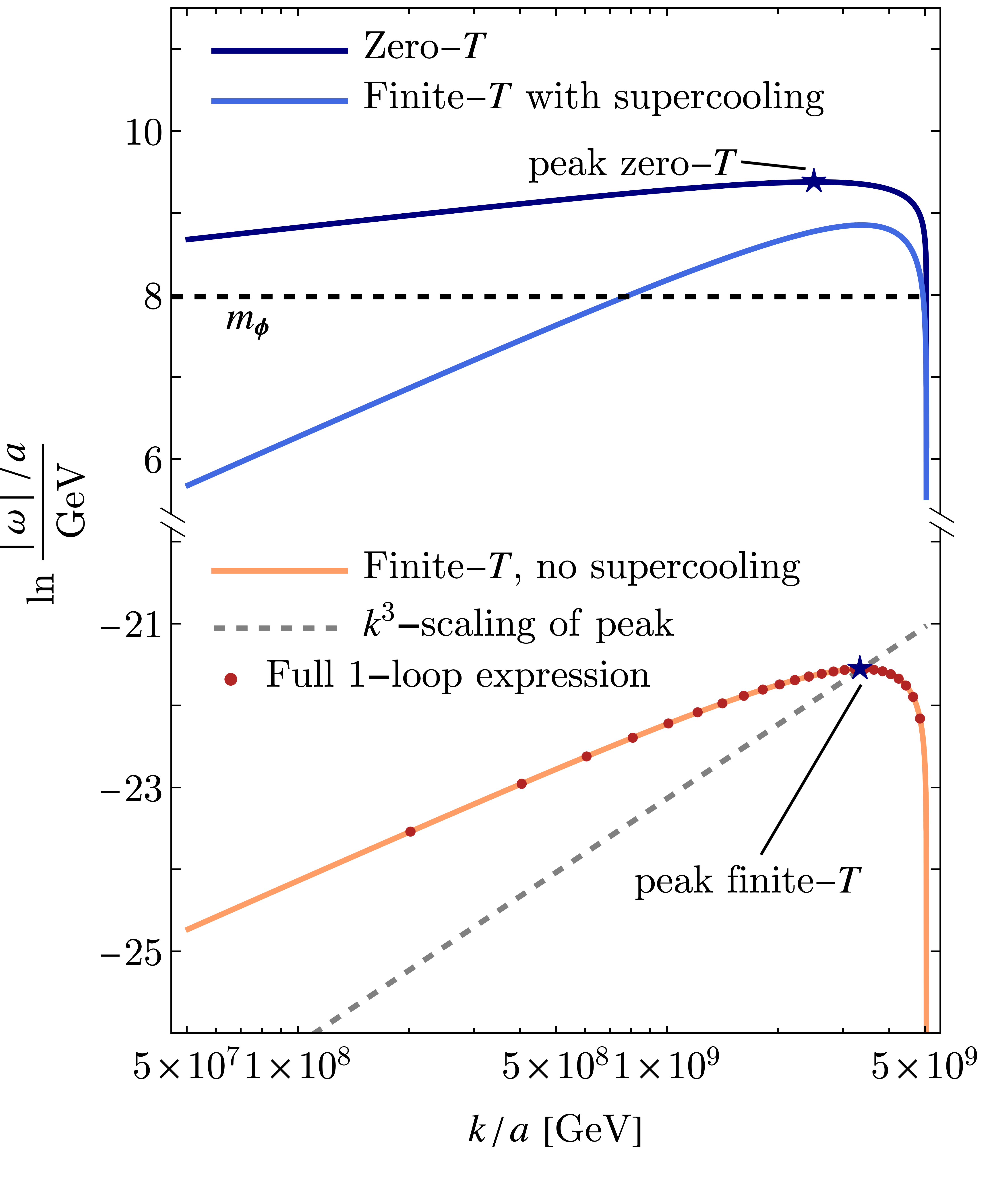}
    \caption{Frequencies in the tachyonic production band for the dark photon with zero-temperature dispersion relation (dark blue, top), SM photon without supercooling with full dispersion relation (red dots), and approximated dispersion relation (first order in $\omega$) without (yellow) as well as with sufficient supercooling (light blue). Also included are the peak positions for zero- and finite-$T$ as well as the $k^3$-scaling of the peak (gray dashed). The presented example is for $\mphi=10^{8}\,\GeV$, $\fphi=10^{18}\,\GeV$. The original finite-$T$ curve is multiple orders below the ALP mass (black horizontal). Once we impose supercooling with $\rsc=2.5 \times 10^{-8}$, the curve is lifted, surpassing the ALP mass and thus opening the tachyonic band.}
    \label{fig:tachyonic_band}
\end{figure}

The treatment of tachyonic instabilities with the SM photon requires us to consider the one-loop dispersion relation of transverse photon modes at finite temperature \cite{Kraemmer:1994az,Kapusta:2006pm,Bellac:2011kqa},
    \begin{align}
        \omega^2 - k^2 \mp k\frac{\alpha}{f_\phi}\phi'
        = \frac{a^2\mD^2}{2}\left[\frac{\omega}{2k}\ln\left(\frac{\omega+k}{\omega-k}\right)\right.
        \\ \nonumber
        \left. -\frac{\omega^3}{2k^3}\ln\left(\frac{\omega+k}{\omega-k}\right)+
        \frac{\omega^2}{k^2}
        \right]\,,
        \label{eq:dispersion_full}
    \end{align}
where $\mD = eT/\sqrt{3}$ is the Debye mass of the photon and $e = 0.3$ denotes the electromagnetic charge. Note that we use comoving quantities here. Following \cite{Hook:2016mqo}, we are interested in solutions yielding imaginary frequencies. With this prerequisite
we simplify the expression by expanding for small comoving frequencies $\omega$,
\begin{equation}
    \omega^2 -k^2 \mp k\frac{\alpha}{f_\phi} \phi' \approx -i\omega \frac{a^2\mD^2\pi}{4k} + \frac{a^2\mD^2\omega^2}{k^2}\,.
    \label{eq:dispersion_quadratic}
\end{equation} 
For tachyonic solutions, we discard the negative solution and consider $\omega$ to be purely imaginary. From now on, we will only consider the absolute value $|\omega|$ of these purely imaginary solutions that parametrizes the growth rate of the tachyonic modes. 
In fig.~\ref{fig:tachyonic_band} we show the tachyonic band for a specific mass and decay constant with and without supercooling compared to the zero-temperature case.

While the cutoff wavenumber at which the band closes, given by $\omega=0$, remains at $\kclose=\alpha\phi'/f_\phi$, the wavenumber that experiences the fastest growth changes compared to the dark photon scenario. 
Neglecting the terms second order in $\omega$, one finds the peak wavenumber
\begin{equation}
    \tilde{k}\approx\frac{2}{3}\frac{\alpha\phi'}{\fphi}\,,
\end{equation}
which is $4/3$ of the value for the dark photon case~\eqref{eq:dark_photon_peak_k}.
The corresponding maximum growth rate at the peak wavenumber is given by
\begin{equation} \label{eq:peak_growth_rate_SM}
    \tilde{\omega}_\rmii{T} \equiv |\omega(\tilde{k})|\approx\frac{16}{27\pi}\frac{\left(\frac{\alpha|\phi'|}{\fphi}\right)^3}{a^2\mD^2} \approx \frac{16}{9 \pi} \frac{(\alpha\theta\mphi)^3}{(e T)^2} a \,,
\end{equation}
where we have used that $|\phi'| \approx \theta \fphi \mphi a$~\cite{Machado:2018nqk}.
This approximation demonstrates that, in contrast to the dark photon scenario, the tachyonic growth rate is suppressed by $\mD^2\sim (eT)^2$. 
Note that the above expression is only valid if $\omega/a \ll k/a \ll T$. In some cases, supercooling is sufficiently strong such that this expansion does not hold. Then, the tachyonic band takes its zero-$T$ form and eq.~\eqref{eq:peak_growth_rate_SM} reduces to eq.~\eqref{eq:dark_photon_peak_k}.

For tachyonic production to be efficient, the growth rate has to be larger than half the axion mass, $\tilde{\omega}_\rmii{T} \gtrsim a\mphi/2$~(see~appendix~\ref{app:minimal_alpha}). 
Applying eq.~\eqref{eq:peak_growth_rate_SM}, this condition dictates the necessary amount of supercooling in terms of eq.~\eqref{eq:r_sc},
\begin{equation}\label{eq:Tosc_min_SM}
    T_{\mathrm{osc}}^2 = r_{\mathrm{sc}}^2
    \Toscaa^2 \lesssim 
    \frac{32}{9 \pi} \frac{(\alpha \theta)^3}{e^2 }\mphi^2\,.
\end{equation}
We see that the required temperature to open the tachyonic band scales with $(\alpha\theta)^{3/2}$ and linear in the ALP mass. Larger masses therefore require less supercooling. In eq.~\eqref{eq:r_sc_thermal_inflation}, we derived that for $\rsci\sim(\fphi/\MPl)^{1/2}$, the axion starts to dominate the Universe, while we now require $\rsc\sim(\mphi/\MPl)^{1/2}$.
{By setting $\rsc < \rsci$, we identify the parameter space where the ALP drives a period of thermal inflation before photon production.\footnote{Note that the $\Neff$ constraint does not apply in the case of the SM photon, since no dark radiation is produced. Therefore a phase of axion domination is allowed, given the Universe can efficiently reheat; see sec.~\ref{sec:SM_cosmo_constraints} and appendix~\ref{app:cosmo_constraints} for details.} This is the case for masses
\begin{equation}\label{eq:SM_mPhi_inflation}
    \mphi \lesssim  
    1.09~ (g^\mathrm{osc}_{\epsilon})^{-\frac{1}{2}}\left(\frac{e}{\alpha^\frac{3}{2}\theta}\right)^2 \fphi \, .
\end{equation}
Only if $\mphi$ is extremely large, the Universe remains radiation-dominated before the tachyonic resonance is effective.}
In the case of axion domination, the number of $e$-folds of thermal inflation reads
\begin{equation}
    N = \ln\left(\frac{T_i}{\Tosc}\right) \approx \ln\left((g_\epsilon^{\rmi{osc},i})^{-\frac{1}{4}}\frac{e}{\alpha^\frac{3}{2} \theta} \left(\frac{\fphi}{\mphi}\right)^\frac{1}{2}\right) \, ,
\end{equation}
where the subscript $i$ indicates the onset of thermal inflation~(cf.~eq.~\eqref{eq:r_sc_thermal_inflation}).
In the most extreme region of the parameter space, we find $N \lesssim 40$.
Therefore, the scales probed via the cosmic microwave background~(CMB) remain super-horizon throughout the evolution~\cite{Planck:2018jri}, i.e., are not affected by the modified cosmic history.

In the numerical parameter scan, we do not compute $\Tosc$ at the peak of the growth rate, but by demanding
\begin{equation}
    |\omega(0.985\,\kclose)| = \frac{\mphi}{2} \, ,
\end{equation}
which secures that the tachyonic band is open for growth rates one order below the maximum rate and therefore for the largest range of modes.
We find that there is no significant difference in the results.

In addition, note that in the presence of the thermal bath, it is important to check that the ALP does not thermalize. Otherwise, the assumption that the ALP remains homogeneous~(cf.~eq.~\eqref{eq:axion_eom}) does not hold.
By comparing the ALP interaction rate with the thermal bath~\cite{Bolz:2000fu,Cadamuro:2011fd}, this translates into an upper bound on the reheating temperature after cosmic inflation. 
We find that only for extremely large $\mphi$ and small $\fphi$, the ALP quanta may thermalize before the onset of the tachyonic resonance. 
However, we will show that this parameter region is irrelevant for GW searches.

\subsection{Timescale of emission}
\label{sec:SM_growth_time}
In contrast to the dark photon, the SM photon is thermalized initially. 
As for the dark photon, we parametrize the growth time in terms of the scale factor ratio between oscillation of the ALP $\aosc$ and the GW emission $\astar$. The detailed derivation is left for appendix~\ref{app:minimal_alpha}. We identify two scenarios: For large ALP masses, less supercooling is required (cf.~eq.~\eqref{eq:Tosc_min_SM}) and the ALP does not necessarily dominate the energy density. The scaling changes in the regime of lower ALP masses, where the Universe needs to be matter-dominated at the time of oscillation, which is possible in the absence of $\Neff$ bounds.
The resulting scale factor ratio therefore depends on whether we have RD,
\begin{equation}\label{eq:SM_photon_growth_time_RD}
    \frac{\astar}{\aosc} =
        1 + \frac{\aosc \mphi}{\langle \tilde{\omega}_\rmii{T}\rangle} \rsc^2 \ln\left(\frac{\theta^2 \mphi^2 \fphi^2}{2 \rho_\gamma(\tauosc)}\right) \,
\end{equation}
or matter domination~(MD),
\begin{equation}\label{eq:SM_photon_growth_time_MD}
    \frac{\astar}{\aosc} = \left[1 + \frac{\aosc {\theta \mphi \fphi}}{2 {\sqrt{6} \MPl}\langle \tilde{\omega}_\rmii{T}\rangle} \ln\left(\frac{\theta^2 \mphi^2 \fphi^2}{2\rho_\gamma(\tauosc)}\right) \right]^2\,.
\end{equation}
Here, $\rho_\gamma(\tau_\mathrm{osc})$ is the initial photon energy density in the instability band and $\langle \tilde{\omega}_\rmii{T}\rangle = 4/(3\pi)\tilde{\omega}_\rmii{T}$ is the average peak growth rate at finite temperature.
Note that due to the large amount of supercooling we have $\langle \tilde{\omega}_\rmii{T} \rangle /\Hosc \gg 1$ in most of the parameter space. Then tachyonic growth happens faster compared to the dark photon case, and $\astar/\aosc \sim 1$.

To determine the minimal $\alpha$ we again demand $a_\rmi{close}~>~\astar$, where the time of band closure is determined by eq.~\eqref{eq:tachyonic_band_closure}.
This is justified, since the cutoff wavenumber is the same as in the dark photon case. Furthermore, most of the parameter space requires significant supercooling, rendering the finite-$T$ photon dispersion relation essentially identical to its zero-$T$ counterpart.

\subsection{Effects of Schwinger pair production}
\label{sec:Schwinger_production}
\begin{figure}
    \centering
    \includegraphics[width=\linewidth]{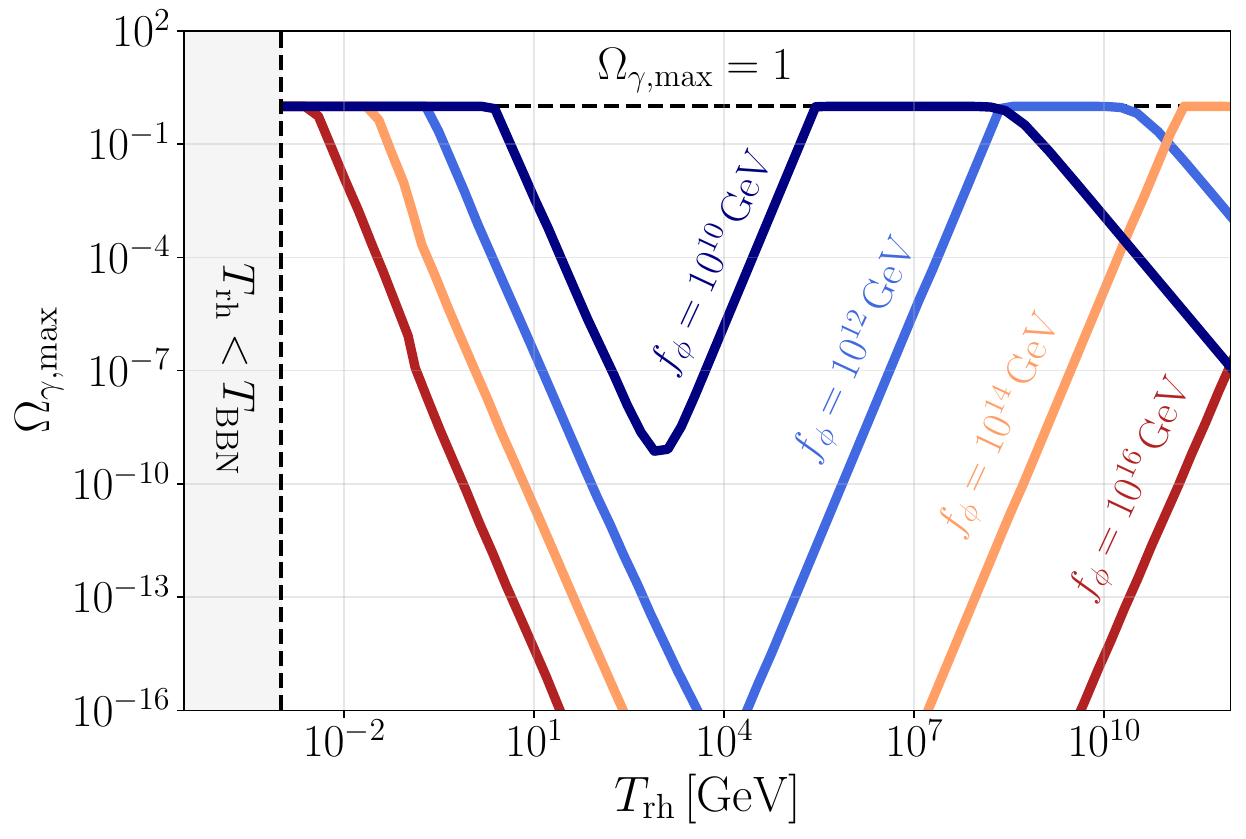}
    \caption{Maximum photon energy density for $\alpha=2\alphamin$ and $\theta=1$ in the presence of Schwinger pair production. The colored curves correspond to different choices of $\fphi$, while the reheating temperature is determined by the axion mass. In the small- (large-)$\mphi$ regime, the Schwinger effect is suppressed by the vacuum~(thermal) mass of the electron.}
    \label{fig:Schwinger_triangle}
\end{figure}

Assuming the Universe is sufficiently supercooled such that the tachyonic band is open for SM photons, Schwinger pair production will limit the amount of energy which can be transferred to the photon. In the Schwinger mechanism, light fermions are created in the presence of a strong electric background field~\cite{Heisenberg:1936nmg,Schwinger:1951nm,Kobayashi:2014zza,Hayashinaka:2016qqn,Gould:2017fve,Lozanov:2018kpk,Domcke:2018eki,Domcke:2019qmm,Domcke:2021yuz}. 
Their subsequent acceleration along the field lines extracts energy from the photon field, reducing the efficiency of the tachyonic resonance.

First, to parametrize the efficiency of photon production, we introduce the efficiency factor 
\begin{equation}\label{eq:Schwinger_efficiency_factor}
    \xi=\frac{\Tilde{\omega}_\rmii{T}}{a H}\,,
\end{equation}
where $\Tilde{\omega}_\rmii{T}$ is the peak growth rate~\eqref{eq:peak_growth_rate_SM}.
As a consequence of the large amount of supercooling, hence the decreased Hubble parameter, we have $\xi \gg 1$ in most of the parameter space.
From the gauge field equation of motion one can derive the energy conservation equation~\cite{Domcke:2019qmm}
\begin{equation}\label{eq:Schwinger_eom}
        \Dot{\rho}_\gamma = -4H \rho_\gamma + 2 \xi H E B - e Q E J_\mathrm{ind} \, .
\end{equation}
Here, $E=|\Vec{E}|$, $B=|\Vec{B}|$, and $\rho_\gamma=\frac{1}{2}(E^2+B^2)$ is the energy of the gauge field.
The first term of the equation corresponds to Hubble friction. The second term describes photon production from ALP oscillations 
and the last term stems from fermion production, where $Q$ is the charge factor.
$J_\mathrm{ind}$ is the induced current, given by~\cite{Domcke:2019qmm} 
\begin{equation} \label{eq:induced_current}
        e Q J_\mathrm{ind} = \frac{(e|Q|)^3}{6\pi^2} \frac{E B}{H} \coth\left(\frac{\pi B}{E}\right) \exp\left(-\frac{\pi m_e^2}{e|Q|E }\right) \, ,
\end{equation}
with electron mass $m_e \approx 511\,\mathrm{keV}$.\footnote{We set $m_e = 0$ if the oscillation temperature $\Tosc$ is above the electroweak scale, $T_\rmii{ew}\approx 150\,\GeV$.} This result assumes that the electron as lightest fermion dominates the Schwinger mechanism.

Interactions with the thermal bath modify the electron dispersion relation. 
In principle, a rigorous analysis therefore necessitates a rederivation of eq.~\eqref{eq:induced_current} in the presence of a thermal bath, taking into account the full finite-$T$ fermionic dispersion relation.
This is approximated by replacing $m_e^2 \to m_e^2 + m_{e,\rmi{th}}^2$ with $m_{e,\mathrm{th}}^2=(eT)^2/8$, which is relevant for larger $\mphi$ where less supercooling is required.
While our treatment is simplified, it allows us to identify the parameter space where sizable GW production is possible.

To analyze the amplification of photon modes in the presence of Schwinger pair production, we assume a dynamical equilibrium between the axion, the photon, and the fermions. 
Since the growth is sufficiently fast, the emission happens well below one Hubble time and we can set $\Dot{\rho}_\gamma$ in eq.~\eqref{eq:Schwinger_eom} to zero, which gives
\begin{equation} \label{eq:constraint_eq}
        E^2 + B^2 - \xi E B + \frac{eQ}{2} \frac{E}{H} J_\mathrm{ind} = 0 \, .
\end{equation}
For a given efficiency parameter, eq.~\eqref{eq:constraint_eq} describes a closed contour in the $E-B$ plane. 
Following the strategy provided in~\cite{Domcke:2019qmm}, we maximize the photon energy density numerically to find its upper bound for given model parameters.
This will eventually provide an upper limit on the GW amplitude from axion-SM photon systems.

In fig.~\ref{fig:Schwinger_triangle} we present the resulting maximal relative photon energy densities as a function of the would-be reheating temperature, 
\begin{equation}
    \Trh = \left(\frac{90}{\pi^2 \gepsrh}\right)^\frac{1}{4} \sqrt{\Hosc \MPl} \, ,
\end{equation}
where the Hubble parameter at the onset of oscillations, $\Hosc$, encodes the relation to the ALP mass. The color coding denotes four different choices of $\fphi$.

Two regions of interest can be identified. For reheating temperatures sufficiently close to BBN, Schwinger pair production is suppressed by the vacuum mass of the electron, and the energy can be transferred to the photon efficiently. The maximal photon energy density then drops rapidly due to the blocking from Schwinger production.
This changes when thermal effects start increasing the effective electron mass, suppressing the Schwinger production. This leads to a range of ALP masses for which the energy can be efficiently deposited in the photon sector again.
For even higher reheating temperatures, the Universe remains radiation-dominated during the tachyonic resonance~(cf.~eq.~\eqref{eq:SM_mPhi_inflation}). Then, $\Omega_{\gamma,\rmi{max}}$ drops due to the reduced axion contribution to the total energy density.

In addition, note that the efficiency of photon production depends strongly on $\fphi$ via $\xi$. 
Smaller $\fphi$ decreases the initial axion energy density $\Omegaphioscaa$, which requires more supercooling until axion domination begins.
This decreases the Hubble parameter~\eqref{eq:Hubble_osc}, implying an larger efficiency parameter for small $\fphi$, hence higher $\Omega_{\gamma,\rmi{max}}$.
Therefore, the cosmologically viable axion mass range is tightly constrained for large decay constants~(see~sec.~\ref{sec:SM_cosmo_constraints}).

Furthermore, fermion production, as can be seen from eq.~\eqref{eq:induced_current}, is exponentially suppressed below the Schwinger limit
\begin{equation}\label{eq:Schwinger_limit}
    E_s=\frac{\pi m_e^2}{e|Q|}\,.
\end{equation}
Note that this limit can be translated to a reheating temperature, below which Schwinger production is guaranteed to be inefficient. Since this yields $T_\mathrm{rh}~\lesssim 2.2\,\mathrm{MeV}/(\gepsrh)^\frac{1}{4}$, it is not relevant
for our parameter range.

\subsection{Cosmological constraints}
\label{sec:SM_cosmo_constraints}
In the absence of dark radiation, the $\Delta \Neff$ constraints from sec.~\ref{sec:dark_photon} do not apply.
In addition to the requirement $\Trh > \TBBN$~(see above), the cosmological viability of the mechanism depends on the relic ALP abundance, which is affected by Schwinger pair production.
To this end, let us again divide the parameter space into cosmologically stable ALPs and ALPs that decay before BBN, $\GammaphitoSM~>~\HBBN$. 

We first consider light ALPs which contribute to the DM density.
If the Schwinger effect is negligible, i.e., energy transfer to the photon is efficient, the ALP abundance after GW emission is given by eq.~\eqref{eq:Omega_phi_after}.
Expressing $\Omega_\phi^\rmii{after}$ in terms of the model parameters, the required suppression to not overproduce DM reads
\begin{equation}
    \begin{split}
    \epssup^\rmii{SM,DM} \lesssim \, & 1.66 \times 10^{-10}\;  \gsrh(\gepsrh)^{-\frac{3}{4}} \left(\frac{\astar}{\aosc}\right)^\frac{3}{4} \\
    &\times \left(\frac{\mathrm{eV}}{\mphi}\right)^\frac{1}{2} \left(\frac{10^{10}\,\GeV}{\theta \fphi}\right)^\frac{1}{2} \, .
    \end{split}
\end{equation}
We refer to appendix~\ref{app:cosmo_constraints} for the computation.
Here, $\astar/\aosc$ is the growth time estimate from sec.~\ref{sec:SM_growth_time}.
This defines the viable parameter space in the small-$\mphi$ range.
Let us stress again that lattice studies of the original setup imply $\epssup\sim 10^{-2}$, which would rule out the light ALP parameter space. 
As argued in sec.~\ref{sec:dark_photon_cosmo_constraints}, the reduced axion-photon coupling in the presence of supercooling and further model building can decrease the suppression of the axion abundance.
Furthermore, in contrast to the dark photon case studied in~\cite{Ratzinger:2020oct}, the SM photon has large couplings to SM fermions, which could additionally reduce the backscattering into ALPs.
In addition, the presence of the thermal bath affects the gauge field backreaction, which is, however, difficult to estimate analytically.
In the absence of lattice studies of the axion-SM photon system, we therefore treat $\epssup$ as a free parameter.

In parameter regions with efficient Schwinger production, photon production is ineffective and $\epssup\sim~1$.
As only a negligible fraction of the axion energy density is transferred during the tachyonic resonance, we have $\rhophiafter \simeq \rhophistar$ and $\Omega_\gamma^\rmii{after} \ll 1$.
In addition, $\Omegaphistar \simeq 1$ in the small-$\mphi$ regime.
Without additional decay channels, the ALP therefore continues to dominate the energy density of the Universe after the onset of oscillations until today.
Thus, this parameter space is excluded.
This mainly affects large decay constants, where the efficiency parameter~\eqref{eq:Schwinger_efficiency_factor} is small.

In the large-$\mphi$ region ALPs decay before BBN, hence the consistency of the cosmic history is ensured.
Since less supercooling is required for efficient photon production~(cf.~eq.~\eqref{eq:Tosc_min_SM}), the Universe does not necessarily inflate before ALP oscillations. 
This leads to different thermal histories depending on the efficiency of Schwinger production, the subsequent suppression of the axion abundance through the tachyonic resonance, and the axion decay rate.
Ultimately, this will only affect our results via the redshift factors of the GW amplitude and frequency.
These are determined numerically; see appendix~\ref{app:cosmo_constraints} for details.

\subsection{Gravitational wave signal}
\begin{figure*}
    \centering
    \includegraphics[width=\linewidth]{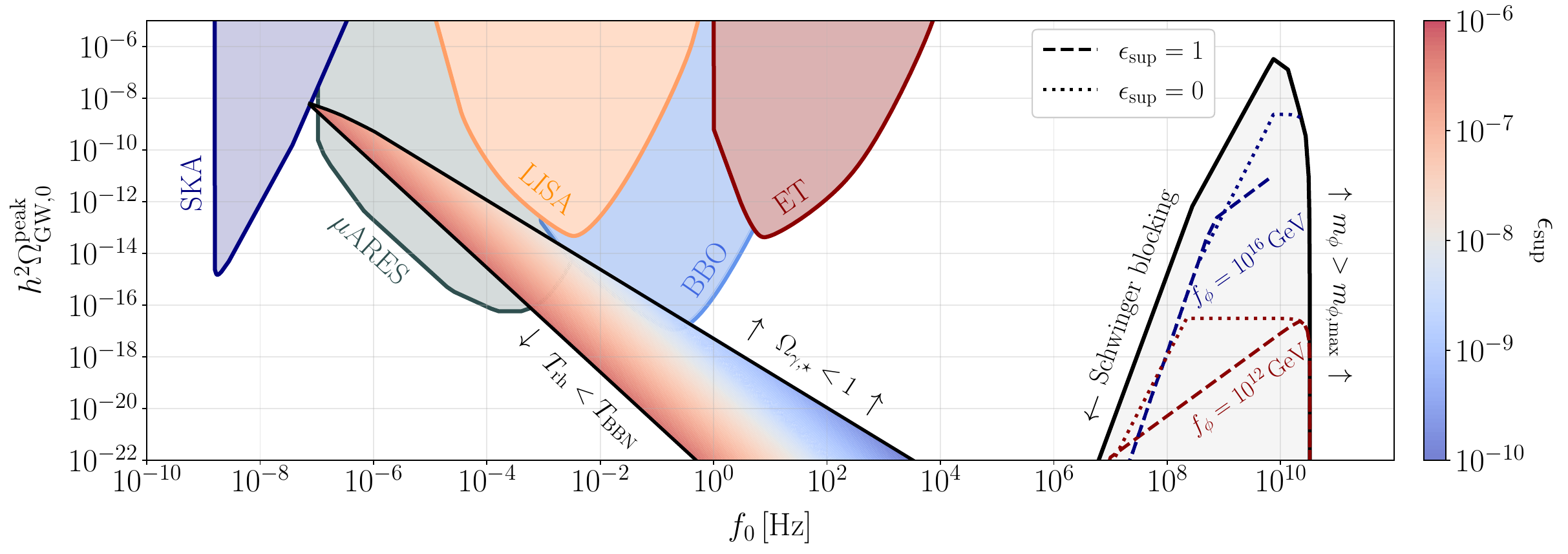}
    \caption{Projected positions of the GW peak in the ALP-SM photon scenario. The colored curves indicate the sensitivity regions of the future observatories SKA, $\mu$ARES, LISA, BBO, and ET. We identify two distinct parameter regions where Schwinger pair production is suppressed, allowing for a sizable GW signal. In the low-frequency (small-$\mphi$) regime, the color coding indicates the required suppression of the abundance for the ALP to constitute DM. In the high-frequency (large-$\mphi$) regime, ALPs decay before BBN, hence evade all cosmological bounds. ALPs undergoing the tachyonic resonance close to BBN show the most promising observational prospects. Since the characteristic scale of the fluctuations is deep inside the Hubble horizon, the corresponding GW peak lies in the sensitivity region of $\mu$ARES.}
    \label{fig:SM_photon_GWs}
\end{figure*}
Regarding the GW signal, we restrict ourselves to estimates of the peak, following the treatment of sec.~\ref{sec:GW_signal_darkphoton}.
To this end, we compute the maximum photon energy density in the presence of Schwinger pair production. Then, the energy available for GW emission reads $\Omega_{\gamma}^\rmii{after} = \chi_\rmi{sp} \Omegaphistar$, where $\chi_\rmi{sp} \in (0,1)$ is computed numerically as outlined in sec.~\ref{sec:Schwinger_production}. 
Furthermore, we distinguish whether we have RD or MD when the tachyonic resonance becomes efficient.
By combining our growth time estimates~\eqref{eq:SM_photon_growth_time_RD} and~\eqref{eq:SM_photon_growth_time_MD} with eq.~\eqref{eq:Hubble_osc}, we find the Hubble parameter at the time of GW emission.  
Together with the finite-$T$ expression of the peak momentum, we then obtain the GW peak frequency and amplitude at the time of production via eq.~\eqref{eq:GW_peak_estimate_production}. Redshifting to today (see~appendix~\ref{app:cosmo_constraints}) and employing our best-fit parameters from table~\ref{tab:GW_fit_parameters}, we have
\begin{widetext}
\begin{gather}
    \displaystyle
    \tilde{f}_0 =  \displaystyle 8.69 \times 10^{-8}\, \mathrm{Hz} \, \left(\frac{100}{\gepsrh}\right)^\frac{1}{12} \alpha\theta \, \frac{\mphi}{\mathrm{eV}} \left(\frac{\aosc}{\astar}\right)^\frac{3}{2} \left(\frac{\mathrm{GeV}}{\Hrh}\right)^\frac{1}{2} \min\left\{1,\frac{\astar}{\adecay} \right\}\,,\\[.4cm]
    \displaystyle h^2\tilde{\Omega}_{\rmii{GW},0}^\rmii{MD} = \displaystyle 4.20 \times 10^{-4}\left(\frac{100}{\gepsstar}\right)^\frac{1}{3}  \chi_\rmi{sp}^2 \left(\frac{\fphi}{\alpha \MPl}\right)^2 \min\left\{1,\frac{\amd}{\adecay}\right\} \,,\\[.4cm]
    h^2\tilde{\Omega}_{\rmii{GW},0}^\rmii{RD}= \displaystyle 7.01 \times 10^{-5}\left(\frac{100}{\gepsstar}\right)^\frac{1}{3} \chi_\rmi{sp}^2 \left(\frac{\theta}{\alpha}\right)^2 \left(\frac{\fphi}{\rsc \MPl}\right)^4  \left(\frac{\astar}{\aosc}\right) \min\left\{1,\frac{\amd}{\adecay}\right\} \, .
\end{gather}
\end{widetext}
Here, $\amd/\adecay$ parametrizes the length of an intermediate MD period after photon production has taken place.
In addition, $\astar/\adecay$ denotes the additional redshift of the GW frequency due to the modified expansion history of the Universe.
Then the Hubble parameter after reheating, $\Hrh$, is either given by $H_\star$ or $\GammaphitoSM$.
This depends on the thermal history; see appendix \ref{app:cosmo_constraints} for details.
From the above expression, we again note the $(\fphi/\MPl)^2$ scaling of the peak amplitude. 
Furthermore, since in most of the parameter space the axion is the dominant energy component in the Universe, we generally find stronger signals compared to the dark photon case, provided $\chi_\rmi{sp} \simeq 1$.

The resulting projected GW peaks are presented in fig.~\ref{fig:SM_photon_GWs}, where we set $\theta=1$ and scan over a large range of values in the $\mphi-\fphi$ plane.
Here we again employ $\alpha=2\alphamin$ as outlined in appendix~\ref{app:minimal_alpha}. 
Hence, our results represent an approximate upper limit on the GW amplitude from ALP-SM photon systems.
The two regions for lower (higher) GW frequency $f_0$ correspond to the two regions identified in sec.~\ref{sec:Schwinger_production} and \ref{sec:SM_cosmo_constraints} for a small (large) axion mass $\mphi$. Both regions are separated by efficient Schwinger pair production blocking the energy injection into photons.

The low-$f_0$ region corresponds to cosmologically stable axions. The color coding indicates the required suppression of the axion energy density to meet the DM relic abundance. Lower frequencies, i.e., smaller axion masses, necessitate less suppression to achieve the same GW amplitude. Towards the left, the window is bounded by the condition $\Trh<\TBBN$.
Towards larger masses, the viable parameter space is constrained by $\Omega_{\gamma}^\rmii{after} < 1$, leading to an inconsistent cosmological evolution in the absence of perturbative axion decays.
This restricts the available parameter space to $\fphi \lesssim 5 \times 10^{16}\,\GeV$. 

We find the most promising observational prospects in the $\mu\mathrm{Hz}$ regime between the pulsar timing arrays and LISA.
Although the reheating temperature is close to $\TBBN \sim \MeV$, the fluctuations lie deep inside the Hubble horizon, shifting the peaks into the $\mu\mathrm{ARES}$ sensitivity region. 
For the same reason, the GW amplitude is suppressed, even though the ALP dominates the energy density of the Universe at the time of production.

The large-$\mphi$ region is bounded by the Planck constraint on the Hubble parameter during inflation~\cite{Planck:2018jri}, i.e., $m_{\phi,\mathrm{max}}~=~6~\times~10^{13}\,\mathrm{GeV}~=~H_\rmi{inf}^\mathrm{max}~\geq~\Hoscaa$.
In this regime, the axion does not constitute DM, but decays before BBN. 
Hence, the suppression of the abundance through the tachyonic instability merely affects the redshift factors of the GW signal.
To exemplify this impact, we choose the most extreme values $\epssup \in \{0,1\}$.
The grey-shaded region shows the envelope of our parameter scan, while the red (blue) curve shows the result for $\fphi = 10^{12}\,(10^{16})\,\GeV$. 
As long as Schwinger pair production is efficient, the dotted and dashed curves agree since the ALP abundance is fixed. 
Once the thermal electron mass enables efficient photon production, the lines start to diverge.
Then, a more efficient suppression of the axion abundance through the tachyonic resonance shortens the MD period.
This induces stronger GW signals with peak frequencies ranging up to $\tilde{f}_0 = \mathcal{O}(10^{10})\,\GeV$ and amplitudes up to $h^2\OmegaGWTildeToday = \mathcal{O}(10^{-6})$.

\section{Conclusion}
We have studied the audible axion mechanism in the case of supercooled axion oscillations, which is realized if the pseudoscalar is trapped in a local minimum initially.
Our main results are presented in figs.~\ref{fig:darkphoton_detectableparamspace} and~\ref{fig:SM_photon_GWs}, which show the GW predictions for ALP-dark photon and ALP-SM photon systems, respectively.

In the case of ALPs coupled to dark photons, a period of supercooling significantly enhances the resulting stochastic GW background.
This allows for GW probes of a large portion of the axion parameter space towards smaller axion decay constants $\fphi$ and axion-photon couplings $\alpha$.
Specifically, $\fphi \sim 10^{12}\,\GeV$ and $\alpha \sim 1$ are sufficient to produce GW signals in the reach of future observatories, while the original setup requires $\alpha \gtrsim 20$ and $\fphi \gtrsim 10^{16}\,\GeV$.

Furthermore, if the onset of oscillations is sufficiently delayed, GW production becomes possible via the axion coupling to the SM photon.
The emergence of a large electromagnetic field leads to Schwinger pair production of electron-positron pairs, which restricts the available parameter space.
We identify a region for small ALP masses and $\fphi \lesssim 5 \times 10^{16}\,\GeV$ that will be tested by $\mu$ARES~\cite{Sesana:2019vho} and BBO~\cite{Crowder:2005nr}.
In addition, heavy ALPs produce sizable GWs at ultra-high frequencies $f_0 = \mathcal{O}(10^8-10^{10})\,\mathrm{Hz}$.

Finally, let us comment on possible future directions. In both scenarios, the light ALP parameter range requires a large suppression of the relic abundance $\epssup = \mathcal{O}(10^{-4}-10^{-6})$ through tachyonic (dark) photon production to not overclose the Universe.
Previous lattice results~\cite{Ratzinger:2020oct} indicate $\epssup \sim 10^{-2}$ for large $\alpha \gtrsim 20$, which may however change for small couplings $\alpha \sim 1$.
In addition, couplings to the SM are expected to modify the backreaction dynamics compared to the dark photon case, further affecting the relic abundance.
To this end, a lattice study of the modified setup is required, which we relegate to the future.
On top, it would be interesting to explore whether the bound on the relic abundance could be alleviated by further model building efforts, such as a time-varying axion mass.

Also, note that we have merely focused on observational signatures in the form of GWs.
In the case of the SM, large helical magnetic fields are produced which can survive until today as intergalactic magnetic fields~\cite{Ando:2010rb,Tavecchio:2010mk,doi:10.1126/science.1184192,Essey:2010nd,Chen:2014rsa,Fermi-LAT:2018jdy}. This may provide a complementary probe of the audible axion model, which we aim to investigate in the future.

\begin{acknowledgments}
The authors thank V.~Domcke, M.~Lewicki, D.~Perri, N.~Ramberg, and P.~Sørensen for useful discussions on a variety of topics.
The authors also thank W.~Ratzinger for helpful comments on the manuscript.
DS acknowledges support by the Deutsche Forschungsgemeinschaft (DFG, German Research Foundation) through the CRC-TR 211 'Strong-interaction matter under extreme conditions'– project number 315477589 – TRR 211.
PS and CG acknowledge support by the Cluster of Excellence “Precision Physics, Fundamental Interactions, and Structure of Matter” (PRISMA+ EXC 2118/1) funded by the Deutsche Forschungsgemeinschaft (DFG, German Research Foundation) within the German Excellence Strategy (Project No. 390831469).
\end{acknowledgments}
\appendix

\section{Trapped misalignment toy model}\label{app:trapped_misalignment}
There exist some models which motivate a delay of ALP oscillations, providing the supercooling we consider in this investigation. Here, we briefly discuss one such model: trapped misalignment ~\cite{Higaki:2016yqk,Kawasaki:2017xwt,Nakagawa:2020zjr,DiLuzio:2021pxd,DiLuzio:2021gos,Kitajima:2023pby,DiLuzio:2024fyt}, where additional $U(1)_\rmii{PQ}$-breaking operators trap the ALP in a false minimum.

Originally introduced for the QCD axion, the constraints for the trapped misalignment mechanism are much simpler for a general ALP. Yet, in contrast to the QCD axion potential, the original ALP potential does not change with the temperature. Therefore, we add an additional temperature-dependent part to the ALP potential,
\begin{equation}
    \begin{split}
        V(\phi,T) = & \mphi^2 \fphi^2 \left(1-\cos \frac{\phi}{\fphi}\right) \\ +  & \Lambda^{4-q} \,T^q\,\left(1-\cos(n\frac{\phi}{\fphi}+\delta)\right)\,,
    \end{split}
\end{equation}
where $\Lambda$ is a newly introduced scale that controls the temperature at which the $U(1)_\rmii{PQ}$ breaking effects become relevant. 
The additional term includes a displacement from the initial potential by $\delta$, while $n$ is an integer and determines the number of false minima in the potential. For now, we keep the temperature-scaling general, parametrized by the $T$-exponent $q$. Depending on $n$ and the initial misalignment, the ALP can be trapped in a false minimum until a certain release temperature $T_\mathrm{rel}$ is reached at which the ALP becomes free to oscillate around the true minimum. 
At the moment of release, two conditions have to be met \cite{DiLuzio:2024fyt},
\begin{equation}\label{eq:release_condition}
    \frac{\partial V(\phi,T)}{\partial \phi}=0\,,\quad \frac{\partial^2 V(\phi,T)}{\partial \phi^2}=0\,.
\end{equation}

\begin{figure}
    \centering
    \includegraphics[width=\textwidth]{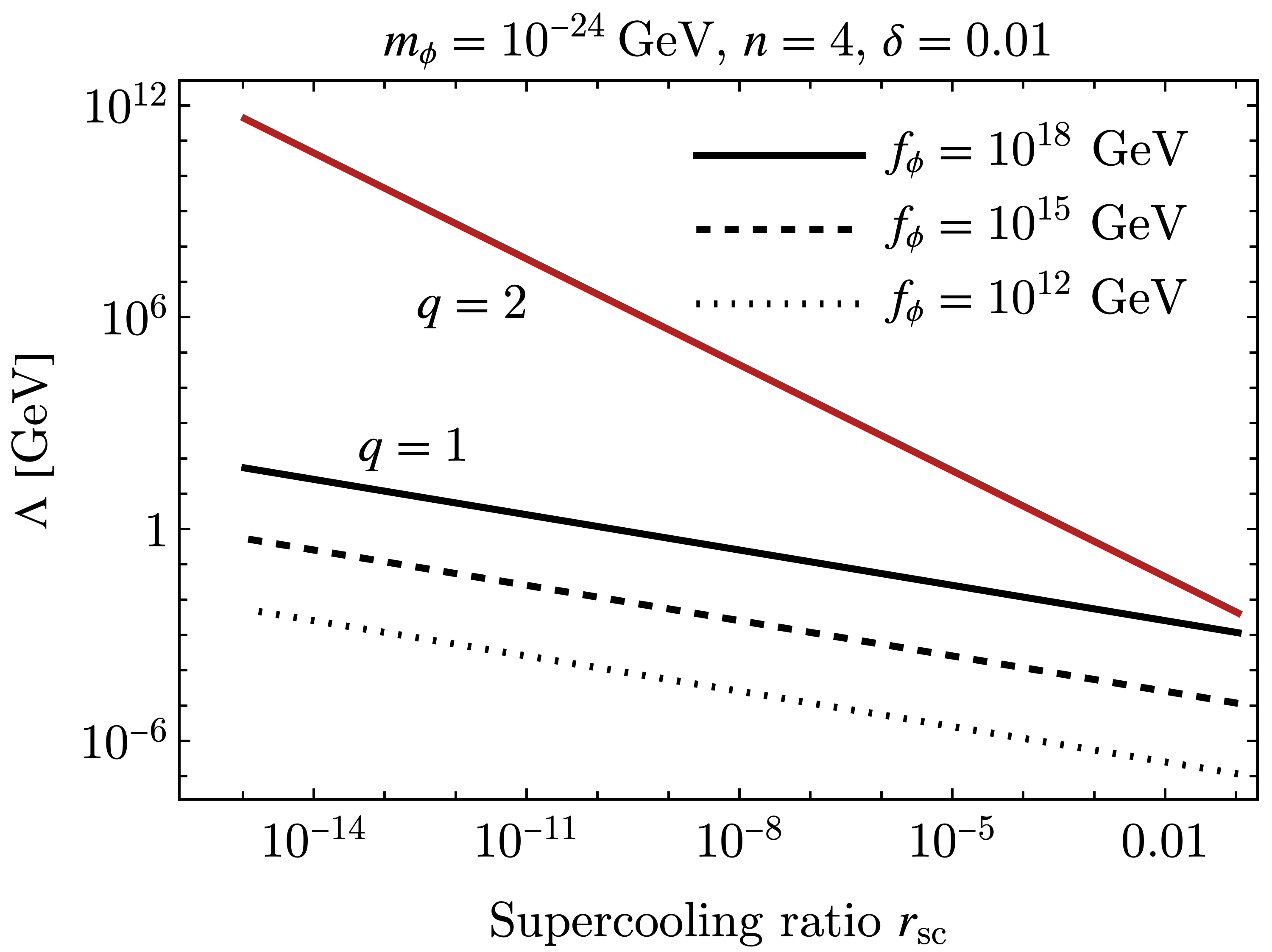}
    \caption{Scale of the necessary trapping potential as a function of the supercooling ratio generated by trapped misalignment.}
    \label{fig:trapped_misalignment_scale}
\end{figure}

These equations can be solved for the scale $\Lambda$, which, in turn, can then be written as a function of the supercooling ratio $\rsc$, if we identify the release temperature with the oscillation temperature in the supercooled scenario,
\begin{equation}
    T_\mathrm{rel}=\Tosc \approx \rsc\sqrt{\mphi \MPl}\,.
\end{equation}
The resulting $\Lambda$ scales as
\begin{equation}
    \Lambda \sim \left(r_{sc}^{-q}\mphi^{2-q/2}\MPl^{-q/2}f_\phi^2
    \right)^{\frac{1}{4-q}}\,.
\end{equation}
In fig.~\ref{fig:trapped_misalignment_scale} we present $\Lambda(\rsc)$ for different model parameters.  
Clearly, there is a large parameter space where we obtain sufficiently large supercooling ratios~$\rsc$.
Only for too much supercooling (small $\rsc$) combined with larger ALP masses and a steeper temperature scaling, the scale would approach the Planck scale.
Note that the ALP acquires a thermal contribution to its effective mass from the $T$-dependent potential. By construction \eqref{eq:release_condition}, thermal and vacuum contribution are of the same order, if $n$ is not too large. After the release, the thermal mass becomes suppressed.

\section{Growth time and minimal $\alpha$} \label{app:minimal_alpha}
In this section, we determine the minimal ALP-photon coupling $\alpha$ in the case of delayed oscillations.\footnote{Note that the possibility of imposing a small $\alpha$ to have efficient dark photon production in the case of trapped misalignment has already been pointed out in~\cite{Kitajima:2023pby}.}
To this end, we derive an expression for the scale factor ratio $\astar/\aosc$ that parametrizes the (dark) photon growth time. We first give a general expression, before specializing to the dark and SM photon, respectively.

We estimate the time when photon production is completed by noting that the (dark) photon energy density grows as
\begin{equation}
    \rho_{X,\gamma}(\tau) \approx \rho_{X,\gamma} (\tau_\mathrm{osc}) \exp(2 \Tilde{\omega} \tau )\, ,
\end{equation}
where $\tau$ denotes conformal time and $\Tilde{\omega}$ is the growth rate of the fastest growing mode. 
The dominant photon helicity only grows when the sign of $\phi'$ takes its initial value, i.e., half of the time. Then the conformal growth time reads
\begin{equation} \label{eq:growth_time_conformal}
   \frac{\delta \tau}{\tau_\mathrm{osc}} = \frac{\taustar - \tauosc}{\tauosc} = \frac{\aosc \Hosc}{\Tilde{\omega}} \ln\left(\frac{\rho_{X,\gamma} (\taustar)}{\rho_{X,\gamma} (\tauosc)}\right) \, ,
\end{equation}
where we neglect the redshift of all quantities between the start of oscillation and dark photon production.
Furthermore, we need to take into account that the peak mode is amplified with the averaged peak growth rate~$\langle \tilde{\omega} \rangle$, which we will shortly specify.
Hence, we replace $\tilde{\omega} \to \langle \tilde{\omega} \rangle$ in eq.~\eqref{eq:growth_time_conformal}.
Since we are interested in the time when the (dark) photon energy density becomes comparable to the initial axion energy density, we further set
\begin{equation}
    \rho_{X,\gamma} (\tau_\star) = \frac{\theta^2}{2} \mphi^2\fphi^2\, .
\end{equation}
Then, the conformal growth time is translated to a scale factor ratio by solving 
\begin{equation}\label{eq:Hubble_conformal}
    H = \frac{a'}{a^2} \, ,
\end{equation}
for the respective background evolution.

\begin{figure}
    \centering
    \includegraphics[width=\linewidth]{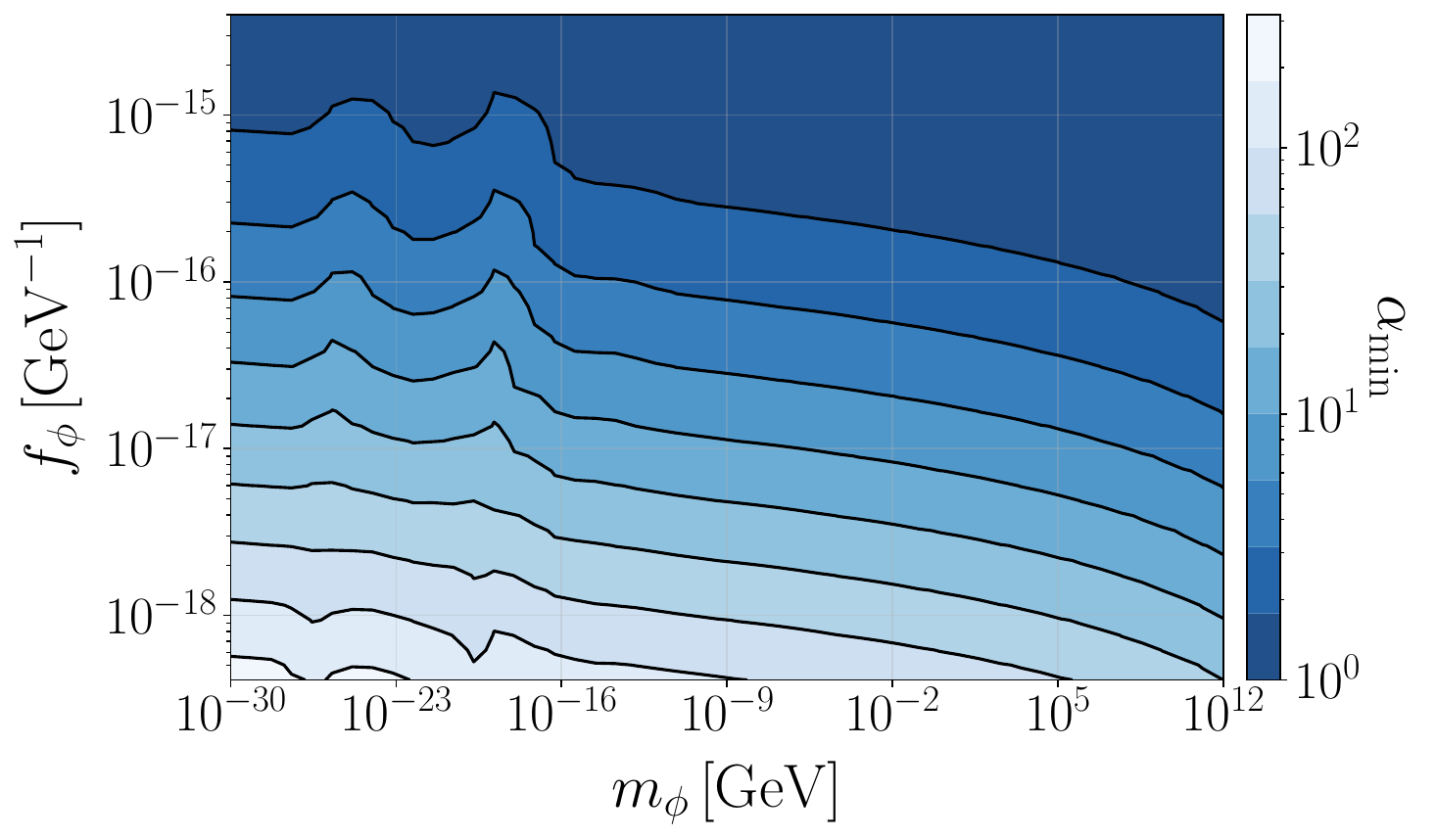}
    \caption{Minimal value for the ALP-dark photon coupling for $\theta=1$, obtained by requiring $\astar < a_\rmi{close}$. For decay constants $\fphi \lesssim 10^{16}\,\mathrm{GeV}$, $\alpha \sim \mathcal{O}(1)$ is sufficient to have efficient tachyonic growth. Note that the irregular behavior in the small-$\mphi$ range is caused by the rapid change of the relativistic degrees of freedom during the QCD epoch.} 
    \label{fig:minimal_alpha}
\end{figure}

In the dark photon case, we assume Bunch-Davies initial conditions~\cite{Bunch:1978yq}, $v_\lambda(k,\tau\ll \tauosc) = \exp(ik\tau)/\sqrt{2k}$. Hence we have
\begin{equation}
    \rho_X(\tauosc) \approx \frac{1}{16\pi^2}\left(\frac{\tilde{k}}{\aosc}\right)^4 \, ,
\end{equation}
where $\tilde{k} = \tilde{\omega} = \alpha \theta \mphi \aosc/2$ is the comoving momentum of the mode which sets the peak of the dark photon energy spectrum.
The average peak growth rate reads $\langle \tilde{\omega} \rangle =\tilde{\omega} \langle |\sin(\mphi t)| \rangle  = 2\tilde{\omega}/\pi$.
Since a MD period is prohibited due to $\Delta \Neff$ constraints, we solve eq.~\eqref{eq:Hubble_conformal} for a radiation-dominated Universe and obtain
\begin{equation} \label{eq:growth_time_app}
    \frac{\astar}{\aosc} = 1 + \frac{\pi}{\alpha \theta} \rsc^2  \ln\left(\frac{128 \pi^2}{\alpha^4 \theta^2}\frac{\fphi^2}{\mphi^2}\right) \, .
\end{equation}
We will shortly verify this estimate with numerical simulations.
The dark photon energy density grows with twice the growth rate of the individual modes, thus the time of tachyonic band closure is estimated by solving 
\begin{equation}
    \omega^2 = -\left(\frac{a \mphi}{2}\right)^2 \, .
\end{equation}
Here, $\omega$ is given by eq.~\eqref{eq:dark_photon_dispersion}.
This gives the range of momenta that exhibit conformal growth rates $\omega$ larger than half the conformal oscillation frequency $am_\phi$. 
Then, the time when dark photon production shuts off reads
\begin{equation}\label{eq:tachyonic_band_closure_app}
    \frac{a_\mathrm{close}}{\aosc} = (\alpha\theta)^\frac{2}{3} \, .
\end{equation}

To find the minimal $\alpha$ for a given set of model parameters $\{\mphi,\fphi,\theta\}$, we demand $\astar < a_\rmi{close}$.
The results are shown in fig.~\ref{fig:minimal_alpha}.
Remarkably, delayed axion oscillations allow for a significantly smaller value of $\alpha$ compared to the original setup, where $\alpha \gtrsim \mathcal{O}(20)$. 
This is due to the decreased Hubble parameter at the time of production, reducing the growth time $\propto \rsc^2$.
Since lower decay constants imply a larger amount of supercooling consistent with the $\Neff$ constraint, we find that for $\fphi \lesssim 10^{16}\,\GeV$, $\alpha \sim \mathcal{O}(1)$ suffices to have efficient GW production.
As a consequence, the GW amplitude~\eqref{eq:GW_peak_estimate_production} is further enhanced.
Note, however, that dark photon production ceases to be efficient when setting $\alpha\lesssim\alphamin$, since the tachyonic band closes shortly after the onset of oscillations. 
Therefore, we employ a more conservative value of $\alpha = 2\alphamin$ to study the upper bound on the GW amplitude.

The computation for the SM photon case proceeds analogously, with a few subtle differences. 
First, SM photons are thermalized initially, hence the Bunch-Davies initial conditions do not apply.
Instead, the initial energy density in the instability band reads
\begin{equation}\label{eq:rho_SM_initial}
    \rho_{\gamma}(\tauosc) = \frac{g_{\gamma}}{2\pi^2 \aosc^4} \int_0^{k_\rmii{close}} dk k^3 \left(\exp\left(\frac{k}{\aosc T}\right) - 1\right)^{-1} \, ,
\end{equation}
where $g_\gamma = 2$.
The cutoff is the same as in the dark photon case, $\kclose = \alpha\theta \mphi \aosc$; see sec.~\ref{sec:thermal_mass}.
In addition, we replace the growth rate $\langle \tilde{\omega}\rangle$ by its average value at finite temperature $\langle \tilde{\omega}_\rmii{T}\rangle = 4/(3\pi) \tilde{\omega}_\rmii{T}$, where $\tilde{\omega}_\rmii{T}$ is given by eq.~\eqref{eq:peak_growth_rate_SM}.
In the absence of $\Neff$ constraints, the ALP is allowed to dominate the energy density of the Universe at the time of production. 
Then, we find via eq.~\eqref{eq:Hubble_osc}
\begin{equation}\label{eq:SM_photon_growth_time}
    \frac{\astar}{\aosc} = \begin{cases}
        \displaystyle
        1 + \frac{\aosc \mphi}{\langle \tilde{\omega}_\rmii{T}\rangle} \rsc^2 \ln\left(\frac{\theta^2 \mphi^2 \fphi^2}{2 \rho_\gamma(\tauosc)}\right) \, , \;&\mathrm{RD}\,,\\
        \displaystyle 
        \left[1 + \frac{\aosc {\theta \mphi \fphi}}{2 {\sqrt{6} \MPl}\langle \tilde{\omega}_\rmii{T}\rangle} \ln\left(\frac{\theta^2 \mphi^2 \fphi^2}{2\rho_\gamma(\tauosc)}\right) \right]^2\, , \;&\mathrm{MD}\,,
    \end{cases} 
\end{equation}
depending on whether the Universe is radiation- or matter-dominated at the time of production. 
This is mainly determined by the ALP mass; see sec.~\ref{sec:thermal_mass}. For small $\mphi$, a large amount of supercooling is required to suppress the Debye mass of the photon, i.e., we enter a phase of thermal inflation.
Conversely, the Universe remains radiation-dominated for large $\mphi$.
In both cases, we impose sufficient supercooling such that (almost) the entire instability band can efficiently grow.
Then the finite-$T$ dispersion relation quickly approaches the zero-$T$ one.
As a consequence, the width of the tachyonic band is the same as in the dark photon scenario and we again employ eq.~\eqref{eq:tachyonic_band_closure_app} to estimate the time of tachyonic band closure and determine the minimal $\alpha$. 

\section{Details on the cosmological constraints} \label{app:cosmo_constraints}
In the following, we show the explicit computations of the cosmological constraints of our model. 
For the ALP-dark photon system, this amounts to deriving the relic abundance and the $\Neff$ constraints in the presence of supercooling.
For ALPs coupled to the SM photon, we compute the length of the MD period after GW production.

\begin{figure*}
    \centering
    \includegraphics[width=\linewidth]{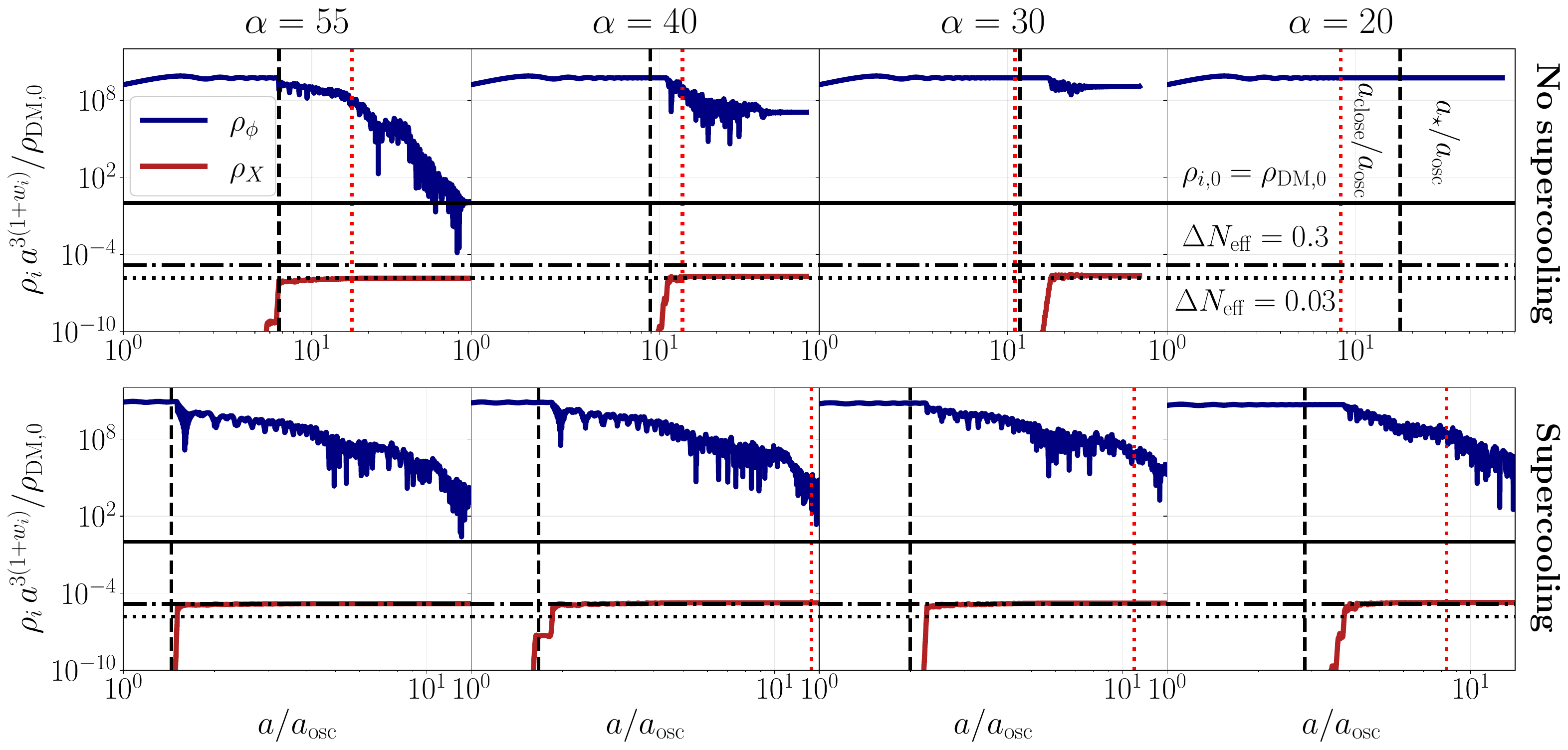}
    \caption{Numerical simulations of the ALP-dark photon system, solving the linearized equations of motion~\eqref{eq:axion_eom} and \eqref{eq:dark_photon_eom} for $\mathcal{O}(10^4)$ dark photon modes. We employ $\mphi = 10^{-11}\,\mathrm{GeV}$, $\fphi = 10^{17}\,\mathrm{GeV}$, and $\theta = 1.2$. The upper row shows the results for the original setup without supercooling, while in the bottom row ALP oscillations are delayed. The dark photon~(red) and axion~(blue) energy densities are redshifted to today and normalized to $\rho_\rmii{{DM},0}$. The black dashed line is the growth time estimate~\eqref{eq:growth_time_estimate}, while the red dotted line indicates our estimate of tachyonic band closure~\eqref{eq:tachyonic_band_closure}. Note that in the case of supercooling, we only display the initial phase of tachyonic growth to highlight the differences in $\astar/\aosc$. 
    We observe an excellent agreement between our analytic estimates and the numerical results, as tachyonic dark photon production quickly shuts off once $\alpha$ becomes small such that $\astar > a_\rmi{close}$.}
    \label{fig:simulation_results}
\end{figure*}
\subsection{Dark photon}
As outlined in the main text, the parameter space can be divided into cosmologically stable axions contributing to the DM energy density and heavy axions that decay before BBN.

The light ALP abundance is computed by redshifting the energy density after production until today,
\begin{equation}\label{eq:axion_DM_constraint}
    h^2 \Omegaphitoday = \epsilon_\mathrm{sup}  \Omegaphistar \left(\frac{H_\star}{H_{100}}\right)^2 \left(\frac{T_0}{T_\star}\right)^3 \frac{g_{s}^0}{g_{s}^\star} \leq 0.12 \, ,
\end{equation}
where $T_0 = 2.73\,\mathrm{K}$ is the temperature of the CMB and $H_{100} = 100\,\mathrm{km}(\mathrm{Mpc}\,\mathrm{s})^{-1}$. 
The suppression of the axion energy density through the production of dark photons is parametrized by $\epssup$.
Given eqs.~\eqref{eq:Hubble_osc},~\eqref{eq:Omegaphistar},~\eqref{eq:growth_time_estimate}, and~\eqref{eq:r_sc_max}, the relic abundance translates to
\begin{equation}
\begin{split}
    h^2 \Omegaphitoday \simeq \, & 4.95\times 10^{-5}\epssup \left(\frac{\theta \fphi}{10^{10}\,\mathrm{GeV}}\right)^2 \left(\frac{\mphi}{\mathrm{eV}}\right)^\frac{1}{2} \\
    &\times\rsc^{-3} (g_{\epsilon}^\mathrm{osc,aa})^\frac{3}{4} (g_s^\star)^{-1} \, .
\end{split}
\end{equation}
Employing the maximum amount of supercooling~\eqref{eq:r_sc_max}, we find the required suppression through the tachyonic instability, 
\begin{equation}
\begin{split}
    \epssup^\rmii{DM} \lesssim  2.89\, &\times \,10^{-9}\left(\frac{g_\epsilon^\star}{g_\epsilon^\rmi{osc}} \frac{\astar}{\aosc}\right)^\frac{3}{4} \\
    &\times \left(\frac{\mathrm{eV}}{\mphi}\right)^\frac{1}{2} \left(\frac{10^{10}\,\mathrm{GeV}}{\theta \fphi}\right)^\frac{1}{2}\, . 
\end{split}
\end{equation}

The heavy axion regime is defined by the requirement that the axion decay rate is larger than the Hubble parameter at BBN. 
After dark photon production through the tachyonic instability, heavy axions follow a matter-like scaling until decaying into dark photons. 
This induces a contribution to the effective relativistic degrees of freedom,
\begin{equation}\label{eq:Neff_decay_constraint}
    \Delta \Neff^\rmi{decay} = \frac{8}{7} \left(\frac{11}{4}\right)^\frac{4}{3} \frac{g^{\star}_{\epsilon}}{g_\gamma} \left(\frac{g_{s}^0}{g_s^{\star}}\right)^\frac{4}{3} \epssup \Omegaphistar \frac{a_\rmi{decay}}{\astar} \, .
\end{equation}
The scale factor ratio reads
\begin{equation}\label{eq:scalefactor_axiondecay}
    \frac{\astar}{a_\rmi{decay}} = \min\left\{1, \left(\frac{\Gamma_{\phi \to XX}}{H_\star}\right)^\frac{1}{2}\right\} \, ,
\end{equation}
hence depends on the timescale of axion decays. Here, the axion decay rate into dark photons is given by eq.~\eqref{eq:GammaPhi_larger_HBBN}.
In terms of our model parameters, we have
\begin{equation}
    \begin{split}
        \frac{\Gamma_{\phi \to XX}}{H_\star} \simeq \, &4.42 \times  10^{-15} \,\frac{(g_s^\star)^\frac{2}{3}}{(g_\epsilon^\star)^\frac{1}{2}} \left(\frac{\astar}{\aosc}\right)^\frac{3}{2}\\
        &\times\frac{\alpha^2}{\theta} \left(\frac{\mphi}{\GeV}\right)^2 \left(\frac{10^{10}\,\mathrm{GeV}}{\fphi}\right)^3\, .
    \end{split}
\end{equation}
From this expression we see that the axion mass needs to be extremely large for decays to become efficient right at the onset of oscillations. 
We do not consider masses in such a regime in the dark photon scenario. 
Therefore, the $\Neff$ contribution from perturbative decays effectively reads
\begin{equation}
    \begin{split}
        \Delta \Neff^\rmi{decay} \simeq \, &4.52\times 10^6\,\epssup (g_\epsilon^\star)^\frac{1}{4} (g_s^\star)^{-\frac{1}{3}} \\ &\times\left(\frac{\aosc}{\astar}\right)^\frac{3}{4} 
        \frac{\theta^\frac{1}{2}}{\alpha} \frac{\GeV}{\mphi} \left(\frac{\fphi}{10^{10}\,\GeV}\right)^\frac{3}{2} \,.
    \end{split}
\end{equation}
Imposing the constraint from the Planck 2018 dataset, $\Delta \Neff^\rmi{decay} < 0.3$~\cite{Planck:2018vyg}, we can derive a condition on the required suppression through the tachyonic resonance in the heavy ALP regime,
\begin{equation}
    \begin{split}
        \epssup^\rmi{decay} \lesssim \,&6.64\times 10^{-8}\,(g_\epsilon^\star)^{-\frac{1}{4}} (g_s^\star)^\frac{1}{3} \\ &\times \left(\frac{\astar}{\aosc}\right)^\frac{3}{4} \frac{\alpha}{\theta^\frac{1}{2}} \,\frac{\mphi}{\GeV}\left(\frac{10^{10}\,\GeV}{\fphi}\right)^\frac{3}{2} \, .
    \end{split}
\end{equation}

\subsection{SM photon}
Since no dark radiation is involved in the ALP-SM photon system, constraints from $\Neff$ do not apply.\footnote{Note that GWs act as dark radiation as well, however, the maximum amplitude we encounter is well below the $\Neff$ constraint.}
Therefore, an axion-driven inflationary period before GW production is not only allowed, but also necessary to sufficiently suppress the Debye mass of the photon in most of the parameter space. 
That is, the Universe is matter-dominated when ALP oscillations start.
It is, however, crucial to ensure that the MD period terminates before BBN, either through the tachyonic resonance or perturbative axion decays into photons.
To this end we again split the parameter space into several regimes, depending on their lifetime and the efficiency of Schwinger pair production. 

Let us start with light axions, where $\Gamma_{\phi \to \gamma\gamma} < \HBBN$, with the decay rate to photons given by eq.~\eqref{eq:GammaPhi_larger_HBBN}.
In this regime, a large amount of supercooling is required, such that the trapped ALP drives a period of thermal inflation and $\Omegaphistar \simeq 1$~(cf.~eq.~\eqref{eq:SM_mPhi_inflation}).
If the Schwinger effect blocks the resonant photon production, i.e., $\Omega_{\gamma}^\rmii{after} < 1$, the ALP would continue to dominate the energy density of the Universe after the onset of oscillations. 
Since perturbative decays to photons are prohibited, we would encounter a matter-domination period during BBN, which is incompatible with observations.
If, on the other hand, the electron mass suppresses Schwinger pair production, hence $\Omega_{\gamma}^\rmii{after} = 1$, a consistent cosmological evolution is possible.
Then, the tachyonic resonance is efficient enough to transition from the ALP-induced thermal inflation to RD.
Assuming quick thermalization, the reheating temperature reads
\begin{equation}\label{eq:T_rh_general}
    \Trh = \left(\frac{90}{\pi^2 \gepsrh}\right)^\frac{1}{4} \sqrt{\Hrh \MPl}\, .
\end{equation}
With
\begin{equation}
    \Hrh = \Hosc \left(\frac{\aosc}{\astar}\right)^\frac{3}{2} = \left(\frac{\rhophiosc}{3\MPl^2}\right)^\frac{1}{2} \left(\frac{\aosc}{\astar}\right)^\frac{3}{2}\, ,
\end{equation}
we have
\begin{equation}
    \begin{split}
        \Trh \simeq \,& 3.51\,\GeV \; (\gepsrh)^{-\frac{1}{4}} \left(\frac{\aosc}{\astar}\right)^\frac{3}{4} \\
        &\times\theta^\frac{1}{2} \left(\frac{\mphi}{\mathrm{eV}}\right)^\frac{1}{2} \left(\frac{\fphi}{10^{10}\,\GeV}\right)^\frac{1}{2}\,,
    \end{split}
\end{equation}
where $\aosc/\astar$ is our growth time estimate from appendix~\ref{app:minimal_alpha}.
Demanding $\Trh > \TBBN$ then gives the first bound on the model parameters.
The second constraint is derived from the present-day ALP abundance~\eqref{eq:axion_DM_constraint}.
In terms of the model parameters, we find
\begin{equation}
    \begin{split}
    h^2\Omegaphitoday =\,& 7.24 \times 10^8 \; \epssup (\gsrh)^{-1}(\gepsrh)^\frac{3}{4} \left(\frac{\aosc}{\astar}\right)^\frac{3}{4} \\
    &\times\theta^\frac{1}{2} \left(\frac{\mphi}{\mathrm{eV}}\right)^\frac{1}{2} \left(\frac{\fphi}{10^{10}\,\GeV}\right)^\frac{1}{2} \, ,
    \end{split}
\end{equation}
where we have used that for light axions, $\Omegaphistar = 1$, hence $\Omega_{\phi}^\rmii{after} = \epssup$~(cf.~eq.~\eqref{eq:Omega_phi_after}). 
Conversely, $h^2\Omegaphitoday \leq 0.12$ yields the required suppression through the tachyonic instability
\begin{equation}
    \begin{split}
    \epssup^\rmii{SM,DM} \lesssim \,& 1.66 \times 10^{-10}\;  \gsrh(\gepsrh)^{-\frac{3}{4}} \left(\frac{\astar}{\aosc}\right)^\frac{3}{4} \\
    &\times \left(\frac{\mathrm{eV}}{\mphi}\right)^\frac{1}{2} \left(\frac{10^{10}\,\GeV}{\theta \fphi}\right)^\frac{1}{2} \, .
    \end{split}
\end{equation}
If this bound is fulfilled, no MD period occurs and the GW spectrum receives the standard redshift factors~\eqref{eq:GW_peak_estimates_today}.

The heavy axion regime is defined by the condition $\Gamma_{\phi \to \gamma\gamma} \geq \HBBN$. 
This corresponds to the large-$\mphi$ region, where less supercooling is required~(cf.~eq.~\eqref{eq:Tosc_min_SM}).
Then the ALP does not necessarily dominate the energy density of the Universe at the time of production and the Hubble parameter at production reads
\begin{equation}
    H_\star = \left[\frac{1}{3\MPl^2} \left(\rho_{\phi,\star} + \rho_{\rmii{rad},\star} \right)\right]^\frac{1}{2} \, ,
\end{equation}
with
\begin{equation}
    \begin{split}
        \rhophistar &= \rhophiosc \left(\frac{\aosc}{\astar}\right)^3 \, , \\
        \rho_{\rmii{rad},\star} &= \rhorad(\Tosc) \left(\frac{\aosc}{\astar}\right)^4 \, ,
    \end{split}
\end{equation}
where $\Tosc$ is computed as outlined in sec.~\ref{sec:thermal_mass}.

\begin{figure*}
    \centering
    \includegraphics[width=\linewidth]{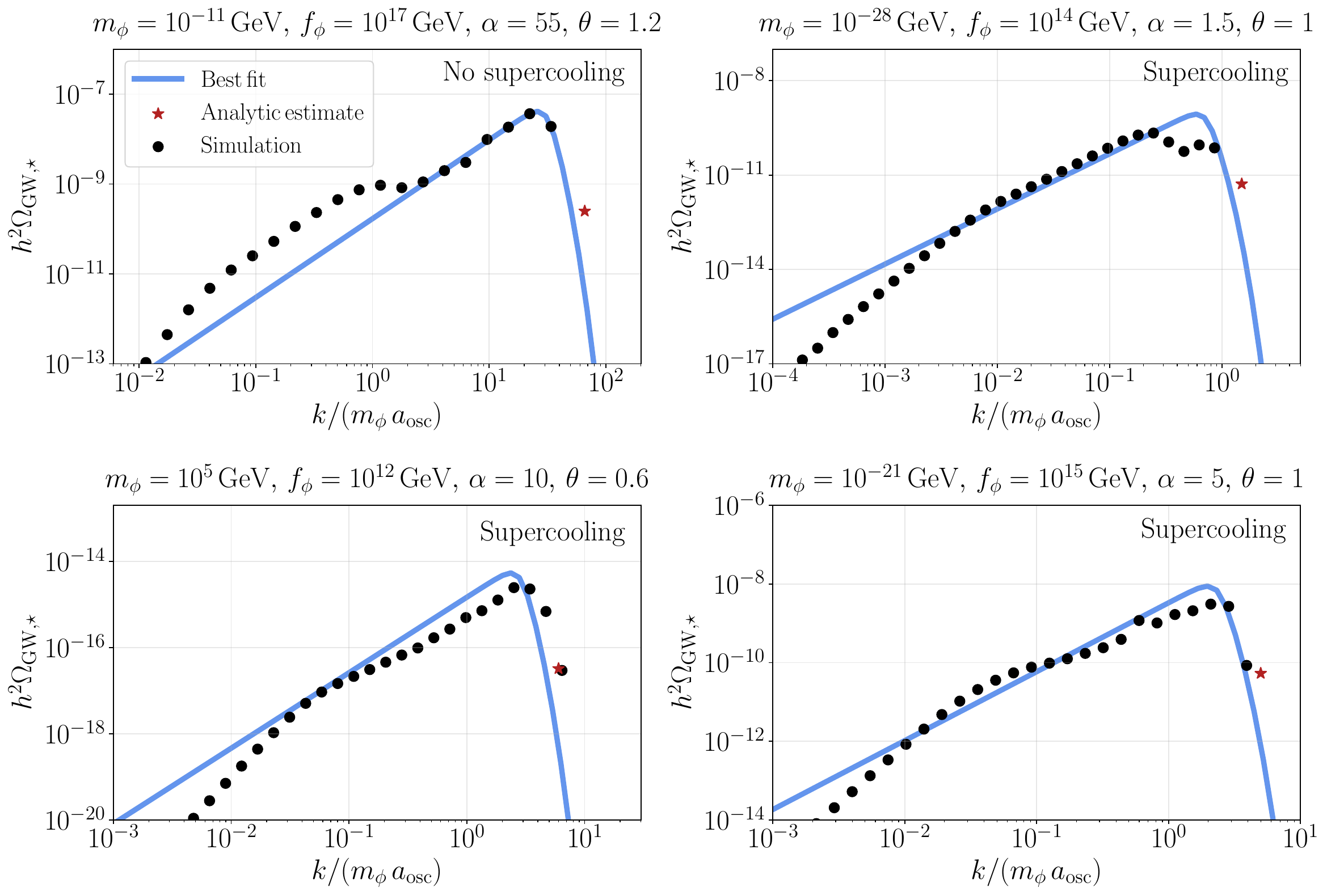}
    \caption{Numerical evaluation of the GW spectra at the time of production~(black dots) along with our analytic peak estimate~\eqref{eq:GW_peak_estimate_production} (red star) and the fit template~\eqref{eq:GW_fit_template}~(blue). To this end, we fit the GW spectrum employing benchmark 2 from ref.~\cite{Machado:2018nqk}~(upper left); see table~\ref{tab:GW_fit_parameters} for the best-fit parameters. Switching on supercooling and varying the model parameters, we observe a good agreement between the fit template and the numerical results.}
    \label{fig:GW_fits}
\end{figure*}

We can distinguish several cases.
If $\Gamma_{\phi \to \gamma\gamma} \geq H_\star$, the ALP decays right after photon production.
In this case the Universe remains radiation-dominated after production, and the GW abundance redshifts in the usual way.

If $\Gamma_{\phi \to \gamma\gamma} < H_\star$, the emergence of a MD period driven by ALP oscillations depends on the efficiency of Schwinger pair production and the resulting suppression of the axion abundance. 
If fermion production is efficient, only a fraction of the axion energy density is transferred to photons, i.e., $\Omega_{\gamma}^\rmii{after} < 1$.
As a consequence, the suppression of the axion abundance is fixed.
If, however, the entire energy density is transferred to the photons, $\Omega_{\gamma}^\rmii{after} = 1$. 
Then the axion suppression is dictated by the efficiency of the tachyonic instability, parametrized by $\epssup$.
Hence, we have
\begin{equation}
    \rhophiafter = \begin{cases}\displaystyle
        \rhophistar - \rho_{\gamma}^\rmii{after} \,, \quad &\rho_{\gamma}^\rmii{after} < \rhophistar \,, \\
        \displaystyle
        \epssup \,\rhophistar \, , \quad &\rho_{\gamma}^\rmii{after} = \rhophistar \, .
    \end{cases}
\end{equation}
The entire radiation energy density is then composed of the thermal bath along with the photons produced through the resonance,
\begin{equation}
    \rhorad^\rmii{after} = 3 H_\star^2 \MPl^2 - \rhophiafter \, .
\end{equation}
Since the axion energy density compared to the background is enhanced $\propto a$, MD sets in when
\begin{equation}
    \frac{\astar}{\amd} = \min\left\{1, \frac{\rhophiafter}{\rhoradafter}\right\} \, .
\end{equation}
The scale factor at the time of axion decay is obtained by solving
\begin{equation}
    \Hrh = \GammaphitoSM \, ,
\end{equation}
where
\begin{equation}
    \begin{split}
    \Hrh = \Biggr[\frac{1}{3\MPl^2} \biggr(\rhophiafter &\left(\frac{\astar}{\adecay}\right)^3 \\
    &+ \rhoradafter \left(\frac{\astar}{\adecay}\right)^4 \biggr)\Biggr]^\frac{1}{2} \, .
    \end{split}
\end{equation}
If $\amd \geq \adecay$, MD is avoided, and the redshift factors of the GW signal reduce to the standard expressions~\eqref{eq:GW_peak_estimates_today}.

If $\amd < \adecay$, the ALP starts to dominate the expansion rate of the Universe before decays become efficient. In that case we compute the length of the MD period by equating the Hubble parameter and the axion decay rate, where
\begin{equation}\label{eq:H_decay_matter_domination}
    \begin{split}
    H_\rmi{decay} = \Biggr[\frac{1}{3\MPl^2} \biggr( \rhophimd &\left(\frac{\amd}{\adecay}\right)^3 \\
    &+ \rhoradmd \left(\frac{\amd}{\adecay}\right)^4 \biggr)\Biggr]^\frac{1}{2} \, .
    \end{split}
\end{equation}
Here,
\begin{equation}
    \rhophimd = \rhophiafter \left(\frac{\astar}{\amd}\right)^3  = \rhoradafter \left(\frac{\astar}{\amd}\right)^4 = \rhoradmd \, .
\end{equation}
We solve for $\amd/\adecay$ and modify the redshift factors for the GW spectrum accordingly,
\begin{equation}\label{eq:redshift_matter_domination}
    \begin{gathered}
        h^2 \OmegaGWTildeToday = 1.67 \times 10^{-5} \left(\frac{100}{g_\epsilon^\rmi{rh} }\right)^\frac{1}{3} \Tilde{\Omega}_{\rmii{GW},\star} \frac{\amd}{\adecay} \, , \\
        \tilde{f}_0 = 1.65 \times 10^{-7} \,\mathrm{Hz}\, \frac{2 \ktildestar}{a_\star \Hrh} \frac{\Trh}{\mathrm{GeV}} \left(\frac{\gepsrh}{100}\right)^\frac{1}{6} \frac{\astar}{\adecay} \, ,
    \end{gathered}
\end{equation}
where $\Hrh = \GammaphitoSM$ and $\Trh$ is given by eq.~\eqref{eq:T_rh_general}.
Hence, the amplitude is only affected by the additional MD period.
The peak frequency, however, receives an additional redshift during a potential RD period between GW production and MD.

We conclude that a suppression of the Schwinger effect enhances the GW amplitude by increasing the energy budget available for GW emission and decreasing the length of the MD epoch.
In the case where Schwinger production is negligible, a large suppression of the axion abundance delays the onset of MD, i.e., further boosts the amplitude.
To study this effect, we choose the extreme values $\epssup \in \{0,1\}$ in fig.~\ref{fig:SM_photon_GWs}. 

\begin{table}
\setlength\tabcolsep{8pt}
\renewcommand{\arraystretch}{1.3}
\begin{tabular}{ |c|c|c|c| }
\hline
$\mathcal{A}_s$
& $f_s$
& $\gamma$ & $p$ \\
\hline
$268.57$ & $0.46$ & $8.34$ & $1.75$
\\
\hline
\end{tabular}
\renewcommand{\arraystretch}{1}
\caption{Best-fit parameters obtained by fitting the GW template~\eqref{eq:GW_fit_template} to the numerical computation in the original setup without supercooling~(see fig.~\ref{fig:GW_fits}, upper left).}
\label{tab:GW_fit_parameters}
\end{table}

\section{Numerical simulations}\label{app:numerics}
In this section, we discuss the numerical simulations of the supercooled ALP-dark photon system.
We use these results to verify our growth time estimate and extract the best-fit parameters for the GW template~\eqref{eq:GW_fit_template}.
To this end, we solve the coupled linearized equations of motion \eqref{eq:axion_eom} and \eqref{eq:dark_photon_eom} for $\mathcal{O}(10^4)$ dark photon modes with Bunch-Davies initial conditions~\cite{Bunch:1978yq}.
Fig.~\ref{fig:simulation_results} shows the evolution of the energy densities of the ALP~(blue) and dark photon~(red) imposing $\mphi = 10^{-11}\,\GeV$, $\fphi=10^{17}\,\GeV$, and $\theta = 1.2$, i.e., benchmark 2 from ref.~\cite{Machado:2018nqk}.
All energy densities are redshifted to today and normalized by the present DM abundance.
In the upper panel, the axion starts oscillating when $\Hosc = \mphi$, while the lower panel corresponds to the case with a finite amount of supercooling, $\Hosc < \mphi$.
We clearly see that after the onset of oscillations, energy transfer from the ALP to the dark photon quickly becomes efficient, exponentially amplifying dark photon modes from the vacuum.
Once $\rho_X \sim \rho_\phi$, backreaction sets in and the axion energy density is suppressed.

Let us note that fig.~\ref{fig:simulation_results} implies a suppression of the axion abundance by many orders of magnitude.
This is an artifact of the linearized approach. 
Lattice studies, taking into account all backreaction effects, have shown that the maximum possible axion suppression is strongly restricted~\cite{Ratzinger:2020oct}. 
Nevertheless, the linearized equations of motion provide reliable insights about the initial phase of tachyonic growth responsible for GW emission.

The black dashed line indicates our growth time estimate~\eqref{eq:growth_time_estimate}, which shows an excellent agreement with the numerics across all panels.
Furthermore, the red dotted line is the estimate for tachyonic band closure~\eqref{eq:tachyonic_band_closure}.
We clearly see that tachyonic production quickly becomes inefficient once $a_\star > a_\rmi{close}$, which verifies our band closure estimate. 
In the case of delayed axion oscillations, we find dark photon production to remain efficient for much smaller values of the ALP-dark photon coupling.
This is expected since the growth time scales $\propto \Hosc/\tilde{\omega}$, where $\Hosc \propto \rsc^2$ is decreased compared to the original setup. 
Hence, less time is required for dark photon growth. 
Therefore, a smaller $\alpha$ can be installed to have a successful energy transfer before band closure.

Given the simulated dark photon mode functions, we numerically compute the associated GW spectrum at the time of production; see ref.~\cite{Machado:2018nqk} for the full expression.
The results are shown in fig.~\ref{fig:GW_fits} by the black dots.
The red stars indicate our analytical estimates~\eqref{eq:GW_peak_estimate_production}.
Note that the deviation between the analytic prediction of the GW peak and the numerical result stems from discarding numerical prefactors~(cf.~ref.~\cite{Giblin:2014gra}).
We parametrize the GW signal by the template~\eqref{eq:GW_fit_template} and fit the peak region for the original setup without supercooling~(upper left).
The best-fit parameters are listed in table~\ref{tab:GW_fit_parameters}.
To verify the validity of our template we repeat the computation of the GW signal in the case of supercooled oscillations for different model parameters.
All benchmarks show excellent agreement between the template (light blue) and the numerical computation, confirming that we have accurately captured the scaling of all model parameters in our analytic calculations. 

\begin{figure}
    \centering
    \includegraphics[width=\linewidth]{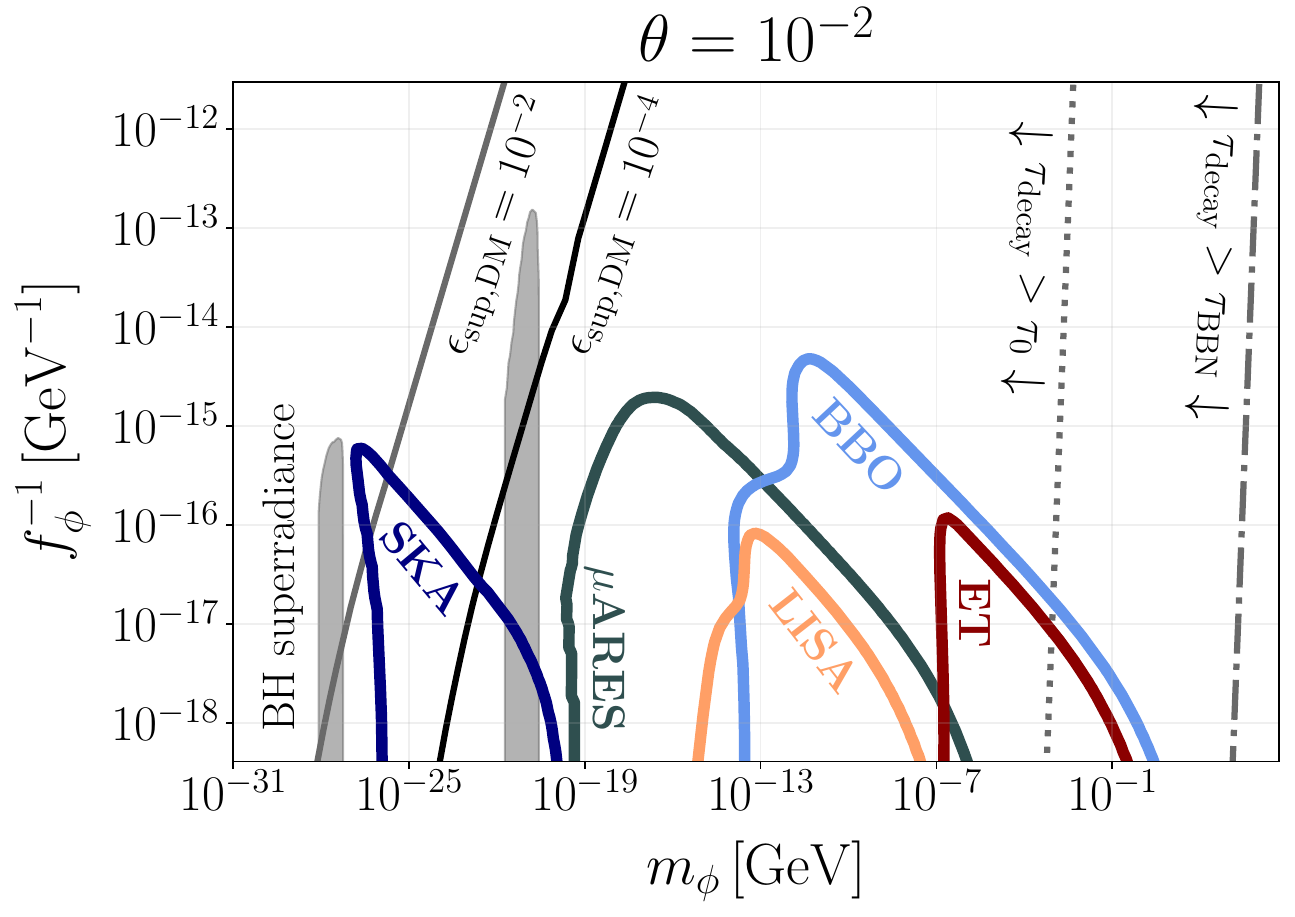}
    \caption{Observable parameter space in the ALP-dark photon scenario, for $\alpha = 2\alphamin$ and $\theta=10^{-2}$. While the parameter space consistent with ALP DM opens up towards larger masses, the GW amplitude is suppressed. Hence, the experimental sensitivities decrease and the testable ALP DM parameter region retains its size compared to the $\theta=1$ case.}
    \label{fig:results_small_alpha}
\end{figure}

\section{Varying $\theta$}\label{app:smaller_theta}
Throughout the main part of this work we have employed an initial misalignment angle $\theta=1$.
In this section, we impose $\theta=10^{-2}$, $\alpha=2\alphamin$ and repeat our analysis for the ALP-dark photon system; see fig.~\ref{fig:results_small_alpha}. 
The colored curves again display the sensitivity regions of several future observatories in the $\mphi - \fphi^{-1}$ plane, and the solid lines indicate the parameter space consistent with ALP DM assuming $\epssup \in \{10^{-2},10^{-4}\}$.

Lowering $\theta$ increases the mass range where the axion can constitute DM.
This is due to the fact that the supercooling ratio \eqref{eq:r_sc_max}, hence $\Tosc$, decreases with $\sqrt{\theta}$.
Then, the enhancement of the axion abundance relative to the background after tachyonic photon production is diminished~(cf.~eq.~\eqref{eq:axion_DM_constraint}).

From the considerations in appendix~\ref{app:minimal_alpha}, in particular eqs.~\eqref{eq:growth_time_app} and~\eqref{eq:tachyonic_band_closure_app}, we see that when lowering $\theta$, $\alpha$ has to be increased by the same factor in order to maintain the efficiency of dark photon production.
As a consequence, the GW amplitude decreases~(cf.~eq.~\eqref{eq:GW_peak_estimate_production}), and the observable parameter space shrinks.
As a consequence, the size of the region that is both observable and realizes ALP DM effectively remains unchanged.

\bibliographystyle{apsrev4-1}
\bibliography{biblio.bib}

\begin{thebibliography}{121}%
\makeatletter
\providecommand \@ifxundefined [1]{%
 \@ifx{#1\undefined}
}%
\providecommand \@ifnum [1]{%
 \ifnum #1\expandafter \@firstoftwo
 \else \expandafter \@secondoftwo
 \fi
}%
\providecommand \@ifx [1]{%
 \ifx #1\expandafter \@firstoftwo
 \else \expandafter \@secondoftwo
 \fi
}%
\providecommand \natexlab [1]{#1}%
\providecommand \enquote  [1]{``#1''}%
\providecommand \bibnamefont  [1]{#1}%
\providecommand \bibfnamefont [1]{#1}%
\providecommand \citenamefont [1]{#1}%
\providecommand \href@noop [0]{\@secondoftwo}%
\providecommand \href [0]{\begingroup \@sanitize@url \@href}%
\providecommand \@href[1]{\@@startlink{#1}\@@href}%
\providecommand \@@href[1]{\endgroup#1\@@endlink}%
\providecommand \@sanitize@url [0]{\catcode `\\12\catcode `\$12\catcode
  `\&12\catcode `\#12\catcode `\^12\catcode `\_12\catcode `\%12\relax}%
\providecommand \@@startlink[1]{}%
\providecommand \@@endlink[0]{}%
\providecommand \url  [0]{\begingroup\@sanitize@url \@url }%
\providecommand \@url [1]{\endgroup\@href {#1}{\urlprefix }}%
\providecommand \urlprefix  [0]{URL }%
\providecommand \Eprint [0]{\href }%
\providecommand \doibase [0]{http://dx.doi.org/}%
\providecommand \selectlanguage [0]{\@gobble}%
\providecommand \bibinfo  [0]{\@secondoftwo}%
\providecommand \bibfield  [0]{\@secondoftwo}%
\providecommand \translation [1]{[#1]}%
\providecommand \BibitemOpen [0]{}%
\providecommand \bibitemStop [0]{}%
\providecommand \bibitemNoStop [0]{.\EOS\space}%
\providecommand \EOS [0]{\spacefactor3000\relax}%
\providecommand \BibitemShut  [1]{\csname bibitem#1\endcsname}%
\let\auto@bib@innerbib\@empty
\bibitem [{\citenamefont {Peccei}\ and\ \citenamefont
  {Quinn}(1977{\natexlab{a}})}]{Peccei:1977hh}%
  \BibitemOpen
  \bibfield  {author} {\bibinfo {author} {\bibfnamefont {R.~D.}\ \bibnamefont
  {Peccei}}\ and\ \bibinfo {author} {\bibfnamefont {H.~R.}\ \bibnamefont
  {Quinn}},\ }\href {\doibase 10.1103/PhysRevLett.38.1440} {\bibfield
  {journal} {\bibinfo  {journal} {Phys. Rev. Lett.}\ }\textbf {\bibinfo
  {volume} {38}},\ \bibinfo {pages} {1440} (\bibinfo {year}
  {1977}{\natexlab{a}})}\BibitemShut {NoStop}%
\bibitem [{\citenamefont {Peccei}\ and\ \citenamefont
  {Quinn}(1977{\natexlab{b}})}]{Peccei:1977ur}%
  \BibitemOpen
  \bibfield  {author} {\bibinfo {author} {\bibfnamefont {R.~D.}\ \bibnamefont
  {Peccei}}\ and\ \bibinfo {author} {\bibfnamefont {H.~R.}\ \bibnamefont
  {Quinn}},\ }\href {\doibase 10.1103/PhysRevD.16.1791} {\bibfield  {journal}
  {\bibinfo  {journal} {Phys. Rev. D}\ }\textbf {\bibinfo {volume} {16}},\
  \bibinfo {pages} {1791} (\bibinfo {year} {1977}{\natexlab{b}})}\BibitemShut
  {NoStop}%
\bibitem [{\citenamefont {Weinberg}(1978)}]{Weinberg:1977ma}%
  \BibitemOpen
  \bibfield  {author} {\bibinfo {author} {\bibfnamefont {S.}~\bibnamefont
  {Weinberg}},\ }\href {\doibase 10.1103/PhysRevLett.40.223} {\bibfield
  {journal} {\bibinfo  {journal} {Phys. Rev. Lett.}\ }\textbf {\bibinfo
  {volume} {40}},\ \bibinfo {pages} {223} (\bibinfo {year} {1978})}\BibitemShut
  {NoStop}%
\bibitem [{\citenamefont {Wilczek}(1978)}]{Wilczek:1977pj}%
  \BibitemOpen
  \bibfield  {author} {\bibinfo {author} {\bibfnamefont {F.}~\bibnamefont
  {Wilczek}},\ }\href {\doibase 10.1103/PhysRevLett.40.279} {\bibfield
  {journal} {\bibinfo  {journal} {Phys. Rev. Lett.}\ }\textbf {\bibinfo
  {volume} {40}},\ \bibinfo {pages} {279} (\bibinfo {year} {1978})}\BibitemShut
  {NoStop}%
\bibitem [{\citenamefont {Preskill}\ \emph {et~al.}(1983)\citenamefont
  {Preskill}, \citenamefont {Wise},\ and\ \citenamefont
  {Wilczek}}]{Preskill:1982cy}%
  \BibitemOpen
  \bibfield  {author} {\bibinfo {author} {\bibfnamefont {J.}~\bibnamefont
  {Preskill}}, \bibinfo {author} {\bibfnamefont {M.~B.}\ \bibnamefont {Wise}},
  \ and\ \bibinfo {author} {\bibfnamefont {F.}~\bibnamefont {Wilczek}},\ }\href
  {\doibase 10.1016/0370-2693(83)90637-8} {\bibfield  {journal} {\bibinfo
  {journal} {Phys. Lett. B}\ }\textbf {\bibinfo {volume} {120}},\ \bibinfo
  {pages} {127} (\bibinfo {year} {1983})}\BibitemShut {NoStop}%
\bibitem [{\citenamefont {Abbott}\ and\ \citenamefont
  {Sikivie}(1983)}]{Abbott:1982af}%
  \BibitemOpen
  \bibfield  {author} {\bibinfo {author} {\bibfnamefont {L.~F.}\ \bibnamefont
  {Abbott}}\ and\ \bibinfo {author} {\bibfnamefont {P.}~\bibnamefont
  {Sikivie}},\ }\href {\doibase 10.1016/0370-2693(83)90638-X} {\bibfield
  {journal} {\bibinfo  {journal} {Phys. Lett. B}\ }\textbf {\bibinfo {volume}
  {120}},\ \bibinfo {pages} {133} (\bibinfo {year} {1983})}\BibitemShut
  {NoStop}%
\bibitem [{\citenamefont {Dine}\ and\ \citenamefont
  {Fischler}(1983)}]{Dine:1982ah}%
  \BibitemOpen
  \bibfield  {author} {\bibinfo {author} {\bibfnamefont {M.}~\bibnamefont
  {Dine}}\ and\ \bibinfo {author} {\bibfnamefont {W.}~\bibnamefont
  {Fischler}},\ }\href {\doibase 10.1016/0370-2693(83)90639-1} {\bibfield
  {journal} {\bibinfo  {journal} {Phys. Lett. B}\ }\textbf {\bibinfo {volume}
  {120}},\ \bibinfo {pages} {137} (\bibinfo {year} {1983})}\BibitemShut
  {NoStop}%
\bibitem [{\citenamefont {Witten}(1984)}]{Witten:1984dg}%
  \BibitemOpen
  \bibfield  {author} {\bibinfo {author} {\bibfnamefont {E.}~\bibnamefont
  {Witten}},\ }\href {\doibase 10.1016/0370-2693(84)90422-2} {\bibfield
  {journal} {\bibinfo  {journal} {Phys. Lett. B}\ }\textbf {\bibinfo {volume}
  {149}},\ \bibinfo {pages} {351} (\bibinfo {year} {1984})}\BibitemShut
  {NoStop}%
\bibitem [{\citenamefont {Svrcek}\ and\ \citenamefont
  {Witten}(2006)}]{Svrcek:2006yi}%
  \BibitemOpen
  \bibfield  {author} {\bibinfo {author} {\bibfnamefont {P.}~\bibnamefont
  {Svrcek}}\ and\ \bibinfo {author} {\bibfnamefont {E.}~\bibnamefont
  {Witten}},\ }\href {\doibase 10.1088/1126-6708/2006/06/051} {\bibfield
  {journal} {\bibinfo  {journal} {JHEP}\ }\textbf {\bibinfo {volume} {06}},\
  \bibinfo {pages} {051} (\bibinfo {year} {2006})},\ \Eprint
  {http://arxiv.org/abs/hep-th/0605206} {arXiv:hep-th/0605206} \BibitemShut
  {NoStop}%
\bibitem [{\citenamefont {Arvanitaki}\ \emph {et~al.}(2010)\citenamefont
  {Arvanitaki}, \citenamefont {Dimopoulos}, \citenamefont {Dubovsky},
  \citenamefont {Kaloper},\ and\ \citenamefont
  {March-Russell}}]{Arvanitaki:2009fg}%
  \BibitemOpen
  \bibfield  {author} {\bibinfo {author} {\bibfnamefont {A.}~\bibnamefont
  {Arvanitaki}}, \bibinfo {author} {\bibfnamefont {S.}~\bibnamefont
  {Dimopoulos}}, \bibinfo {author} {\bibfnamefont {S.}~\bibnamefont
  {Dubovsky}}, \bibinfo {author} {\bibfnamefont {N.}~\bibnamefont {Kaloper}}, \
  and\ \bibinfo {author} {\bibfnamefont {J.}~\bibnamefont {March-Russell}},\
  }\href {\doibase 10.1103/PhysRevD.81.123530} {\bibfield  {journal} {\bibinfo
  {journal} {Phys. Rev. D}\ }\textbf {\bibinfo {volume} {81}},\ \bibinfo
  {pages} {123530} (\bibinfo {year} {2010})},\ \Eprint
  {http://arxiv.org/abs/0905.4720} {arXiv:0905.4720 [hep-th]} \BibitemShut
  {NoStop}%
\bibitem [{\citenamefont {Marsh}(2016)}]{Marsh:2015xka}%
  \BibitemOpen
  \bibfield  {author} {\bibinfo {author} {\bibfnamefont {D.~J.~E.}\
  \bibnamefont {Marsh}},\ }\href {\doibase 10.1016/j.physrep.2016.06.005}
  {\bibfield  {journal} {\bibinfo  {journal} {Phys. Rept.}\ }\textbf {\bibinfo
  {volume} {643}},\ \bibinfo {pages} {1} (\bibinfo {year} {2016})},\ \Eprint
  {http://arxiv.org/abs/1510.07633} {arXiv:1510.07633 [astro-ph.CO]}
  \BibitemShut {NoStop}%
\bibitem [{\citenamefont {Freese}\ \emph {et~al.}(1990)\citenamefont {Freese},
  \citenamefont {Frieman},\ and\ \citenamefont {Olinto}}]{Freese:1990rb}%
  \BibitemOpen
  \bibfield  {author} {\bibinfo {author} {\bibfnamefont {K.}~\bibnamefont
  {Freese}}, \bibinfo {author} {\bibfnamefont {J.~A.}\ \bibnamefont {Frieman}},
  \ and\ \bibinfo {author} {\bibfnamefont {A.~V.}\ \bibnamefont {Olinto}},\
  }\href {\doibase 10.1103/PhysRevLett.65.3233} {\bibfield  {journal} {\bibinfo
   {journal} {Phys. Rev. Lett.}\ }\textbf {\bibinfo {volume} {65}},\ \bibinfo
  {pages} {3233} (\bibinfo {year} {1990})}\BibitemShut {NoStop}%
\bibitem [{\citenamefont {Pajer}\ and\ \citenamefont
  {Peloso}(2013)}]{Pajer:2013fsa}%
  \BibitemOpen
  \bibfield  {author} {\bibinfo {author} {\bibfnamefont {E.}~\bibnamefont
  {Pajer}}\ and\ \bibinfo {author} {\bibfnamefont {M.}~\bibnamefont {Peloso}},\
  }\href {\doibase 10.1088/0264-9381/30/21/214002} {\bibfield  {journal}
  {\bibinfo  {journal} {Class. Quant. Grav.}\ }\textbf {\bibinfo {volume}
  {30}},\ \bibinfo {pages} {214002} (\bibinfo {year} {2013})},\ \Eprint
  {http://arxiv.org/abs/1305.3557} {arXiv:1305.3557 [hep-th]} \BibitemShut
  {NoStop}%
\bibitem [{\citenamefont {Adshead}\ \emph {et~al.}(2015)\citenamefont
  {Adshead}, \citenamefont {Giblin}, \citenamefont {Scully},\ and\
  \citenamefont {Sfakianakis}}]{Adshead:2015pva}%
  \BibitemOpen
  \bibfield  {author} {\bibinfo {author} {\bibfnamefont {P.}~\bibnamefont
  {Adshead}}, \bibinfo {author} {\bibfnamefont {J.~T.}\ \bibnamefont {Giblin}},
  \bibinfo {author} {\bibfnamefont {T.~R.}\ \bibnamefont {Scully}}, \ and\
  \bibinfo {author} {\bibfnamefont {E.~I.}\ \bibnamefont {Sfakianakis}},\
  }\href {\doibase 10.1088/1475-7516/2015/12/034} {\bibfield  {journal}
  {\bibinfo  {journal} {JCAP}\ }\textbf {\bibinfo {volume} {12}},\ \bibinfo
  {pages} {034} (\bibinfo {year} {2015})},\ \Eprint
  {http://arxiv.org/abs/1502.06506} {arXiv:1502.06506 [astro-ph.CO]}
  \BibitemShut {NoStop}%
\bibitem [{\citenamefont {Daido}\ \emph {et~al.}(2017)\citenamefont {Daido},
  \citenamefont {Takahashi},\ and\ \citenamefont {Yin}}]{Daido:2017wwb}%
  \BibitemOpen
  \bibfield  {author} {\bibinfo {author} {\bibfnamefont {R.}~\bibnamefont
  {Daido}}, \bibinfo {author} {\bibfnamefont {F.}~\bibnamefont {Takahashi}}, \
  and\ \bibinfo {author} {\bibfnamefont {W.}~\bibnamefont {Yin}},\ }\href
  {\doibase 10.1088/1475-7516/2017/05/044} {\bibfield  {journal} {\bibinfo
  {journal} {JCAP}\ }\textbf {\bibinfo {volume} {05}},\ \bibinfo {pages} {044}
  (\bibinfo {year} {2017})},\ \Eprint {http://arxiv.org/abs/1702.03284}
  {arXiv:1702.03284 [hep-ph]} \BibitemShut {NoStop}%
\bibitem [{\citenamefont {Takahashi}\ and\ \citenamefont
  {Yin}(2021)}]{Takahashi:2021tff}%
  \BibitemOpen
  \bibfield  {author} {\bibinfo {author} {\bibfnamefont {F.}~\bibnamefont
  {Takahashi}}\ and\ \bibinfo {author} {\bibfnamefont {W.}~\bibnamefont
  {Yin}},\ }\href {\doibase 10.1088/1475-7516/2021/10/057} {\bibfield
  {journal} {\bibinfo  {journal} {JCAP}\ }\textbf {\bibinfo {volume} {10}},\
  \bibinfo {pages} {057} (\bibinfo {year} {2021})},\ \Eprint
  {http://arxiv.org/abs/2105.10493} {arXiv:2105.10493 [hep-ph]} \BibitemShut
  {NoStop}%
\bibitem [{\citenamefont {Domcke}\ \emph {et~al.}(2020)\citenamefont {Domcke},
  \citenamefont {Ema},\ and\ \citenamefont {Mukaida}}]{Domcke:2019qmm}%
  \BibitemOpen
  \bibfield  {author} {\bibinfo {author} {\bibfnamefont {V.}~\bibnamefont
  {Domcke}}, \bibinfo {author} {\bibfnamefont {Y.}~\bibnamefont {Ema}}, \ and\
  \bibinfo {author} {\bibfnamefont {K.}~\bibnamefont {Mukaida}},\ }\href
  {\doibase 10.1007/JHEP02(2020)055} {\bibfield  {journal} {\bibinfo  {journal}
  {JHEP}\ }\textbf {\bibinfo {volume} {02}},\ \bibinfo {pages} {055} (\bibinfo
  {year} {2020})},\ \Eprint {http://arxiv.org/abs/1910.01205} {arXiv:1910.01205
  [hep-ph]} \BibitemShut {NoStop}%
\bibitem [{\citenamefont {Kitajima}\ \emph {et~al.}(2022)\citenamefont
  {Kitajima}, \citenamefont {Nakagawa},\ and\ \citenamefont
  {Takahashi}}]{Kitajima:2021bjq}%
  \BibitemOpen
  \bibfield  {author} {\bibinfo {author} {\bibfnamefont {N.}~\bibnamefont
  {Kitajima}}, \bibinfo {author} {\bibfnamefont {S.}~\bibnamefont {Nakagawa}},
  \ and\ \bibinfo {author} {\bibfnamefont {F.}~\bibnamefont {Takahashi}},\
  }\href {\doibase 10.1103/PhysRevD.105.103011} {\bibfield  {journal} {\bibinfo
   {journal} {Phys. Rev. D}\ }\textbf {\bibinfo {volume} {105}},\ \bibinfo
  {pages} {103011} (\bibinfo {year} {2022})},\ \Eprint
  {http://arxiv.org/abs/2111.06696} {arXiv:2111.06696 [hep-ph]} \BibitemShut
  {NoStop}%
\bibitem [{\citenamefont {O'Hare}(2024)}]{OHare:2024nmr}%
  \BibitemOpen
  \bibfield  {author} {\bibinfo {author} {\bibfnamefont {C.~A.~J.}\
  \bibnamefont {O'Hare}},\ }\href {\doibase 10.22323/1.454.0040} {\bibfield
  {journal} {\bibinfo  {journal} {PoS}\ }\textbf {\bibinfo {volume}
  {COSMICWISPers}},\ \bibinfo {pages} {040} (\bibinfo {year} {2024})},\ \Eprint
  {http://arxiv.org/abs/2403.17697} {arXiv:2403.17697 [hep-ph]} \BibitemShut
  {NoStop}%
\bibitem [{\citenamefont {Abbott}\ \emph {et~al.}(2016)\citenamefont {Abbott}
  \emph {et~al.}}]{LIGOScientific:2016aoc}%
  \BibitemOpen
  \bibfield  {author} {\bibinfo {author} {\bibfnamefont {B.~P.}\ \bibnamefont
  {Abbott}} \emph {et~al.} (\bibinfo {collaboration} {LIGO Scientific,
  Virgo}),\ }\href {\doibase 10.1103/PhysRevLett.116.061102} {\bibfield
  {journal} {\bibinfo  {journal} {Phys. Rev. Lett.}\ }\textbf {\bibinfo
  {volume} {116}},\ \bibinfo {pages} {061102} (\bibinfo {year} {2016})},\
  \Eprint {http://arxiv.org/abs/1602.03837} {arXiv:1602.03837 [gr-qc]}
  \BibitemShut {NoStop}%
\bibitem [{\citenamefont {Arzoumanian}\ \emph {et~al.}(2020)\citenamefont
  {Arzoumanian} \emph {et~al.}}]{NANOGrav:2020bcs}%
  \BibitemOpen
  \bibfield  {author} {\bibinfo {author} {\bibfnamefont {Z.}~\bibnamefont
  {Arzoumanian}} \emph {et~al.} (\bibinfo {collaboration} {NANOGrav}),\ }\href
  {\doibase 10.3847/2041-8213/abd401} {\bibfield  {journal} {\bibinfo
  {journal} {Astrophys. J. Lett.}\ }\textbf {\bibinfo {volume} {905}},\
  \bibinfo {pages} {L34} (\bibinfo {year} {2020})},\ \Eprint
  {http://arxiv.org/abs/2009.04496} {arXiv:2009.04496 [astro-ph.HE]}
  \BibitemShut {NoStop}%
\bibitem [{\citenamefont {Arzoumanian}\ \emph {et~al.}(2021)\citenamefont
  {Arzoumanian} \emph {et~al.}}]{NANOGrav:2021flc}%
  \BibitemOpen
  \bibfield  {author} {\bibinfo {author} {\bibfnamefont {Z.}~\bibnamefont
  {Arzoumanian}} \emph {et~al.} (\bibinfo {collaboration} {NANOGrav}),\ }\href
  {\doibase 10.1103/PhysRevLett.127.251302} {\bibfield  {journal} {\bibinfo
  {journal} {Phys. Rev. Lett.}\ }\textbf {\bibinfo {volume} {127}},\ \bibinfo
  {pages} {251302} (\bibinfo {year} {2021})},\ \Eprint
  {http://arxiv.org/abs/2104.13930} {arXiv:2104.13930 [astro-ph.CO]}
  \BibitemShut {NoStop}%
\bibitem [{\citenamefont {Goncharov}\ \emph {et~al.}(2021)\citenamefont
  {Goncharov} \emph {et~al.}}]{Goncharov:2021oub}%
  \BibitemOpen
  \bibfield  {author} {\bibinfo {author} {\bibfnamefont {B.}~\bibnamefont
  {Goncharov}} \emph {et~al.},\ }\href {\doibase 10.3847/2041-8213/ac17f4}
  {\bibfield  {journal} {\bibinfo  {journal} {Astrophys. J. Lett.}\ }\textbf
  {\bibinfo {volume} {917}},\ \bibinfo {pages} {L19} (\bibinfo {year}
  {2021})},\ \Eprint {http://arxiv.org/abs/2107.12112} {arXiv:2107.12112
  [astro-ph.HE]} \BibitemShut {NoStop}%
\bibitem [{\citenamefont {Chen}\ \emph {et~al.}(2021)\citenamefont {Chen} \emph
  {et~al.}}]{EPTA:2021crs}%
  \BibitemOpen
  \bibfield  {author} {\bibinfo {author} {\bibfnamefont {S.}~\bibnamefont
  {Chen}} \emph {et~al.} (\bibinfo {collaboration} {EPTA}),\ }\href {\doibase
  10.1093/mnras/stab2833} {\bibfield  {journal} {\bibinfo  {journal} {Mon. Not.
  Roy. Astron. Soc.}\ }\textbf {\bibinfo {volume} {508}},\ \bibinfo {pages}
  {4970} (\bibinfo {year} {2021})},\ \Eprint {http://arxiv.org/abs/2110.13184}
  {arXiv:2110.13184 [astro-ph.HE]} \BibitemShut {NoStop}%
\bibitem [{\citenamefont {Antoniadis}\ \emph {et~al.}(2022)\citenamefont
  {Antoniadis} \emph {et~al.}}]{Antoniadis:2022pcn}%
  \BibitemOpen
  \bibfield  {author} {\bibinfo {author} {\bibfnamefont {J.}~\bibnamefont
  {Antoniadis}} \emph {et~al.},\ }\href {\doibase 10.1093/mnras/stab3418}
  {\bibfield  {journal} {\bibinfo  {journal} {Mon. Not. Roy. Astron. Soc.}\
  }\textbf {\bibinfo {volume} {510}},\ \bibinfo {pages} {4873} (\bibinfo {year}
  {2022})},\ \Eprint {http://arxiv.org/abs/2201.03980} {arXiv:2201.03980
  [astro-ph.HE]} \BibitemShut {NoStop}%
\bibitem [{\citenamefont {Agazie}\ \emph {et~al.}(2023)\citenamefont {Agazie}
  \emph {et~al.}}]{NANOGrav:2023gor}%
  \BibitemOpen
  \bibfield  {author} {\bibinfo {author} {\bibfnamefont {G.}~\bibnamefont
  {Agazie}} \emph {et~al.} (\bibinfo {collaboration} {NANOGrav}),\ }\href
  {\doibase 10.3847/2041-8213/acdac6} {\bibfield  {journal} {\bibinfo
  {journal} {Astrophys. J. Lett.}\ }\textbf {\bibinfo {volume} {951}},\
  \bibinfo {pages} {L8} (\bibinfo {year} {2023})},\ \Eprint
  {http://arxiv.org/abs/2306.16213} {arXiv:2306.16213 [astro-ph.HE]}
  \BibitemShut {NoStop}%
\bibitem [{\citenamefont {Antoniadis}\ \emph {et~al.}(2023)\citenamefont
  {Antoniadis} \emph {et~al.}}]{EPTA:2023fyk}%
  \BibitemOpen
  \bibfield  {author} {\bibinfo {author} {\bibfnamefont {J.}~\bibnamefont
  {Antoniadis}} \emph {et~al.} (\bibinfo {collaboration} {EPTA, InPTA:}),\
  }\href {\doibase 10.1051/0004-6361/202346844} {\bibfield  {journal} {\bibinfo
   {journal} {Astron. Astrophys.}\ }\textbf {\bibinfo {volume} {678}},\
  \bibinfo {pages} {A50} (\bibinfo {year} {2023})},\ \Eprint
  {http://arxiv.org/abs/2306.16214} {arXiv:2306.16214 [astro-ph.HE]}
  \BibitemShut {NoStop}%
\bibitem [{\citenamefont {Reardon}\ \emph {et~al.}(2023)\citenamefont {Reardon}
  \emph {et~al.}}]{Reardon:2023gzh}%
  \BibitemOpen
  \bibfield  {author} {\bibinfo {author} {\bibfnamefont {D.~J.}\ \bibnamefont
  {Reardon}} \emph {et~al.},\ }\href {\doibase 10.3847/2041-8213/acdd02}
  {\bibfield  {journal} {\bibinfo  {journal} {Astrophys. J. Lett.}\ }\textbf
  {\bibinfo {volume} {951}},\ \bibinfo {pages} {L6} (\bibinfo {year} {2023})},\
  \Eprint {http://arxiv.org/abs/2306.16215} {arXiv:2306.16215 [astro-ph.HE]}
  \BibitemShut {NoStop}%
\bibitem [{\citenamefont {Xu}\ \emph {et~al.}(2023)\citenamefont {Xu} \emph
  {et~al.}}]{Xu:2023wog}%
  \BibitemOpen
  \bibfield  {author} {\bibinfo {author} {\bibfnamefont {H.}~\bibnamefont {Xu}}
  \emph {et~al.},\ }\href {\doibase 10.1088/1674-4527/acdfa5} {\bibfield
  {journal} {\bibinfo  {journal} {Res. Astron. Astrophys.}\ }\textbf {\bibinfo
  {volume} {23}},\ \bibinfo {pages} {075024} (\bibinfo {year} {2023})},\
  \Eprint {http://arxiv.org/abs/2306.16216} {arXiv:2306.16216 [astro-ph.HE]}
  \BibitemShut {NoStop}%
\bibitem [{\citenamefont {Miles}\ \emph {et~al.}(2024)\citenamefont {Miles}
  \emph {et~al.}}]{Miles:2024seg}%
  \BibitemOpen
  \bibfield  {author} {\bibinfo {author} {\bibfnamefont {M.~T.}\ \bibnamefont
  {Miles}} \emph {et~al.},\ }\href {\doibase 10.1093/mnras/stae2571} {\bibfield
   {journal} {\bibinfo  {journal} {Mon. Not. Roy. Astron. Soc.}\ }\textbf
  {\bibinfo {volume} {536}},\ \bibinfo {pages} {1489} (\bibinfo {year}
  {2024})},\ \Eprint {http://arxiv.org/abs/2412.01153} {arXiv:2412.01153
  [astro-ph.HE]} \BibitemShut {NoStop}%
\bibitem [{\citenamefont {Anber}\ and\ \citenamefont
  {Sorbo}(2010)}]{Anber:2009ua}%
  \BibitemOpen
  \bibfield  {author} {\bibinfo {author} {\bibfnamefont {M.~M.}\ \bibnamefont
  {Anber}}\ and\ \bibinfo {author} {\bibfnamefont {L.}~\bibnamefont {Sorbo}},\
  }\href {\doibase 10.1103/PhysRevD.81.043534} {\bibfield  {journal} {\bibinfo
  {journal} {Phys. Rev. D}\ }\textbf {\bibinfo {volume} {81}},\ \bibinfo
  {pages} {043534} (\bibinfo {year} {2010})},\ \Eprint
  {http://arxiv.org/abs/0908.4089} {arXiv:0908.4089 [hep-th]} \BibitemShut
  {NoStop}%
\bibitem [{\citenamefont {Anber}\ and\ \citenamefont
  {Sorbo}(2012)}]{Anber:2012du}%
  \BibitemOpen
  \bibfield  {author} {\bibinfo {author} {\bibfnamefont {M.~M.}\ \bibnamefont
  {Anber}}\ and\ \bibinfo {author} {\bibfnamefont {L.}~\bibnamefont {Sorbo}},\
  }\href {\doibase 10.1103/PhysRevD.85.123537} {\bibfield  {journal} {\bibinfo
  {journal} {Phys. Rev. D}\ }\textbf {\bibinfo {volume} {85}},\ \bibinfo
  {pages} {123537} (\bibinfo {year} {2012})},\ \Eprint
  {http://arxiv.org/abs/1203.5849} {arXiv:1203.5849 [astro-ph.CO]} \BibitemShut
  {NoStop}%
\bibitem [{\citenamefont {Barnaby}\ \emph {et~al.}(2012)\citenamefont
  {Barnaby}, \citenamefont {Moxon}, \citenamefont {Namba}, \citenamefont
  {Peloso}, \citenamefont {Shiu},\ and\ \citenamefont {Zhou}}]{Barnaby:2012xt}%
  \BibitemOpen
  \bibfield  {author} {\bibinfo {author} {\bibfnamefont {N.}~\bibnamefont
  {Barnaby}}, \bibinfo {author} {\bibfnamefont {J.}~\bibnamefont {Moxon}},
  \bibinfo {author} {\bibfnamefont {R.}~\bibnamefont {Namba}}, \bibinfo
  {author} {\bibfnamefont {M.}~\bibnamefont {Peloso}}, \bibinfo {author}
  {\bibfnamefont {G.}~\bibnamefont {Shiu}}, \ and\ \bibinfo {author}
  {\bibfnamefont {P.}~\bibnamefont {Zhou}},\ }\href {\doibase
  10.1103/PhysRevD.86.103508} {\bibfield  {journal} {\bibinfo  {journal} {Phys.
  Rev. D}\ }\textbf {\bibinfo {volume} {86}},\ \bibinfo {pages} {103508}
  (\bibinfo {year} {2012})},\ \Eprint {http://arxiv.org/abs/1206.6117}
  {arXiv:1206.6117 [astro-ph.CO]} \BibitemShut {NoStop}%
\bibitem [{\citenamefont {Domcke}\ \emph {et~al.}(2016)\citenamefont {Domcke},
  \citenamefont {Pieroni},\ and\ \citenamefont {Bin\'etruy}}]{Domcke:2016bkh}%
  \BibitemOpen
  \bibfield  {author} {\bibinfo {author} {\bibfnamefont {V.}~\bibnamefont
  {Domcke}}, \bibinfo {author} {\bibfnamefont {M.}~\bibnamefont {Pieroni}}, \
  and\ \bibinfo {author} {\bibfnamefont {P.}~\bibnamefont {Bin\'etruy}},\
  }\href {\doibase 10.1088/1475-7516/2016/06/031} {\bibfield  {journal}
  {\bibinfo  {journal} {JCAP}\ }\textbf {\bibinfo {volume} {06}},\ \bibinfo
  {pages} {031} (\bibinfo {year} {2016})},\ \Eprint
  {http://arxiv.org/abs/1603.01287} {arXiv:1603.01287 [astro-ph.CO]}
  \BibitemShut {NoStop}%
\bibitem [{\citenamefont {Greene}\ \emph {et~al.}(1997)\citenamefont {Greene},
  \citenamefont {Kofman}, \citenamefont {Linde},\ and\ \citenamefont
  {Starobinsky}}]{Greene:1997fu}%
  \BibitemOpen
  \bibfield  {author} {\bibinfo {author} {\bibfnamefont {P.~B.}\ \bibnamefont
  {Greene}}, \bibinfo {author} {\bibfnamefont {L.}~\bibnamefont {Kofman}},
  \bibinfo {author} {\bibfnamefont {A.~D.}\ \bibnamefont {Linde}}, \ and\
  \bibinfo {author} {\bibfnamefont {A.~A.}\ \bibnamefont {Starobinsky}},\
  }\href {\doibase 10.1103/PhysRevD.56.6175} {\bibfield  {journal} {\bibinfo
  {journal} {Phys. Rev. D}\ }\textbf {\bibinfo {volume} {56}},\ \bibinfo
  {pages} {6175} (\bibinfo {year} {1997})},\ \Eprint
  {http://arxiv.org/abs/hep-ph/9705347} {arXiv:hep-ph/9705347} \BibitemShut
  {NoStop}%
\bibitem [{\citenamefont {Kofman}\ \emph {et~al.}(1994)\citenamefont {Kofman},
  \citenamefont {Linde},\ and\ \citenamefont {Starobinsky}}]{Kofman:1994rk}%
  \BibitemOpen
  \bibfield  {author} {\bibinfo {author} {\bibfnamefont {L.}~\bibnamefont
  {Kofman}}, \bibinfo {author} {\bibfnamefont {A.~D.}\ \bibnamefont {Linde}}, \
  and\ \bibinfo {author} {\bibfnamefont {A.~A.}\ \bibnamefont {Starobinsky}},\
  }\href {\doibase 10.1103/PhysRevLett.73.3195} {\bibfield  {journal} {\bibinfo
   {journal} {Phys. Rev. Lett.}\ }\textbf {\bibinfo {volume} {73}},\ \bibinfo
  {pages} {3195} (\bibinfo {year} {1994})},\ \Eprint
  {http://arxiv.org/abs/hep-th/9405187} {arXiv:hep-th/9405187} \BibitemShut
  {NoStop}%
\bibitem [{\citenamefont {Shtanov}\ \emph {et~al.}(1995)\citenamefont
  {Shtanov}, \citenamefont {Traschen},\ and\ \citenamefont
  {Brandenberger}}]{Shtanov:1994ce}%
  \BibitemOpen
  \bibfield  {author} {\bibinfo {author} {\bibfnamefont {Y.}~\bibnamefont
  {Shtanov}}, \bibinfo {author} {\bibfnamefont {J.~H.}\ \bibnamefont
  {Traschen}}, \ and\ \bibinfo {author} {\bibfnamefont {R.~H.}\ \bibnamefont
  {Brandenberger}},\ }\href {\doibase 10.1103/PhysRevD.51.5438} {\bibfield
  {journal} {\bibinfo  {journal} {Phys. Rev. D}\ }\textbf {\bibinfo {volume}
  {51}},\ \bibinfo {pages} {5438} (\bibinfo {year} {1995})},\ \Eprint
  {http://arxiv.org/abs/hep-ph/9407247} {arXiv:hep-ph/9407247} \BibitemShut
  {NoStop}%
\bibitem [{\citenamefont {Kofman}\ \emph {et~al.}(1997)\citenamefont {Kofman},
  \citenamefont {Linde},\ and\ \citenamefont {Starobinsky}}]{Kofman:1997yn}%
  \BibitemOpen
  \bibfield  {author} {\bibinfo {author} {\bibfnamefont {L.}~\bibnamefont
  {Kofman}}, \bibinfo {author} {\bibfnamefont {A.~D.}\ \bibnamefont {Linde}}, \
  and\ \bibinfo {author} {\bibfnamefont {A.~A.}\ \bibnamefont {Starobinsky}},\
  }\href {\doibase 10.1103/PhysRevD.56.3258} {\bibfield  {journal} {\bibinfo
  {journal} {Phys. Rev. D}\ }\textbf {\bibinfo {volume} {56}},\ \bibinfo
  {pages} {3258} (\bibinfo {year} {1997})},\ \Eprint
  {http://arxiv.org/abs/hep-ph/9704452} {arXiv:hep-ph/9704452} \BibitemShut
  {NoStop}%
\bibitem [{\citenamefont {Dufaux}\ \emph {et~al.}(2007)\citenamefont {Dufaux},
  \citenamefont {Bergman}, \citenamefont {Felder}, \citenamefont {Kofman},\
  and\ \citenamefont {Uzan}}]{Dufaux:2007pt}%
  \BibitemOpen
  \bibfield  {author} {\bibinfo {author} {\bibfnamefont {J.~F.}\ \bibnamefont
  {Dufaux}}, \bibinfo {author} {\bibfnamefont {A.}~\bibnamefont {Bergman}},
  \bibinfo {author} {\bibfnamefont {G.~N.}\ \bibnamefont {Felder}}, \bibinfo
  {author} {\bibfnamefont {L.}~\bibnamefont {Kofman}}, \ and\ \bibinfo {author}
  {\bibfnamefont {J.-P.}\ \bibnamefont {Uzan}},\ }\href {\doibase
  10.1103/PhysRevD.76.123517} {\bibfield  {journal} {\bibinfo  {journal} {Phys.
  Rev. D}\ }\textbf {\bibinfo {volume} {76}},\ \bibinfo {pages} {123517}
  (\bibinfo {year} {2007})},\ \Eprint {http://arxiv.org/abs/0707.0875}
  {arXiv:0707.0875 [astro-ph]} \BibitemShut {NoStop}%
\bibitem [{\citenamefont {Maleknejad}(2016)}]{Maleknejad:2016qjz}%
  \BibitemOpen
  \bibfield  {author} {\bibinfo {author} {\bibfnamefont {A.}~\bibnamefont
  {Maleknejad}},\ }\href {\doibase 10.1007/JHEP07(2016)104} {\bibfield
  {journal} {\bibinfo  {journal} {JHEP}\ }\textbf {\bibinfo {volume} {07}},\
  \bibinfo {pages} {104} (\bibinfo {year} {2016})},\ \Eprint
  {http://arxiv.org/abs/1604.03327} {arXiv:1604.03327 [hep-ph]} \BibitemShut
  {NoStop}%
\bibitem [{\citenamefont {Figueroa}\ and\ \citenamefont
  {Torrenti}(2017)}]{Figueroa:2016wxr}%
  \BibitemOpen
  \bibfield  {author} {\bibinfo {author} {\bibfnamefont {D.~G.}\ \bibnamefont
  {Figueroa}}\ and\ \bibinfo {author} {\bibfnamefont {F.}~\bibnamefont
  {Torrenti}},\ }\href {\doibase 10.1088/1475-7516/2017/02/001} {\bibfield
  {journal} {\bibinfo  {journal} {JCAP}\ }\textbf {\bibinfo {volume} {02}},\
  \bibinfo {pages} {001} (\bibinfo {year} {2017})},\ \Eprint
  {http://arxiv.org/abs/1609.05197} {arXiv:1609.05197 [astro-ph.CO]}
  \BibitemShut {NoStop}%
\bibitem [{\citenamefont {Adshead}\ \emph {et~al.}(2018)\citenamefont
  {Adshead}, \citenamefont {Giblin},\ and\ \citenamefont
  {Weiner}}]{Adshead:2018doq}%
  \BibitemOpen
  \bibfield  {author} {\bibinfo {author} {\bibfnamefont {P.}~\bibnamefont
  {Adshead}}, \bibinfo {author} {\bibfnamefont {J.~T.}\ \bibnamefont {Giblin}},
  \ and\ \bibinfo {author} {\bibfnamefont {Z.~J.}\ \bibnamefont {Weiner}},\
  }\href {\doibase 10.1103/PhysRevD.98.043525} {\bibfield  {journal} {\bibinfo
  {journal} {Phys. Rev. D}\ }\textbf {\bibinfo {volume} {98}},\ \bibinfo
  {pages} {043525} (\bibinfo {year} {2018})},\ \Eprint
  {http://arxiv.org/abs/1805.04550} {arXiv:1805.04550 [astro-ph.CO]}
  \BibitemShut {NoStop}%
\bibitem [{\citenamefont {Cuissa}\ and\ \citenamefont
  {Figueroa}(2019)}]{Cuissa:2018oiw}%
  \BibitemOpen
  \bibfield  {author} {\bibinfo {author} {\bibfnamefont {J.~R.~C.}\
  \bibnamefont {Cuissa}}\ and\ \bibinfo {author} {\bibfnamefont {D.~G.}\
  \bibnamefont {Figueroa}},\ }\href {\doibase 10.1088/1475-7516/2019/06/002}
  {\bibfield  {journal} {\bibinfo  {journal} {JCAP}\ }\textbf {\bibinfo
  {volume} {06}},\ \bibinfo {pages} {002} (\bibinfo {year} {2019})},\ \Eprint
  {http://arxiv.org/abs/1812.03132} {arXiv:1812.03132 [astro-ph.CO]}
  \BibitemShut {NoStop}%
\bibitem [{\citenamefont {Cyncynates}\ and\ \citenamefont
  {Weiner}(2023)}]{Cyncynates:2023zwj}%
  \BibitemOpen
  \bibfield  {author} {\bibinfo {author} {\bibfnamefont {D.}~\bibnamefont
  {Cyncynates}}\ and\ \bibinfo {author} {\bibfnamefont {Z.~J.}\ \bibnamefont
  {Weiner}},\ }\href@noop {} {\  (\bibinfo {year} {2023})},\ \Eprint
  {http://arxiv.org/abs/2310.18397} {arXiv:2310.18397 [hep-ph]} \BibitemShut
  {NoStop}%
\bibitem [{\citenamefont {Cyncynates}\ and\ \citenamefont
  {Weiner}(2024)}]{Cyncynates:2024yxm}%
  \BibitemOpen
  \bibfield  {author} {\bibinfo {author} {\bibfnamefont {D.}~\bibnamefont
  {Cyncynates}}\ and\ \bibinfo {author} {\bibfnamefont {Z.~J.}\ \bibnamefont
  {Weiner}},\ }\href@noop {} {\  (\bibinfo {year} {2024})},\ \Eprint
  {http://arxiv.org/abs/2410.14774} {arXiv:2410.14774 [hep-ph]} \BibitemShut
  {NoStop}%
\bibitem [{\citenamefont {Machado}\ \emph {et~al.}(2019)\citenamefont
  {Machado}, \citenamefont {Ratzinger}, \citenamefont {Schwaller},\ and\
  \citenamefont {Stefanek}}]{Machado:2018nqk}%
  \BibitemOpen
  \bibfield  {author} {\bibinfo {author} {\bibfnamefont {C.~S.}\ \bibnamefont
  {Machado}}, \bibinfo {author} {\bibfnamefont {W.}~\bibnamefont {Ratzinger}},
  \bibinfo {author} {\bibfnamefont {P.}~\bibnamefont {Schwaller}}, \ and\
  \bibinfo {author} {\bibfnamefont {B.~A.}\ \bibnamefont {Stefanek}},\ }\href
  {\doibase 10.1007/JHEP01(2019)053} {\bibfield  {journal} {\bibinfo  {journal}
  {JHEP}\ }\textbf {\bibinfo {volume} {01}},\ \bibinfo {pages} {053} (\bibinfo
  {year} {2019})},\ \Eprint {http://arxiv.org/abs/1811.01950} {arXiv:1811.01950
  [hep-ph]} \BibitemShut {NoStop}%
\bibitem [{\citenamefont {Chatrchyan}\ and\ \citenamefont
  {Jaeckel}(2021)}]{Chatrchyan:2020pzh}%
  \BibitemOpen
  \bibfield  {author} {\bibinfo {author} {\bibfnamefont {A.}~\bibnamefont
  {Chatrchyan}}\ and\ \bibinfo {author} {\bibfnamefont {J.}~\bibnamefont
  {Jaeckel}},\ }\href {\doibase 10.1088/1475-7516/2021/02/003} {\bibfield
  {journal} {\bibinfo  {journal} {JCAP}\ }\textbf {\bibinfo {volume} {02}},\
  \bibinfo {pages} {003} (\bibinfo {year} {2021})},\ \Eprint
  {http://arxiv.org/abs/2004.07844} {arXiv:2004.07844 [hep-ph]} \BibitemShut
  {NoStop}%
\bibitem [{\citenamefont {Salehian}\ \emph {et~al.}(2021)\citenamefont
  {Salehian}, \citenamefont {Gorji}, \citenamefont {Mukohyama},\ and\
  \citenamefont {Firouzjahi}}]{Salehian:2020dsf}%
  \BibitemOpen
  \bibfield  {author} {\bibinfo {author} {\bibfnamefont {B.}~\bibnamefont
  {Salehian}}, \bibinfo {author} {\bibfnamefont {M.~A.}\ \bibnamefont {Gorji}},
  \bibinfo {author} {\bibfnamefont {S.}~\bibnamefont {Mukohyama}}, \ and\
  \bibinfo {author} {\bibfnamefont {H.}~\bibnamefont {Firouzjahi}},\ }\href
  {\doibase 10.1007/JHEP05(2021)043} {\bibfield  {journal} {\bibinfo  {journal}
  {JHEP}\ }\textbf {\bibinfo {volume} {05}},\ \bibinfo {pages} {043} (\bibinfo
  {year} {2021})},\ \Eprint {http://arxiv.org/abs/2007.08148} {arXiv:2007.08148
  [hep-ph]} \BibitemShut {NoStop}%
\bibitem [{\citenamefont {Namba}\ and\ \citenamefont
  {Suzuki}(2020)}]{Namba:2020kij}%
  \BibitemOpen
  \bibfield  {author} {\bibinfo {author} {\bibfnamefont {R.}~\bibnamefont
  {Namba}}\ and\ \bibinfo {author} {\bibfnamefont {M.}~\bibnamefont {Suzuki}},\
  }\href {\doibase 10.1103/PhysRevD.102.123527} {\bibfield  {journal} {\bibinfo
   {journal} {Phys. Rev. D}\ }\textbf {\bibinfo {volume} {102}},\ \bibinfo
  {pages} {123527} (\bibinfo {year} {2020})},\ \Eprint
  {http://arxiv.org/abs/2009.13909} {arXiv:2009.13909 [astro-ph.CO]}
  \BibitemShut {NoStop}%
\bibitem [{\citenamefont {Kite}\ \emph {et~al.}(2021)\citenamefont {Kite},
  \citenamefont {Ravenni}, \citenamefont {Patil},\ and\ \citenamefont
  {Chluba}}]{Kite:2020uix}%
  \BibitemOpen
  \bibfield  {author} {\bibinfo {author} {\bibfnamefont {T.}~\bibnamefont
  {Kite}}, \bibinfo {author} {\bibfnamefont {A.}~\bibnamefont {Ravenni}},
  \bibinfo {author} {\bibfnamefont {S.~P.}\ \bibnamefont {Patil}}, \ and\
  \bibinfo {author} {\bibfnamefont {J.}~\bibnamefont {Chluba}},\ }\href
  {\doibase 10.1093/mnras/stab1558} {\bibfield  {journal} {\bibinfo  {journal}
  {Mon. Not. Roy. Astron. Soc.}\ }\textbf {\bibinfo {volume} {505}},\ \bibinfo
  {pages} {4396} (\bibinfo {year} {2021})},\ \Eprint
  {http://arxiv.org/abs/2010.00040} {arXiv:2010.00040 [astro-ph.CO]}
  \BibitemShut {NoStop}%
\bibitem [{\citenamefont {Kitajima}\ \emph {et~al.}(2021)\citenamefont
  {Kitajima}, \citenamefont {Soda},\ and\ \citenamefont
  {Urakawa}}]{Kitajima:2020rpm}%
  \BibitemOpen
  \bibfield  {author} {\bibinfo {author} {\bibfnamefont {N.}~\bibnamefont
  {Kitajima}}, \bibinfo {author} {\bibfnamefont {J.}~\bibnamefont {Soda}}, \
  and\ \bibinfo {author} {\bibfnamefont {Y.}~\bibnamefont {Urakawa}},\ }\href
  {\doibase 10.1103/PhysRevLett.126.121301} {\bibfield  {journal} {\bibinfo
  {journal} {Phys. Rev. Lett.}\ }\textbf {\bibinfo {volume} {126}},\ \bibinfo
  {pages} {121301} (\bibinfo {year} {2021})},\ \Eprint
  {http://arxiv.org/abs/2010.10990} {arXiv:2010.10990 [astro-ph.CO]}
  \BibitemShut {NoStop}%
\bibitem [{\citenamefont {Ratzinger}\ \emph {et~al.}(2021)\citenamefont
  {Ratzinger}, \citenamefont {Schwaller},\ and\ \citenamefont
  {Stefanek}}]{Ratzinger:2020oct}%
  \BibitemOpen
  \bibfield  {author} {\bibinfo {author} {\bibfnamefont {W.}~\bibnamefont
  {Ratzinger}}, \bibinfo {author} {\bibfnamefont {P.}~\bibnamefont
  {Schwaller}}, \ and\ \bibinfo {author} {\bibfnamefont {B.~A.}\ \bibnamefont
  {Stefanek}},\ }\href {\doibase 10.21468/SciPostPhys.11.1.001} {\bibfield
  {journal} {\bibinfo  {journal} {SciPost Phys.}\ }\textbf {\bibinfo {volume}
  {11}},\ \bibinfo {pages} {001} (\bibinfo {year} {2021})},\ \Eprint
  {http://arxiv.org/abs/2012.11584} {arXiv:2012.11584 [astro-ph.CO]}
  \BibitemShut {NoStop}%
\bibitem [{\citenamefont {Banerjee}\ \emph {et~al.}(2021)\citenamefont
  {Banerjee}, \citenamefont {Madge}, \citenamefont {Perez}, \citenamefont
  {Ratzinger},\ and\ \citenamefont {Schwaller}}]{Banerjee:2021oeu}%
  \BibitemOpen
  \bibfield  {author} {\bibinfo {author} {\bibfnamefont {A.}~\bibnamefont
  {Banerjee}}, \bibinfo {author} {\bibfnamefont {E.}~\bibnamefont {Madge}},
  \bibinfo {author} {\bibfnamefont {G.}~\bibnamefont {Perez}}, \bibinfo
  {author} {\bibfnamefont {W.}~\bibnamefont {Ratzinger}}, \ and\ \bibinfo
  {author} {\bibfnamefont {P.}~\bibnamefont {Schwaller}},\ }\href {\doibase
  10.1103/PhysRevD.104.055026} {\bibfield  {journal} {\bibinfo  {journal}
  {Phys. Rev. D}\ }\textbf {\bibinfo {volume} {104}},\ \bibinfo {pages}
  {055026} (\bibinfo {year} {2021})},\ \Eprint
  {http://arxiv.org/abs/2105.12135} {arXiv:2105.12135 [hep-ph]} \BibitemShut
  {NoStop}%
\bibitem [{\citenamefont {Madge}\ \emph {et~al.}(2022)\citenamefont {Madge},
  \citenamefont {Ratzinger}, \citenamefont {Schmitt},\ and\ \citenamefont
  {Schwaller}}]{Madge:2021abk}%
  \BibitemOpen
  \bibfield  {author} {\bibinfo {author} {\bibfnamefont {E.}~\bibnamefont
  {Madge}}, \bibinfo {author} {\bibfnamefont {W.}~\bibnamefont {Ratzinger}},
  \bibinfo {author} {\bibfnamefont {D.}~\bibnamefont {Schmitt}}, \ and\
  \bibinfo {author} {\bibfnamefont {P.}~\bibnamefont {Schwaller}},\ }\href
  {\doibase 10.21468/SciPostPhys.12.5.171} {\bibfield  {journal} {\bibinfo
  {journal} {SciPost Phys.}\ }\textbf {\bibinfo {volume} {12}},\ \bibinfo
  {pages} {171} (\bibinfo {year} {2022})},\ \Eprint
  {http://arxiv.org/abs/2111.12730} {arXiv:2111.12730 [hep-ph]} \BibitemShut
  {NoStop}%
\bibitem [{\citenamefont {Er\"oncel}\ \emph {et~al.}(2022)\citenamefont
  {Er\"oncel}, \citenamefont {Sato}, \citenamefont {Servant},\ and\
  \citenamefont {S\o{}rensen}}]{Eroncel:2022vjg}%
  \BibitemOpen
  \bibfield  {author} {\bibinfo {author} {\bibfnamefont {C.}~\bibnamefont
  {Er\"oncel}}, \bibinfo {author} {\bibfnamefont {R.}~\bibnamefont {Sato}},
  \bibinfo {author} {\bibfnamefont {G.}~\bibnamefont {Servant}}, \ and\
  \bibinfo {author} {\bibfnamefont {P.}~\bibnamefont {S\o{}rensen}},\ }\href
  {\doibase 10.1088/1475-7516/2022/10/053} {\bibfield  {journal} {\bibinfo
  {journal} {JCAP}\ }\textbf {\bibinfo {volume} {10}},\ \bibinfo {pages} {053}
  (\bibinfo {year} {2022})},\ \Eprint {http://arxiv.org/abs/2206.14259}
  {arXiv:2206.14259 [hep-ph]} \BibitemShut {NoStop}%
\bibitem [{\citenamefont {Madge}\ \emph {et~al.}(2023)\citenamefont {Madge},
  \citenamefont {Morgante}, \citenamefont {Puchades-Ib\'a\~nez}, \citenamefont
  {Ramberg}, \citenamefont {Ratzinger}, \citenamefont {Schenk},\ and\
  \citenamefont {Schwaller}}]{Madge:2023dxc}%
  \BibitemOpen
  \bibfield  {author} {\bibinfo {author} {\bibfnamefont {E.}~\bibnamefont
  {Madge}}, \bibinfo {author} {\bibfnamefont {E.}~\bibnamefont {Morgante}},
  \bibinfo {author} {\bibfnamefont {C.}~\bibnamefont {Puchades-Ib\'a\~nez}},
  \bibinfo {author} {\bibfnamefont {N.}~\bibnamefont {Ramberg}}, \bibinfo
  {author} {\bibfnamefont {W.}~\bibnamefont {Ratzinger}}, \bibinfo {author}
  {\bibfnamefont {S.}~\bibnamefont {Schenk}}, \ and\ \bibinfo {author}
  {\bibfnamefont {P.}~\bibnamefont {Schwaller}},\ }\href {\doibase
  10.1007/JHEP10(2023)171} {\bibfield  {journal} {\bibinfo  {journal} {JHEP}\
  }\textbf {\bibinfo {volume} {10}},\ \bibinfo {pages} {171} (\bibinfo {year}
  {2023})},\ \Eprint {http://arxiv.org/abs/2306.14856} {arXiv:2306.14856
  [hep-ph]} \BibitemShut {NoStop}%
\bibitem [{\citenamefont {Su}\ \emph {et~al.}(2025)\citenamefont {Su},
  \citenamefont {Xu}, \citenamefont {Chen}, \citenamefont {Liu},\ and\
  \citenamefont {Zhang}}]{Su:2025nkl}%
  \BibitemOpen
  \bibfield  {author} {\bibinfo {author} {\bibfnamefont {H.}~\bibnamefont
  {Su}}, \bibinfo {author} {\bibfnamefont {B.}~\bibnamefont {Xu}}, \bibinfo
  {author} {\bibfnamefont {J.}~\bibnamefont {Chen}}, \bibinfo {author}
  {\bibfnamefont {C.}~\bibnamefont {Liu}}, \ and\ \bibinfo {author}
  {\bibfnamefont {Y.-L.}\ \bibnamefont {Zhang}},\ }\href@noop {} {\  (\bibinfo
  {year} {2025})},\ \Eprint {http://arxiv.org/abs/2503.20778} {arXiv:2503.20778
  [astro-ph.CO]} \BibitemShut {NoStop}%
\bibitem [{\citenamefont {Hook}\ and\ \citenamefont
  {Marques-Tavares}(2016)}]{Hook:2016mqo}%
  \BibitemOpen
  \bibfield  {author} {\bibinfo {author} {\bibfnamefont {A.}~\bibnamefont
  {Hook}}\ and\ \bibinfo {author} {\bibfnamefont {G.}~\bibnamefont
  {Marques-Tavares}},\ }\href {\doibase 10.1007/JHEP12(2016)101} {\bibfield
  {journal} {\bibinfo  {journal} {JHEP}\ }\textbf {\bibinfo {volume} {12}},\
  \bibinfo {pages} {101} (\bibinfo {year} {2016})},\ \Eprint
  {http://arxiv.org/abs/1607.01786} {arXiv:1607.01786 [hep-ph]} \BibitemShut
  {NoStop}%
\bibitem [{\citenamefont {Agrawal}\ \emph {et~al.}(2018)\citenamefont
  {Agrawal}, \citenamefont {Marques-Tavares},\ and\ \citenamefont
  {Xue}}]{Agrawal:2017eqm}%
  \BibitemOpen
  \bibfield  {author} {\bibinfo {author} {\bibfnamefont {P.}~\bibnamefont
  {Agrawal}}, \bibinfo {author} {\bibfnamefont {G.}~\bibnamefont
  {Marques-Tavares}}, \ and\ \bibinfo {author} {\bibfnamefont {W.}~\bibnamefont
  {Xue}},\ }\href {\doibase 10.1007/JHEP03(2018)049} {\bibfield  {journal}
  {\bibinfo  {journal} {JHEP}\ }\textbf {\bibinfo {volume} {03}},\ \bibinfo
  {pages} {049} (\bibinfo {year} {2018})},\ \Eprint
  {http://arxiv.org/abs/1708.05008} {arXiv:1708.05008 [hep-ph]} \BibitemShut
  {NoStop}%
\bibitem [{\citenamefont {Co}\ \emph {et~al.}(2020)\citenamefont {Co},
  \citenamefont {Hall},\ and\ \citenamefont {Harigaya}}]{Co:2019jts}%
  \BibitemOpen
  \bibfield  {author} {\bibinfo {author} {\bibfnamefont {R.~T.}\ \bibnamefont
  {Co}}, \bibinfo {author} {\bibfnamefont {L.~J.}\ \bibnamefont {Hall}}, \ and\
  \bibinfo {author} {\bibfnamefont {K.}~\bibnamefont {Harigaya}},\ }\href
  {\doibase 10.1103/PhysRevLett.124.251802} {\bibfield  {journal} {\bibinfo
  {journal} {Phys. Rev. Lett.}\ }\textbf {\bibinfo {volume} {124}},\ \bibinfo
  {pages} {251802} (\bibinfo {year} {2020})},\ \Eprint
  {http://arxiv.org/abs/1910.14152} {arXiv:1910.14152 [hep-ph]} \BibitemShut
  {NoStop}%
\bibitem [{\citenamefont {Chang}\ and\ \citenamefont
  {Cui}(2020)}]{Chang:2019tvx}%
  \BibitemOpen
  \bibfield  {author} {\bibinfo {author} {\bibfnamefont {C.-F.}\ \bibnamefont
  {Chang}}\ and\ \bibinfo {author} {\bibfnamefont {Y.}~\bibnamefont {Cui}},\
  }\href {\doibase 10.1103/PhysRevD.102.015003} {\bibfield  {journal} {\bibinfo
   {journal} {Phys. Rev. D}\ }\textbf {\bibinfo {volume} {102}},\ \bibinfo
  {pages} {015003} (\bibinfo {year} {2020})},\ \Eprint
  {http://arxiv.org/abs/1911.11885} {arXiv:1911.11885 [hep-ph]} \BibitemShut
  {NoStop}%
\bibitem [{\citenamefont {Co}\ \emph {et~al.}(2021)\citenamefont {Co},
  \citenamefont {Harigaya},\ and\ \citenamefont {Pierce}}]{Co:2021rhi}%
  \BibitemOpen
  \bibfield  {author} {\bibinfo {author} {\bibfnamefont {R.~T.}\ \bibnamefont
  {Co}}, \bibinfo {author} {\bibfnamefont {K.}~\bibnamefont {Harigaya}}, \ and\
  \bibinfo {author} {\bibfnamefont {A.}~\bibnamefont {Pierce}},\ }\href
  {\doibase 10.1007/JHEP12(2021)099} {\bibfield  {journal} {\bibinfo  {journal}
  {JHEP}\ }\textbf {\bibinfo {volume} {12}},\ \bibinfo {pages} {099} (\bibinfo
  {year} {2021})},\ \Eprint {http://arxiv.org/abs/2104.02077} {arXiv:2104.02077
  [hep-ph]} \BibitemShut {NoStop}%
\bibitem [{\citenamefont {Xu}\ \emph {et~al.}(2024)\citenamefont {Xu},
  \citenamefont {Ding}, \citenamefont {Su}, \citenamefont {Chen},\ and\
  \citenamefont {Zhang}}]{Xu:2024kwy}%
  \BibitemOpen
  \bibfield  {author} {\bibinfo {author} {\bibfnamefont {B.}~\bibnamefont
  {Xu}}, \bibinfo {author} {\bibfnamefont {K.}~\bibnamefont {Ding}}, \bibinfo
  {author} {\bibfnamefont {H.}~\bibnamefont {Su}}, \bibinfo {author}
  {\bibfnamefont {J.}~\bibnamefont {Chen}}, \ and\ \bibinfo {author}
  {\bibfnamefont {Y.-L.}\ \bibnamefont {Zhang}},\ }\href@noop {} {\  (\bibinfo
  {year} {2024})},\ \Eprint {http://arxiv.org/abs/2411.08691} {arXiv:2411.08691
  [hep-ph]} \BibitemShut {NoStop}%
\bibitem [{\citenamefont {Ramberg}\ \emph {et~al.}(2023)\citenamefont
  {Ramberg}, \citenamefont {Ratzinger},\ and\ \citenamefont
  {Schwaller}}]{Ramberg:2022irf}%
  \BibitemOpen
  \bibfield  {author} {\bibinfo {author} {\bibfnamefont {N.}~\bibnamefont
  {Ramberg}}, \bibinfo {author} {\bibfnamefont {W.}~\bibnamefont {Ratzinger}},
  \ and\ \bibinfo {author} {\bibfnamefont {P.}~\bibnamefont {Schwaller}},\
  }\href {\doibase 10.1088/1475-7516/2023/02/039} {\bibfield  {journal}
  {\bibinfo  {journal} {JCAP}\ }\textbf {\bibinfo {volume} {02}},\ \bibinfo
  {pages} {039} (\bibinfo {year} {2023})},\ \Eprint
  {http://arxiv.org/abs/2209.14313} {arXiv:2209.14313 [hep-ph]} \BibitemShut
  {NoStop}%
\bibitem [{\citenamefont {Cui}\ \emph {et~al.}(2024)\citenamefont {Cui},
  \citenamefont {Saha},\ and\ \citenamefont {Sfakianakis}}]{Cui:2023fbg}%
  \BibitemOpen
  \bibfield  {author} {\bibinfo {author} {\bibfnamefont {Y.}~\bibnamefont
  {Cui}}, \bibinfo {author} {\bibfnamefont {P.}~\bibnamefont {Saha}}, \ and\
  \bibinfo {author} {\bibfnamefont {E.~I.}\ \bibnamefont {Sfakianakis}},\
  }\href {\doibase 10.1103/PhysRevLett.133.021004} {\bibfield  {journal}
  {\bibinfo  {journal} {Phys. Rev. Lett.}\ }\textbf {\bibinfo {volume} {133}},\
  \bibinfo {pages} {021004} (\bibinfo {year} {2024})},\ \Eprint
  {http://arxiv.org/abs/2310.13060} {arXiv:2310.13060 [hep-ph]} \BibitemShut
  {NoStop}%
\bibitem [{\citenamefont {Machado}\ \emph {et~al.}(2020)\citenamefont
  {Machado}, \citenamefont {Ratzinger}, \citenamefont {Schwaller},\ and\
  \citenamefont {Stefanek}}]{Machado:2019xuc}%
  \BibitemOpen
  \bibfield  {author} {\bibinfo {author} {\bibfnamefont {C.~S.}\ \bibnamefont
  {Machado}}, \bibinfo {author} {\bibfnamefont {W.}~\bibnamefont {Ratzinger}},
  \bibinfo {author} {\bibfnamefont {P.}~\bibnamefont {Schwaller}}, \ and\
  \bibinfo {author} {\bibfnamefont {B.~A.}\ \bibnamefont {Stefanek}},\ }\href
  {\doibase 10.1103/PhysRevD.102.075033} {\bibfield  {journal} {\bibinfo
  {journal} {Phys. Rev. D}\ }\textbf {\bibinfo {volume} {102}},\ \bibinfo
  {pages} {075033} (\bibinfo {year} {2020})},\ \Eprint
  {http://arxiv.org/abs/1912.01007} {arXiv:1912.01007 [hep-ph]} \BibitemShut
  {NoStop}%
\bibitem [{\citenamefont {Higaki}\ \emph {et~al.}(2016)\citenamefont {Higaki},
  \citenamefont {Jeong}, \citenamefont {Kitajima},\ and\ \citenamefont
  {Takahashi}}]{Higaki:2016yqk}%
  \BibitemOpen
  \bibfield  {author} {\bibinfo {author} {\bibfnamefont {T.}~\bibnamefont
  {Higaki}}, \bibinfo {author} {\bibfnamefont {K.~S.}\ \bibnamefont {Jeong}},
  \bibinfo {author} {\bibfnamefont {N.}~\bibnamefont {Kitajima}}, \ and\
  \bibinfo {author} {\bibfnamefont {F.}~\bibnamefont {Takahashi}},\ }\href
  {\doibase 10.1007/JHEP06(2016)150} {\bibfield  {journal} {\bibinfo  {journal}
  {JHEP}\ }\textbf {\bibinfo {volume} {06}},\ \bibinfo {pages} {150} (\bibinfo
  {year} {2016})},\ \Eprint {http://arxiv.org/abs/1603.02090} {arXiv:1603.02090
  [hep-ph]} \BibitemShut {NoStop}%
\bibitem [{\citenamefont {Kawasaki}\ \emph {et~al.}(2018)\citenamefont
  {Kawasaki}, \citenamefont {Takahashi},\ and\ \citenamefont
  {Yamada}}]{Kawasaki:2017xwt}%
  \BibitemOpen
  \bibfield  {author} {\bibinfo {author} {\bibfnamefont {M.}~\bibnamefont
  {Kawasaki}}, \bibinfo {author} {\bibfnamefont {F.}~\bibnamefont {Takahashi}},
  \ and\ \bibinfo {author} {\bibfnamefont {M.}~\bibnamefont {Yamada}},\ }\href
  {\doibase 10.1007/JHEP01(2018)053} {\bibfield  {journal} {\bibinfo  {journal}
  {JHEP}\ }\textbf {\bibinfo {volume} {01}},\ \bibinfo {pages} {053} (\bibinfo
  {year} {2018})},\ \Eprint {http://arxiv.org/abs/1708.06047} {arXiv:1708.06047
  [hep-ph]} \BibitemShut {NoStop}%
\bibitem [{\citenamefont {Nakagawa}\ \emph {et~al.}(2021)\citenamefont
  {Nakagawa}, \citenamefont {Takahashi},\ and\ \citenamefont
  {Yamada}}]{Nakagawa:2020zjr}%
  \BibitemOpen
  \bibfield  {author} {\bibinfo {author} {\bibfnamefont {S.}~\bibnamefont
  {Nakagawa}}, \bibinfo {author} {\bibfnamefont {F.}~\bibnamefont {Takahashi}},
  \ and\ \bibinfo {author} {\bibfnamefont {M.}~\bibnamefont {Yamada}},\ }\href
  {\doibase 10.1088/1475-7516/2021/05/062} {\bibfield  {journal} {\bibinfo
  {journal} {JCAP}\ }\textbf {\bibinfo {volume} {05}},\ \bibinfo {pages} {062}
  (\bibinfo {year} {2021})},\ \Eprint {http://arxiv.org/abs/2012.13592}
  {arXiv:2012.13592 [hep-ph]} \BibitemShut {NoStop}%
\bibitem [{\citenamefont {Di~Luzio}\ \emph
  {et~al.}(2021{\natexlab{a}})\citenamefont {Di~Luzio}, \citenamefont {Gavela},
  \citenamefont {Quilez},\ and\ \citenamefont {Ringwald}}]{DiLuzio:2021pxd}%
  \BibitemOpen
  \bibfield  {author} {\bibinfo {author} {\bibfnamefont {L.}~\bibnamefont
  {Di~Luzio}}, \bibinfo {author} {\bibfnamefont {B.}~\bibnamefont {Gavela}},
  \bibinfo {author} {\bibfnamefont {P.}~\bibnamefont {Quilez}}, \ and\ \bibinfo
  {author} {\bibfnamefont {A.}~\bibnamefont {Ringwald}},\ }\href {\doibase
  10.1007/JHEP05(2021)184} {\bibfield  {journal} {\bibinfo  {journal} {JHEP}\
  }\textbf {\bibinfo {volume} {05}},\ \bibinfo {pages} {184} (\bibinfo {year}
  {2021}{\natexlab{a}})},\ \Eprint {http://arxiv.org/abs/2102.00012}
  {arXiv:2102.00012 [hep-ph]} \BibitemShut {NoStop}%
\bibitem [{\citenamefont {Di~Luzio}\ \emph
  {et~al.}(2021{\natexlab{b}})\citenamefont {Di~Luzio}, \citenamefont {Gavela},
  \citenamefont {Quilez},\ and\ \citenamefont {Ringwald}}]{DiLuzio:2021gos}%
  \BibitemOpen
  \bibfield  {author} {\bibinfo {author} {\bibfnamefont {L.}~\bibnamefont
  {Di~Luzio}}, \bibinfo {author} {\bibfnamefont {B.}~\bibnamefont {Gavela}},
  \bibinfo {author} {\bibfnamefont {P.}~\bibnamefont {Quilez}}, \ and\ \bibinfo
  {author} {\bibfnamefont {A.}~\bibnamefont {Ringwald}},\ }\href {\doibase
  10.1088/1475-7516/2021/10/001} {\bibfield  {journal} {\bibinfo  {journal}
  {JCAP}\ }\textbf {\bibinfo {volume} {10}},\ \bibinfo {pages} {001} (\bibinfo
  {year} {2021}{\natexlab{b}})},\ \Eprint {http://arxiv.org/abs/2102.01082}
  {arXiv:2102.01082 [hep-ph]} \BibitemShut {NoStop}%
\bibitem [{\citenamefont {Kitajima}\ and\ \citenamefont
  {Takahashi}(2023)}]{Kitajima:2023pby}%
  \BibitemOpen
  \bibfield  {author} {\bibinfo {author} {\bibfnamefont {N.}~\bibnamefont
  {Kitajima}}\ and\ \bibinfo {author} {\bibfnamefont {F.}~\bibnamefont
  {Takahashi}},\ }\href {\doibase 10.1103/PhysRevD.107.123518} {\bibfield
  {journal} {\bibinfo  {journal} {Phys. Rev. D}\ }\textbf {\bibinfo {volume}
  {107}},\ \bibinfo {pages} {123518} (\bibinfo {year} {2023})},\ \Eprint
  {http://arxiv.org/abs/2303.05492} {arXiv:2303.05492 [hep-ph]} \BibitemShut
  {NoStop}%
\bibitem [{\citenamefont {Di~Luzio}\ and\ \citenamefont
  {S\o{}rensen}(2024)}]{DiLuzio:2024fyt}%
  \BibitemOpen
  \bibfield  {author} {\bibinfo {author} {\bibfnamefont {L.}~\bibnamefont
  {Di~Luzio}}\ and\ \bibinfo {author} {\bibfnamefont {P.}~\bibnamefont
  {S\o{}rensen}},\ }\href {\doibase 10.1007/JHEP10(2024)239} {\bibfield
  {journal} {\bibinfo  {journal} {JHEP}\ }\textbf {\bibinfo {volume} {10}},\
  \bibinfo {pages} {239} (\bibinfo {year} {2024})},\ \Eprint
  {http://arxiv.org/abs/2408.04623} {arXiv:2408.04623 [hep-ph]} \BibitemShut
  {NoStop}%
\bibitem [{\citenamefont {Aghanim}\ \emph {et~al.}(2020)\citenamefont {Aghanim}
  \emph {et~al.}}]{Planck:2018vyg}%
  \BibitemOpen
  \bibfield  {author} {\bibinfo {author} {\bibfnamefont {N.}~\bibnamefont
  {Aghanim}} \emph {et~al.} (\bibinfo {collaboration} {Planck}),\ }\href
  {\doibase 10.1051/0004-6361/201833910} {\bibfield  {journal} {\bibinfo
  {journal} {Astron. Astrophys.}\ }\textbf {\bibinfo {volume} {641}},\ \bibinfo
  {pages} {A6} (\bibinfo {year} {2020})},\ \bibinfo {note} {[Erratum:
  Astron.Astrophys. 652, C4 (2021)]},\ \Eprint
  {http://arxiv.org/abs/1807.06209} {arXiv:1807.06209 [astro-ph.CO]}
  \BibitemShut {NoStop}%
\bibitem [{\citenamefont {O'Hare}(2020)}]{AxionLimits}%
  \BibitemOpen
  \bibfield  {author} {\bibinfo {author} {\bibfnamefont {C.}~\bibnamefont
  {O'Hare}},\ }\href {\doibase 10.5281/zenodo.3932430} {\enquote {\bibinfo
  {title} {cajohare/axionlimits: Axionlimits},}\ }\bibinfo {howpublished}
  {\url{https://cajohare.github.io/AxionLimits/}} (\bibinfo {year}
  {2020})\BibitemShut {NoStop}%
\bibitem [{\citenamefont {Baryakhtar}\ \emph {et~al.}(2021)\citenamefont
  {Baryakhtar}, \citenamefont {Galanis}, \citenamefont {Lasenby},\ and\
  \citenamefont {Simon}}]{Baryakhtar:2020gao}%
  \BibitemOpen
  \bibfield  {author} {\bibinfo {author} {\bibfnamefont {M.}~\bibnamefont
  {Baryakhtar}}, \bibinfo {author} {\bibfnamefont {M.}~\bibnamefont {Galanis}},
  \bibinfo {author} {\bibfnamefont {R.}~\bibnamefont {Lasenby}}, \ and\
  \bibinfo {author} {\bibfnamefont {O.}~\bibnamefont {Simon}},\ }\href
  {\doibase 10.1103/PhysRevD.103.095019} {\bibfield  {journal} {\bibinfo
  {journal} {Phys. Rev. D}\ }\textbf {\bibinfo {volume} {103}},\ \bibinfo
  {pages} {095019} (\bibinfo {year} {2021})},\ \Eprint
  {http://arxiv.org/abs/2011.11646} {arXiv:2011.11646 [hep-ph]} \BibitemShut
  {NoStop}%
\bibitem [{\citenamefont {Stott}(2020)}]{Stott:2020gjj}%
  \BibitemOpen
  \bibfield  {author} {\bibinfo {author} {\bibfnamefont {M.~J.}\ \bibnamefont
  {Stott}},\ }\href@noop {} {\  (\bibinfo {year} {2020})},\ \Eprint
  {http://arxiv.org/abs/2009.07206} {arXiv:2009.07206 [hep-ph]} \BibitemShut
  {NoStop}%
\bibitem [{\citenamefont {Hoof}\ \emph {et~al.}(2024)\citenamefont {Hoof},
  \citenamefont {Marsh}, \citenamefont {Sisk-Reyn\'es}, \citenamefont
  {Matthews},\ and\ \citenamefont {Reynolds}}]{Hoof:2024quk}%
  \BibitemOpen
  \bibfield  {author} {\bibinfo {author} {\bibfnamefont {S.}~\bibnamefont
  {Hoof}}, \bibinfo {author} {\bibfnamefont {D.~J.~E.}\ \bibnamefont {Marsh}},
  \bibinfo {author} {\bibfnamefont {J.}~\bibnamefont {Sisk-Reyn\'es}}, \bibinfo
  {author} {\bibfnamefont {J.~H.}\ \bibnamefont {Matthews}}, \ and\ \bibinfo
  {author} {\bibfnamefont {C.}~\bibnamefont {Reynolds}},\ }\href@noop {} {\
  (\bibinfo {year} {2024})},\ \Eprint {http://arxiv.org/abs/2406.10337}
  {arXiv:2406.10337 [hep-ph]} \BibitemShut {NoStop}%
\bibitem [{\citenamefont {\"Unal}\ \emph {et~al.}(2021)\citenamefont {\"Unal},
  \citenamefont {Pacucci},\ and\ \citenamefont {Loeb}}]{Unal:2020jiy}%
  \BibitemOpen
  \bibfield  {author} {\bibinfo {author} {\bibfnamefont {C.}~\bibnamefont
  {\"Unal}}, \bibinfo {author} {\bibfnamefont {F.}~\bibnamefont {Pacucci}}, \
  and\ \bibinfo {author} {\bibfnamefont {A.}~\bibnamefont {Loeb}},\ }\href
  {\doibase 10.1088/1475-7516/2021/05/007} {\bibfield  {journal} {\bibinfo
  {journal} {JCAP}\ }\textbf {\bibinfo {volume} {05}},\ \bibinfo {pages} {007}
  (\bibinfo {year} {2021})},\ \Eprint {http://arxiv.org/abs/2012.12790}
  {arXiv:2012.12790 [hep-ph]} \BibitemShut {NoStop}%
\bibitem [{\citenamefont {Witte}\ and\ \citenamefont
  {Mummery}(2024)}]{Witte:2024drg}%
  \BibitemOpen
  \bibfield  {author} {\bibinfo {author} {\bibfnamefont {S.~J.}\ \bibnamefont
  {Witte}}\ and\ \bibinfo {author} {\bibfnamefont {A.}~\bibnamefont
  {Mummery}},\ }\href@noop {} {\  (\bibinfo {year} {2024})},\ \Eprint
  {http://arxiv.org/abs/2412.03655} {arXiv:2412.03655 [hep-ph]} \BibitemShut
  {NoStop}%
\bibitem [{\citenamefont {Cardoso}\ \emph {et~al.}(2018)\citenamefont
  {Cardoso}, \citenamefont {Dias}, \citenamefont {Hartnett}, \citenamefont
  {Middleton}, \citenamefont {Pani},\ and\ \citenamefont
  {Santos}}]{Cardoso:2018tly}%
  \BibitemOpen
  \bibfield  {author} {\bibinfo {author} {\bibfnamefont {V.}~\bibnamefont
  {Cardoso}}, \bibinfo {author} {\bibfnamefont {O.~J.~C.}\ \bibnamefont
  {Dias}}, \bibinfo {author} {\bibfnamefont {G.~S.}\ \bibnamefont {Hartnett}},
  \bibinfo {author} {\bibfnamefont {M.}~\bibnamefont {Middleton}}, \bibinfo
  {author} {\bibfnamefont {P.}~\bibnamefont {Pani}}, \ and\ \bibinfo {author}
  {\bibfnamefont {J.~E.}\ \bibnamefont {Santos}},\ }\href {\doibase
  10.1088/1475-7516/2018/03/043} {\bibfield  {journal} {\bibinfo  {journal}
  {JCAP}\ }\textbf {\bibinfo {volume} {03}},\ \bibinfo {pages} {043} (\bibinfo
  {year} {2018})},\ \Eprint {http://arxiv.org/abs/1801.01420} {arXiv:1801.01420
  [gr-qc]} \BibitemShut {NoStop}%
\bibitem [{\citenamefont {Buchm\"uller}\ \emph {et~al.}(2013)\citenamefont
  {Buchm\"uller}, \citenamefont {Domcke}, \citenamefont {Kamada},\ and\
  \citenamefont {Schmitz}}]{Buchmuller:2013lra}%
  \BibitemOpen
  \bibfield  {author} {\bibinfo {author} {\bibfnamefont {W.}~\bibnamefont
  {Buchm\"uller}}, \bibinfo {author} {\bibfnamefont {V.}~\bibnamefont
  {Domcke}}, \bibinfo {author} {\bibfnamefont {K.}~\bibnamefont {Kamada}}, \
  and\ \bibinfo {author} {\bibfnamefont {K.}~\bibnamefont {Schmitz}},\ }\href
  {\doibase 10.1088/1475-7516/2013/10/003} {\bibfield  {journal} {\bibinfo
  {journal} {JCAP}\ }\textbf {\bibinfo {volume} {10}},\ \bibinfo {pages} {003}
  (\bibinfo {year} {2013})},\ \Eprint {http://arxiv.org/abs/1305.3392}
  {arXiv:1305.3392 [hep-ph]} \BibitemShut {NoStop}%
\bibitem [{\citenamefont {Giblin}\ and\ \citenamefont
  {Thrane}(2014)}]{Giblin:2014gra}%
  \BibitemOpen
  \bibfield  {author} {\bibinfo {author} {\bibfnamefont {J.~T.}\ \bibnamefont
  {Giblin}}\ and\ \bibinfo {author} {\bibfnamefont {E.}~\bibnamefont
  {Thrane}},\ }\href {\doibase 10.1103/PhysRevD.90.107502} {\bibfield
  {journal} {\bibinfo  {journal} {Phys. Rev. D}\ }\textbf {\bibinfo {volume}
  {90}},\ \bibinfo {pages} {107502} (\bibinfo {year} {2014})},\ \Eprint
  {http://arxiv.org/abs/1410.4779} {arXiv:1410.4779 [gr-qc]} \BibitemShut
  {NoStop}%
\bibitem [{\citenamefont {Kamionkowski}\ \emph {et~al.}(1994)\citenamefont
  {Kamionkowski}, \citenamefont {Kosowsky},\ and\ \citenamefont
  {Turner}}]{Kamionkowski:1993fg}%
  \BibitemOpen
  \bibfield  {author} {\bibinfo {author} {\bibfnamefont {M.}~\bibnamefont
  {Kamionkowski}}, \bibinfo {author} {\bibfnamefont {A.}~\bibnamefont
  {Kosowsky}}, \ and\ \bibinfo {author} {\bibfnamefont {M.~S.}\ \bibnamefont
  {Turner}},\ }\href {\doibase 10.1103/PhysRevD.49.2837} {\bibfield  {journal}
  {\bibinfo  {journal} {Phys. Rev. D}\ }\textbf {\bibinfo {volume} {49}},\
  \bibinfo {pages} {2837} (\bibinfo {year} {1994})},\ \Eprint
  {http://arxiv.org/abs/astro-ph/9310044} {arXiv:astro-ph/9310044} \BibitemShut
  {NoStop}%
\bibitem [{\citenamefont {Caprini}\ and\ \citenamefont
  {Figueroa}(2018)}]{Caprini:2018mtu}%
  \BibitemOpen
  \bibfield  {author} {\bibinfo {author} {\bibfnamefont {C.}~\bibnamefont
  {Caprini}}\ and\ \bibinfo {author} {\bibfnamefont {D.~G.}\ \bibnamefont
  {Figueroa}},\ }\href {\doibase 10.1088/1361-6382/aac608} {\bibfield
  {journal} {\bibinfo  {journal} {Class. Quant. Grav.}\ }\textbf {\bibinfo
  {volume} {35}},\ \bibinfo {pages} {163001} (\bibinfo {year} {2018})},\
  \Eprint {http://arxiv.org/abs/1801.04268} {arXiv:1801.04268 [astro-ph.CO]}
  \BibitemShut {NoStop}%
\bibitem [{\citenamefont {Janssen}\ \emph {et~al.}(2015)\citenamefont {Janssen}
  \emph {et~al.}}]{Janssen:2014dka}%
  \BibitemOpen
  \bibfield  {author} {\bibinfo {author} {\bibfnamefont {G.}~\bibnamefont
  {Janssen}} \emph {et~al.},\ }\href {\doibase 10.22323/1.215.0037} {\bibfield
  {journal} {\bibinfo  {journal} {PoS}\ }\textbf {\bibinfo {volume}
  {AASKA14}},\ \bibinfo {pages} {037} (\bibinfo {year} {2015})},\ \Eprint
  {http://arxiv.org/abs/1501.00127} {arXiv:1501.00127 [astro-ph.IM]}
  \BibitemShut {NoStop}%
\bibitem [{\citenamefont {Sesana}\ \emph {et~al.}(2021)\citenamefont {Sesana}
  \emph {et~al.}}]{Sesana:2019vho}%
  \BibitemOpen
  \bibfield  {author} {\bibinfo {author} {\bibfnamefont {A.}~\bibnamefont
  {Sesana}} \emph {et~al.},\ }\href {\doibase 10.1007/s10686-021-09709-9}
  {\bibfield  {journal} {\bibinfo  {journal} {Exper. Astron.}\ }\textbf
  {\bibinfo {volume} {51}},\ \bibinfo {pages} {1333} (\bibinfo {year}
  {2021})},\ \Eprint {http://arxiv.org/abs/1908.11391} {arXiv:1908.11391
  [astro-ph.IM]} \BibitemShut {NoStop}%
\bibitem [{\citenamefont {{LISA collaboration}}(2017)}]{2017arXiv170200786A}%
  \BibitemOpen
  \bibfield  {author} {\bibinfo {author} {\bibnamefont {{LISA
  collaboration}}},\ }\href@noop {} {\bibfield  {journal} {\bibinfo  {journal}
  {arXiv e-prints}\ ,\ \bibinfo {eid} {arXiv:1702.00786}} (\bibinfo {year}
  {2017})},\ \Eprint {http://arxiv.org/abs/1702.00786} {arXiv:1702.00786
  [astro-ph.IM]} \BibitemShut {NoStop}%
\bibitem [{\citenamefont {Robson}\ \emph {et~al.}(2019)\citenamefont {Robson},
  \citenamefont {Cornish},\ and\ \citenamefont {Liu}}]{Robson:2018ifk}%
  \BibitemOpen
  \bibfield  {author} {\bibinfo {author} {\bibfnamefont {T.}~\bibnamefont
  {Robson}}, \bibinfo {author} {\bibfnamefont {N.~J.}\ \bibnamefont {Cornish}},
  \ and\ \bibinfo {author} {\bibfnamefont {C.}~\bibnamefont {Liu}},\ }\href
  {\doibase 10.1088/1361-6382/ab1101} {\bibfield  {journal} {\bibinfo
  {journal} {Class. Quant. Grav.}\ }\textbf {\bibinfo {volume} {36}},\ \bibinfo
  {pages} {105011} (\bibinfo {year} {2019})},\ \Eprint
  {http://arxiv.org/abs/1803.01944} {arXiv:1803.01944 [astro-ph.HE]}
  \BibitemShut {NoStop}%
\bibitem [{\citenamefont {Auclair}\ \emph {et~al.}(2023)\citenamefont {Auclair}
  \emph {et~al.}}]{LISACosmologyWorkingGroup:2022jok}%
  \BibitemOpen
  \bibfield  {author} {\bibinfo {author} {\bibfnamefont {P.}~\bibnamefont
  {Auclair}} \emph {et~al.} (\bibinfo {collaboration} {LISA Cosmology Working
  Group}),\ }\href {\doibase 10.1007/s41114-023-00045-2} {\bibfield  {journal}
  {\bibinfo  {journal} {Living Rev. Rel.}\ }\textbf {\bibinfo {volume} {26}},\
  \bibinfo {pages} {5} (\bibinfo {year} {2023})},\ \Eprint
  {http://arxiv.org/abs/2204.05434} {arXiv:2204.05434 [astro-ph.CO]}
  \BibitemShut {NoStop}%
\bibitem [{\citenamefont {Crowder}\ and\ \citenamefont
  {Cornish}(2005)}]{Crowder:2005nr}%
  \BibitemOpen
  \bibfield  {author} {\bibinfo {author} {\bibfnamefont {J.}~\bibnamefont
  {Crowder}}\ and\ \bibinfo {author} {\bibfnamefont {N.~J.}\ \bibnamefont
  {Cornish}},\ }\href {\doibase 10.1103/PhysRevD.72.083005} {\bibfield
  {journal} {\bibinfo  {journal} {Phys. Rev. D}\ }\textbf {\bibinfo {volume}
  {72}},\ \bibinfo {pages} {083005} (\bibinfo {year} {2005})},\ \Eprint
  {http://arxiv.org/abs/gr-qc/0506015} {arXiv:gr-qc/0506015} \BibitemShut
  {NoStop}%
\bibitem [{\citenamefont {Sathyaprakash}\ \emph {et~al.}(2012)\citenamefont
  {Sathyaprakash} \emph {et~al.}}]{Sathyaprakash:2012jk}%
  \BibitemOpen
  \bibfield  {author} {\bibinfo {author} {\bibfnamefont {B.}~\bibnamefont
  {Sathyaprakash}} \emph {et~al.},\ }\href {\doibase
  10.1088/0264-9381/29/12/124013} {\bibfield  {journal} {\bibinfo  {journal}
  {Class. Quant. Grav.}\ }\textbf {\bibinfo {volume} {29}},\ \bibinfo {pages}
  {124013} (\bibinfo {year} {2012})},\ \bibinfo {note} {[Erratum:
  Class.Quant.Grav. 30, 079501 (2013)]},\ \Eprint
  {http://arxiv.org/abs/1206.0331} {arXiv:1206.0331 [gr-qc]} \BibitemShut
  {NoStop}%
\bibitem [{\citenamefont {Breitbach}\ \emph {et~al.}(2019)\citenamefont
  {Breitbach}, \citenamefont {Kopp}, \citenamefont {Madge}, \citenamefont
  {Opferkuch},\ and\ \citenamefont {Schwaller}}]{Breitbach:2018ddu}%
  \BibitemOpen
  \bibfield  {author} {\bibinfo {author} {\bibfnamefont {M.}~\bibnamefont
  {Breitbach}}, \bibinfo {author} {\bibfnamefont {J.}~\bibnamefont {Kopp}},
  \bibinfo {author} {\bibfnamefont {E.}~\bibnamefont {Madge}}, \bibinfo
  {author} {\bibfnamefont {T.}~\bibnamefont {Opferkuch}}, \ and\ \bibinfo
  {author} {\bibfnamefont {P.}~\bibnamefont {Schwaller}},\ }\href {\doibase
  10.1088/1475-7516/2019/07/007} {\bibfield  {journal} {\bibinfo  {journal}
  {JCAP}\ }\textbf {\bibinfo {volume} {07}},\ \bibinfo {pages} {007} (\bibinfo
  {year} {2019})},\ \Eprint {http://arxiv.org/abs/1811.11175} {arXiv:1811.11175
  [hep-ph]} \BibitemShut {NoStop}%
\bibitem [{\citenamefont {Fox}\ \emph {et~al.}(2023)\citenamefont {Fox},
  \citenamefont {Weiner},\ and\ \citenamefont {Xiao}}]{Fox:2023xgx}%
  \BibitemOpen
  \bibfield  {author} {\bibinfo {author} {\bibfnamefont {P.~J.}\ \bibnamefont
  {Fox}}, \bibinfo {author} {\bibfnamefont {N.}~\bibnamefont {Weiner}}, \ and\
  \bibinfo {author} {\bibfnamefont {H.}~\bibnamefont {Xiao}},\ }\href {\doibase
  10.1103/PhysRevD.108.095043} {\bibfield  {journal} {\bibinfo  {journal}
  {Phys. Rev. D}\ }\textbf {\bibinfo {volume} {108}},\ \bibinfo {pages}
  {095043} (\bibinfo {year} {2023})},\ \Eprint
  {http://arxiv.org/abs/2302.00685} {arXiv:2302.00685 [hep-ph]} \BibitemShut
  {NoStop}%
\bibitem [{\citenamefont {McAllister}\ \emph {et~al.}(2010)\citenamefont
  {McAllister}, \citenamefont {Silverstein},\ and\ \citenamefont
  {Westphal}}]{McAllister:2008hb}%
  \BibitemOpen
  \bibfield  {author} {\bibinfo {author} {\bibfnamefont {L.}~\bibnamefont
  {McAllister}}, \bibinfo {author} {\bibfnamefont {E.}~\bibnamefont
  {Silverstein}}, \ and\ \bibinfo {author} {\bibfnamefont {A.}~\bibnamefont
  {Westphal}},\ }\href {\doibase 10.1103/PhysRevD.82.046003} {\bibfield
  {journal} {\bibinfo  {journal} {Phys. Rev. D}\ }\textbf {\bibinfo {volume}
  {82}},\ \bibinfo {pages} {046003} (\bibinfo {year} {2010})},\ \Eprint
  {http://arxiv.org/abs/0808.0706} {arXiv:0808.0706 [hep-th]} \BibitemShut
  {NoStop}%
\bibitem [{\citenamefont {Silverstein}\ and\ \citenamefont
  {Westphal}(2008)}]{Silverstein:2008sg}%
  \BibitemOpen
  \bibfield  {author} {\bibinfo {author} {\bibfnamefont {E.}~\bibnamefont
  {Silverstein}}\ and\ \bibinfo {author} {\bibfnamefont {A.}~\bibnamefont
  {Westphal}},\ }\href {\doibase 10.1103/PhysRevD.78.106003} {\bibfield
  {journal} {\bibinfo  {journal} {Phys. Rev. D}\ }\textbf {\bibinfo {volume}
  {78}},\ \bibinfo {pages} {106003} (\bibinfo {year} {2008})},\ \Eprint
  {http://arxiv.org/abs/0803.3085} {arXiv:0803.3085 [hep-th]} \BibitemShut
  {NoStop}%
\bibitem [{\citenamefont {Hebecker}\ \emph {et~al.}(2014)\citenamefont
  {Hebecker}, \citenamefont {Kraus},\ and\ \citenamefont
  {Witkowski}}]{Hebecker:2014eua}%
  \BibitemOpen
  \bibfield  {author} {\bibinfo {author} {\bibfnamefont {A.}~\bibnamefont
  {Hebecker}}, \bibinfo {author} {\bibfnamefont {S.~C.}\ \bibnamefont {Kraus}},
  \ and\ \bibinfo {author} {\bibfnamefont {L.~T.}\ \bibnamefont {Witkowski}},\
  }\href {\doibase 10.1016/j.physletb.2014.08.028} {\bibfield  {journal}
  {\bibinfo  {journal} {Phys. Lett. B}\ }\textbf {\bibinfo {volume} {737}},\
  \bibinfo {pages} {16} (\bibinfo {year} {2014})},\ \Eprint
  {http://arxiv.org/abs/1404.3711} {arXiv:1404.3711 [hep-th]} \BibitemShut
  {NoStop}%
\bibitem [{\citenamefont {McAllister}\ \emph {et~al.}(2014)\citenamefont
  {McAllister}, \citenamefont {Silverstein}, \citenamefont {Westphal},\ and\
  \citenamefont {Wrase}}]{McAllister:2014mpa}%
  \BibitemOpen
  \bibfield  {author} {\bibinfo {author} {\bibfnamefont {L.}~\bibnamefont
  {McAllister}}, \bibinfo {author} {\bibfnamefont {E.}~\bibnamefont
  {Silverstein}}, \bibinfo {author} {\bibfnamefont {A.}~\bibnamefont
  {Westphal}}, \ and\ \bibinfo {author} {\bibfnamefont {T.}~\bibnamefont
  {Wrase}},\ }\href {\doibase 10.1007/JHEP09(2014)123} {\bibfield  {journal}
  {\bibinfo  {journal} {JHEP}\ }\textbf {\bibinfo {volume} {09}},\ \bibinfo
  {pages} {123} (\bibinfo {year} {2014})},\ \Eprint
  {http://arxiv.org/abs/1405.3652} {arXiv:1405.3652 [hep-th]} \BibitemShut
  {NoStop}%
\bibitem [{\citenamefont {Blumenhagen}\ and\ \citenamefont
  {Plauschinn}(2014)}]{Blumenhagen:2014gta}%
  \BibitemOpen
  \bibfield  {author} {\bibinfo {author} {\bibfnamefont {R.}~\bibnamefont
  {Blumenhagen}}\ and\ \bibinfo {author} {\bibfnamefont {E.}~\bibnamefont
  {Plauschinn}},\ }\href {\doibase 10.1016/j.physletb.2014.08.007} {\bibfield
  {journal} {\bibinfo  {journal} {Phys. Lett. B}\ }\textbf {\bibinfo {volume}
  {736}},\ \bibinfo {pages} {482} (\bibinfo {year} {2014})},\ \Eprint
  {http://arxiv.org/abs/1404.3542} {arXiv:1404.3542 [hep-th]} \BibitemShut
  {NoStop}%
\bibitem [{\citenamefont {Marchesano}\ \emph {et~al.}(2014)\citenamefont
  {Marchesano}, \citenamefont {Shiu},\ and\ \citenamefont
  {Uranga}}]{Marchesano:2014mla}%
  \BibitemOpen
  \bibfield  {author} {\bibinfo {author} {\bibfnamefont {F.}~\bibnamefont
  {Marchesano}}, \bibinfo {author} {\bibfnamefont {G.}~\bibnamefont {Shiu}}, \
  and\ \bibinfo {author} {\bibfnamefont {A.~M.}\ \bibnamefont {Uranga}},\
  }\href {\doibase 10.1007/JHEP09(2014)184} {\bibfield  {journal} {\bibinfo
  {journal} {JHEP}\ }\textbf {\bibinfo {volume} {09}},\ \bibinfo {pages} {184}
  (\bibinfo {year} {2014})},\ \Eprint {http://arxiv.org/abs/1404.3040}
  {arXiv:1404.3040 [hep-th]} \BibitemShut {NoStop}%
\bibitem [{\citenamefont {Kraemmer}\ \emph {et~al.}(1995)\citenamefont
  {Kraemmer}, \citenamefont {Rebhan},\ and\ \citenamefont
  {Schulz}}]{Kraemmer:1994az}%
  \BibitemOpen
  \bibfield  {author} {\bibinfo {author} {\bibfnamefont {U.}~\bibnamefont
  {Kraemmer}}, \bibinfo {author} {\bibfnamefont {A.~K.}\ \bibnamefont
  {Rebhan}}, \ and\ \bibinfo {author} {\bibfnamefont {H.}~\bibnamefont
  {Schulz}},\ }\href {\doibase 10.1006/aphy.1995.1023} {\bibfield  {journal}
  {\bibinfo  {journal} {Annals Phys.}\ }\textbf {\bibinfo {volume} {238}},\
  \bibinfo {pages} {286} (\bibinfo {year} {1995})},\ \Eprint
  {http://arxiv.org/abs/hep-ph/9403301} {arXiv:hep-ph/9403301} \BibitemShut
  {NoStop}%
\bibitem [{\citenamefont {Kapusta}\ and\ \citenamefont
  {Gale}(2011)}]{Kapusta:2006pm}%
  \BibitemOpen
  \bibfield  {author} {\bibinfo {author} {\bibfnamefont {J.~I.}\ \bibnamefont
  {Kapusta}}\ and\ \bibinfo {author} {\bibfnamefont {C.}~\bibnamefont {Gale}},\
  }\href {\doibase 10.1017/CBO9780511535130} {\emph {\bibinfo {title}
  {{Finite-temperature field theory: Principles and applications}}}},\
  Cambridge Monographs on Mathematical Physics\ (\bibinfo  {publisher}
  {Cambridge University Press},\ \bibinfo {year} {2011})\BibitemShut {NoStop}%
\bibitem [{\citenamefont {Bellac}(2011)}]{Bellac:2011kqa}%
  \BibitemOpen
  \bibfield  {author} {\bibinfo {author} {\bibfnamefont {M.~L.}\ \bibnamefont
  {Bellac}},\ }\href {\doibase 10.1017/CBO9780511721700} {\emph {\bibinfo
  {title} {{Thermal Field Theory}}}},\ Cambridge Monographs on Mathematical
  Physics\ (\bibinfo  {publisher} {Cambridge University Press},\ \bibinfo
  {year} {2011})\BibitemShut {NoStop}%
\bibitem [{\citenamefont {Akrami}\ \emph {et~al.}(2020)\citenamefont {Akrami}
  \emph {et~al.}}]{Planck:2018jri}%
  \BibitemOpen
  \bibfield  {author} {\bibinfo {author} {\bibfnamefont {Y.}~\bibnamefont
  {Akrami}} \emph {et~al.} (\bibinfo {collaboration} {Planck}),\ }\href
  {\doibase 10.1051/0004-6361/201833887} {\bibfield  {journal} {\bibinfo
  {journal} {Astron. Astrophys.}\ }\textbf {\bibinfo {volume} {641}},\ \bibinfo
  {pages} {A10} (\bibinfo {year} {2020})},\ \Eprint
  {http://arxiv.org/abs/1807.06211} {arXiv:1807.06211 [astro-ph.CO]}
  \BibitemShut {NoStop}%
\bibitem [{\citenamefont {Bolz}\ \emph {et~al.}(2001)\citenamefont {Bolz},
  \citenamefont {Brandenburg},\ and\ \citenamefont {Buchmuller}}]{Bolz:2000fu}%
  \BibitemOpen
  \bibfield  {author} {\bibinfo {author} {\bibfnamefont {M.}~\bibnamefont
  {Bolz}}, \bibinfo {author} {\bibfnamefont {A.}~\bibnamefont {Brandenburg}}, \
  and\ \bibinfo {author} {\bibfnamefont {W.}~\bibnamefont {Buchmuller}},\
  }\href {\doibase 10.1016/S0550-3213(01)00132-8} {\bibfield  {journal}
  {\bibinfo  {journal} {Nucl. Phys. B}\ }\textbf {\bibinfo {volume} {606}},\
  \bibinfo {pages} {518} (\bibinfo {year} {2001})},\ \bibinfo {note} {[Erratum:
  Nucl.Phys.B 790, 336--337 (2008)]},\ \Eprint
  {http://arxiv.org/abs/hep-ph/0012052} {arXiv:hep-ph/0012052} \BibitemShut
  {NoStop}%
\bibitem [{\citenamefont {Cadamuro}\ and\ \citenamefont
  {Redondo}(2012)}]{Cadamuro:2011fd}%
  \BibitemOpen
  \bibfield  {author} {\bibinfo {author} {\bibfnamefont {D.}~\bibnamefont
  {Cadamuro}}\ and\ \bibinfo {author} {\bibfnamefont {J.}~\bibnamefont
  {Redondo}},\ }\href {\doibase 10.1088/1475-7516/2012/02/032} {\bibfield
  {journal} {\bibinfo  {journal} {JCAP}\ }\textbf {\bibinfo {volume} {02}},\
  \bibinfo {pages} {032} (\bibinfo {year} {2012})},\ \Eprint
  {http://arxiv.org/abs/1110.2895} {arXiv:1110.2895 [hep-ph]} \BibitemShut
  {NoStop}%
\bibitem [{\citenamefont {Heisenberg}\ and\ \citenamefont
  {Euler}(1936)}]{Heisenberg:1936nmg}%
  \BibitemOpen
  \bibfield  {author} {\bibinfo {author} {\bibfnamefont {W.}~\bibnamefont
  {Heisenberg}}\ and\ \bibinfo {author} {\bibfnamefont {H.}~\bibnamefont
  {Euler}},\ }\href {\doibase 10.1007/BF01343663} {\bibfield  {journal}
  {\bibinfo  {journal} {Z. Phys.}\ }\textbf {\bibinfo {volume} {98}},\ \bibinfo
  {pages} {714} (\bibinfo {year} {1936})},\ \Eprint
  {http://arxiv.org/abs/physics/0605038} {arXiv:physics/0605038} \BibitemShut
  {NoStop}%
\bibitem [{\citenamefont {Schwinger}(1951)}]{Schwinger:1951nm}%
  \BibitemOpen
  \bibfield  {author} {\bibinfo {author} {\bibfnamefont {J.~S.}\ \bibnamefont
  {Schwinger}},\ }\href {\doibase 10.1103/PhysRev.82.664} {\bibfield  {journal}
  {\bibinfo  {journal} {Phys. Rev.}\ }\textbf {\bibinfo {volume} {82}},\
  \bibinfo {pages} {664} (\bibinfo {year} {1951})}\BibitemShut {NoStop}%
\bibitem [{\citenamefont {Kobayashi}\ and\ \citenamefont
  {Afshordi}(2014)}]{Kobayashi:2014zza}%
  \BibitemOpen
  \bibfield  {author} {\bibinfo {author} {\bibfnamefont {T.}~\bibnamefont
  {Kobayashi}}\ and\ \bibinfo {author} {\bibfnamefont {N.}~\bibnamefont
  {Afshordi}},\ }\href {\doibase 10.1007/JHEP10(2014)166} {\bibfield  {journal}
  {\bibinfo  {journal} {JHEP}\ }\textbf {\bibinfo {volume} {10}},\ \bibinfo
  {pages} {166} (\bibinfo {year} {2014})},\ \Eprint
  {http://arxiv.org/abs/1408.4141} {arXiv:1408.4141 [hep-th]} \BibitemShut
  {NoStop}%
\bibitem [{\citenamefont {Hayashinaka}\ \emph {et~al.}(2016)\citenamefont
  {Hayashinaka}, \citenamefont {Fujita},\ and\ \citenamefont
  {Yokoyama}}]{Hayashinaka:2016qqn}%
  \BibitemOpen
  \bibfield  {author} {\bibinfo {author} {\bibfnamefont {T.}~\bibnamefont
  {Hayashinaka}}, \bibinfo {author} {\bibfnamefont {T.}~\bibnamefont {Fujita}},
  \ and\ \bibinfo {author} {\bibfnamefont {J.}~\bibnamefont {Yokoyama}},\
  }\href {\doibase 10.1088/1475-7516/2016/07/010} {\bibfield  {journal}
  {\bibinfo  {journal} {JCAP}\ }\textbf {\bibinfo {volume} {07}},\ \bibinfo
  {pages} {010} (\bibinfo {year} {2016})},\ \Eprint
  {http://arxiv.org/abs/1603.04165} {arXiv:1603.04165 [hep-th]} \BibitemShut
  {NoStop}%
\bibitem [{\citenamefont {Gould}\ and\ \citenamefont
  {Rajantie}(2017)}]{Gould:2017fve}%
  \BibitemOpen
  \bibfield  {author} {\bibinfo {author} {\bibfnamefont {O.}~\bibnamefont
  {Gould}}\ and\ \bibinfo {author} {\bibfnamefont {A.}~\bibnamefont
  {Rajantie}},\ }\href {\doibase 10.1103/PhysRevD.96.076002} {\bibfield
  {journal} {\bibinfo  {journal} {Phys. Rev. D}\ }\textbf {\bibinfo {volume}
  {96}},\ \bibinfo {pages} {076002} (\bibinfo {year} {2017})},\ \Eprint
  {http://arxiv.org/abs/1704.04801} {arXiv:1704.04801 [hep-th]} \BibitemShut
  {NoStop}%
\bibitem [{\citenamefont {Lozanov}\ \emph {et~al.}(2019)\citenamefont
  {Lozanov}, \citenamefont {Maleknejad},\ and\ \citenamefont
  {Komatsu}}]{Lozanov:2018kpk}%
  \BibitemOpen
  \bibfield  {author} {\bibinfo {author} {\bibfnamefont {K.~D.}\ \bibnamefont
  {Lozanov}}, \bibinfo {author} {\bibfnamefont {A.}~\bibnamefont {Maleknejad}},
  \ and\ \bibinfo {author} {\bibfnamefont {E.}~\bibnamefont {Komatsu}},\ }\href
  {\doibase 10.1007/JHEP02(2019)041} {\bibfield  {journal} {\bibinfo  {journal}
  {JHEP}\ }\textbf {\bibinfo {volume} {02}},\ \bibinfo {pages} {041} (\bibinfo
  {year} {2019})},\ \Eprint {http://arxiv.org/abs/1805.09318} {arXiv:1805.09318
  [hep-th]} \BibitemShut {NoStop}%
\bibitem [{\citenamefont {Domcke}\ and\ \citenamefont
  {Mukaida}(2018)}]{Domcke:2018eki}%
  \BibitemOpen
  \bibfield  {author} {\bibinfo {author} {\bibfnamefont {V.}~\bibnamefont
  {Domcke}}\ and\ \bibinfo {author} {\bibfnamefont {K.}~\bibnamefont
  {Mukaida}},\ }\href {\doibase 10.1088/1475-7516/2018/11/020} {\bibfield
  {journal} {\bibinfo  {journal} {JCAP}\ }\textbf {\bibinfo {volume} {11}},\
  \bibinfo {pages} {020} (\bibinfo {year} {2018})},\ \Eprint
  {http://arxiv.org/abs/1806.08769} {arXiv:1806.08769 [hep-ph]} \BibitemShut
  {NoStop}%
\bibitem [{\citenamefont {Domcke}\ \emph {et~al.}(2022)\citenamefont {Domcke},
  \citenamefont {Schmitz},\ and\ \citenamefont {You}}]{Domcke:2021yuz}%
  \BibitemOpen
  \bibfield  {author} {\bibinfo {author} {\bibfnamefont {V.}~\bibnamefont
  {Domcke}}, \bibinfo {author} {\bibfnamefont {K.}~\bibnamefont {Schmitz}}, \
  and\ \bibinfo {author} {\bibfnamefont {T.}~\bibnamefont {You}},\ }\href
  {\doibase 10.1007/JHEP07(2022)126} {\bibfield  {journal} {\bibinfo  {journal}
  {JHEP}\ }\textbf {\bibinfo {volume} {07}},\ \bibinfo {pages} {126} (\bibinfo
  {year} {2022})},\ \Eprint {http://arxiv.org/abs/2108.11295} {arXiv:2108.11295
  [hep-ph]} \BibitemShut {NoStop}%
\bibitem [{\citenamefont {Ando}\ and\ \citenamefont
  {Kusenko}(2010)}]{Ando:2010rb}%
  \BibitemOpen
  \bibfield  {author} {\bibinfo {author} {\bibfnamefont {S.}~\bibnamefont
  {Ando}}\ and\ \bibinfo {author} {\bibfnamefont {A.}~\bibnamefont {Kusenko}},\
  }\href {\doibase 10.1088/2041-8205/722/1/L39} {\bibfield  {journal} {\bibinfo
   {journal} {Astrophys. J. Lett.}\ }\textbf {\bibinfo {volume} {722}},\
  \bibinfo {pages} {L39} (\bibinfo {year} {2010})},\ \Eprint
  {http://arxiv.org/abs/1005.1924} {arXiv:1005.1924 [astro-ph.HE]} \BibitemShut
  {NoStop}%
\bibitem [{\citenamefont {Tavecchio}\ \emph {et~al.}(2010)\citenamefont
  {Tavecchio}, \citenamefont {Ghisellini}, \citenamefont {Foschini},
  \citenamefont {Bonnoli}, \citenamefont {Ghirlanda},\ and\ \citenamefont
  {Coppi}}]{Tavecchio:2010mk}%
  \BibitemOpen
  \bibfield  {author} {\bibinfo {author} {\bibfnamefont {F.}~\bibnamefont
  {Tavecchio}}, \bibinfo {author} {\bibfnamefont {G.}~\bibnamefont
  {Ghisellini}}, \bibinfo {author} {\bibfnamefont {L.}~\bibnamefont
  {Foschini}}, \bibinfo {author} {\bibfnamefont {G.}~\bibnamefont {Bonnoli}},
  \bibinfo {author} {\bibfnamefont {G.}~\bibnamefont {Ghirlanda}}, \ and\
  \bibinfo {author} {\bibfnamefont {P.}~\bibnamefont {Coppi}},\ }\href
  {\doibase 10.1111/j.1745-3933.2010.00884.x} {\bibfield  {journal} {\bibinfo
  {journal} {Mon. Not. Roy. Astron. Soc.}\ }\textbf {\bibinfo {volume} {406}},\
  \bibinfo {pages} {L70} (\bibinfo {year} {2010})},\ \Eprint
  {http://arxiv.org/abs/1004.1329} {arXiv:1004.1329 [astro-ph.CO]} \BibitemShut
  {NoStop}%
\bibitem [{\citenamefont {Neronov}\ and\ \citenamefont
  {Vovk}(2010)}]{doi:10.1126/science.1184192}%
  \BibitemOpen
  \bibfield  {author} {\bibinfo {author} {\bibfnamefont {A.}~\bibnamefont
  {Neronov}}\ and\ \bibinfo {author} {\bibfnamefont {I.}~\bibnamefont {Vovk}},\
  }\href {\doibase 10.1126/science.1184192} {\bibfield  {journal} {\bibinfo
  {journal} {Science}\ }\textbf {\bibinfo {volume} {328}},\ \bibinfo {pages}
  {73} (\bibinfo {year} {2010})},\ \Eprint
  {http://arxiv.org/abs/https://www.science.org/doi/pdf/10.1126/science.1184192}
  {https://www.science.org/doi/pdf/10.1126/science.1184192} \BibitemShut
  {NoStop}%
\bibitem [{\citenamefont {Essey}\ \emph {et~al.}(2011)\citenamefont {Essey},
  \citenamefont {Ando},\ and\ \citenamefont {Kusenko}}]{Essey:2010nd}%
  \BibitemOpen
  \bibfield  {author} {\bibinfo {author} {\bibfnamefont {W.}~\bibnamefont
  {Essey}}, \bibinfo {author} {\bibfnamefont {S.}~\bibnamefont {Ando}}, \ and\
  \bibinfo {author} {\bibfnamefont {A.}~\bibnamefont {Kusenko}},\ }\href
  {\doibase 10.1016/j.astropartphys.2011.06.010} {\bibfield  {journal}
  {\bibinfo  {journal} {Astropart. Phys.}\ }\textbf {\bibinfo {volume} {35}},\
  \bibinfo {pages} {135} (\bibinfo {year} {2011})},\ \Eprint
  {http://arxiv.org/abs/1012.5313} {arXiv:1012.5313 [astro-ph.HE]} \BibitemShut
  {NoStop}%
\bibitem [{\citenamefont {Chen}\ \emph {et~al.}(2015)\citenamefont {Chen},
  \citenamefont {Buckley},\ and\ \citenamefont {Ferrer}}]{Chen:2014rsa}%
  \BibitemOpen
  \bibfield  {author} {\bibinfo {author} {\bibfnamefont {W.}~\bibnamefont
  {Chen}}, \bibinfo {author} {\bibfnamefont {J.~H.}\ \bibnamefont {Buckley}}, \
  and\ \bibinfo {author} {\bibfnamefont {F.}~\bibnamefont {Ferrer}},\ }\href
  {\doibase 10.1103/PhysRevLett.115.211103} {\bibfield  {journal} {\bibinfo
  {journal} {Phys. Rev. Lett.}\ }\textbf {\bibinfo {volume} {115}},\ \bibinfo
  {pages} {211103} (\bibinfo {year} {2015})},\ \Eprint
  {http://arxiv.org/abs/1410.7717} {arXiv:1410.7717 [astro-ph.HE]} \BibitemShut
  {NoStop}%
\bibitem [{\citenamefont {Ackermann}\ \emph {et~al.}(2018)\citenamefont
  {Ackermann} \emph {et~al.}}]{Fermi-LAT:2018jdy}%
  \BibitemOpen
  \bibfield  {author} {\bibinfo {author} {\bibfnamefont {M.}~\bibnamefont
  {Ackermann}} \emph {et~al.} (\bibinfo {collaboration} {Fermi-LAT}),\ }\href
  {\doibase 10.3847/1538-4365/aacdf7} {\bibfield  {journal} {\bibinfo
  {journal} {Astrophys. J. Suppl.}\ }\textbf {\bibinfo {volume} {237}},\
  \bibinfo {pages} {32} (\bibinfo {year} {2018})},\ \Eprint
  {http://arxiv.org/abs/1804.08035} {arXiv:1804.08035 [astro-ph.HE]}
  \BibitemShut {NoStop}%
\bibitem [{\citenamefont {Bunch}\ and\ \citenamefont
  {Davies}(1978)}]{Bunch:1978yq}%
  \BibitemOpen
  \bibfield  {author} {\bibinfo {author} {\bibfnamefont {T.~S.}\ \bibnamefont
  {Bunch}}\ and\ \bibinfo {author} {\bibfnamefont {P.~C.~W.}\ \bibnamefont
  {Davies}},\ }\href {\doibase 10.1098/rspa.1978.0060} {\bibfield  {journal}
  {\bibinfo  {journal} {Proc. Roy. Soc. Lond. A}\ }\textbf {\bibinfo {volume}
  {360}},\ \bibinfo {pages} {117} (\bibinfo {year} {1978})}\BibitemShut
  {NoStop}%
\end{thebibliography}%
\end{document}